%% file: prd.tex
\documentclass[twocolumn,showpacs,aps,prd,floatfix,superscriptaddress]{revtex4}

\usepackage{graphicx} 
\usepackage{dcolumn} 
\usepackage{amsmath} 
\usepackage{epsfig} 
\usepackage{hhline} 
\usepackage{soul} 
\usepackage{tabularx,pifont,multirow} 
\usepackage{enumitem} 
\usepackage{colordvi} 
 
\RequirePackage{xspace} 
 
\newcommand{\BABARPubYear}    {08} 
\newcommand{\BABARPubNumber}  {006} 
\newcommand{\SLACPubNumber} {13209} 
\newcommand{\LANLNumber} {0804.2089}

\input babarsym 
\newcommand{\npa}       [1]  {\npBase\ A~{\bf #1}} 
 
\newcommand{\pvec}{{\bf p}}

\def\beq{\begin{equation}} 
\def\eeq{\end{equation}} 
\def\bea{\begin{eqnarray}} 
\def\eea{\end{eqnarray}} 
\def\bq{\begin{quote}} 
\def\eq{\end{quote}} 
\def\ben{\begin{enumerate}} 
\def\een{\end{enumerate}} 
\def\nn{\nonumber}

\newcommand{\phm}{\ensuremath{\phantom{-}}} 
 
\newcommand{\phz}{\ensuremath{\phantom{0}}} 

\newcommand{\real}{\Re} 
\newcommand{\imag}{\Im} 
\def\rms   {\ensuremath {r.m.s.}\xspace} 
\def\Dztilde   {\ensuremath {\tilde{D}^0}\xspace} 
\def\Dstarztilde   {\ensuremath {\tilde{D}^{\ast 0}}\xspace} 
\def\kspipi{\ensuremath{\KS \pip \pim}\xspace} 
\def\kspipib{\ensuremath{\KS \pim \pip}\xspace} 
\def\kskk{\ensuremath{\KS \Kp \Km}\xspace} 
 
\def\kshh{\ensuremath{\KS h^+ h^-}\xspace} 

\def\Dztokspipi{\ensuremath{\Dz \to \kspipi}\xspace} 
\def\Dztokspipib{\ensuremath{\Dzb \to \kspipib}\xspace} 
\def\Dztokskk{\ensuremath{\Dz \to \kskk}\xspace} 
 
\def\Dztokshh{\ensuremath{\Dz \to \kshh}\xspace} 

\def\Dztildetokspipi{\ensuremath{\Dztilde \to \kspipi}\xspace} 
\def\Dztildetokskk{\ensuremath{\Dztilde \to \kskk}\xspace} 
\def\D       {\ensuremath{D}\xspace} 
\def\K  {\ensuremath{K}\xspace} 
\def\DzDstarz {\ensuremath {\D^{(\ast)0}}\xspace} 
\def\DzDstarzb {\ensuremath {\Db^{(\ast)0}}\xspace} 
\def\DzDstarztilde {\ensuremath {\tilde{\D}^{(\ast)0}}\xspace} 
\def\mtwo{{\ensuremath{\bf m}}\xspace} 
\def\mtwoj{{\ensuremath{{\bf m}_j}}\xspace} 
\def\mtwob{{\ensuremath{\bf \overline m}}\xspace} 
\def \rb {\ensuremath {r_\B}\xspace} 
\def \rbst {\ensuremath {r^\ast_\B}\xspace} 
\def \rbrbst {\ensuremath {r^{(\ast)}_\B}\xspace} 
\def \rs {\ensuremath {r_s}\xspace} 
\def \rstwo {\ensuremath {r_s^2}\xspace} 
\def \krs {\ensuremath {\kappa \rs}\xspace} 
\def \deltab {\ensuremath {\delta_\B}\xspace} 
\def \deltabst {\ensuremath {\delta^\ast_\B}\xspace} 
\def \deltabdeltabst {\ensuremath {\delta^{(\ast)}_\B}\xspace} 
\def \deltas {\ensuremath {\delta_s}\xspace} 
\def \rbm {\ensuremath {r_{\Bm}}\xspace} 
\def \rbp {\ensuremath {r_{\Bp}}\xspace}

\def \xbp {\ensuremath {x_+}\xspace} 
\def \xbm {\ensuremath {x_-}\xspace} 
\def \xbpm {\ensuremath {x_\pm}\xspace} 
\def \xbmp {\ensuremath {x_\mp}\xspace} 
\def \ybp {\ensuremath {y_+}\xspace} 
\def \ybm {\ensuremath {y_-}\xspace} 
\def \ybpm {\ensuremath {y_\pm}\xspace} 
\def \ybmp {\ensuremath {y_\mp}\xspace} 
\def \xbstp {\ensuremath {x_+^\ast}\xspace} 
\def \xbstm {\ensuremath {x_-^\ast}\xspace} 
 
\def \xbstmp {\ensuremath {x_\mp^\ast}\xspace} 
\def \ybstp {\ensuremath {y_+^\ast}\xspace} 
\def \ybstm {\ensuremath {y_-^\ast}\xspace} 
 
\def \ybstmp {\ensuremath {y_\mp^\ast}\xspace} 

\def \xbxbstmp {\ensuremath {x_\mp^{(\ast)}}\xspace}

\def \ybybstmp {\ensuremath {y_\mp^{(\ast)}}\xspace} 
\def \xsp {\ensuremath {x_{s+}}\xspace} 
\def \xsm {\ensuremath {x_{s-}}\xspace} 
 
\def \xsmp {\ensuremath {x_{s\mp}}\xspace} 
\def \xsmptwo {\ensuremath {x_{s\mp}^2}\xspace} 
\def \ysp {\ensuremath {y_{s+}}\xspace} 
\def \ysm {\ensuremath {y_{s-}}\xspace} 
 
\def \ysmp {\ensuremath {y_{s\mp}}\xspace} 
\def \ysmptwo {\ensuremath {y_{s\mp}^2}\xspace} 
\def \z {\ensuremath {{\bf{z}}}\xspace} 
\def \zbest {\ensuremath {{\bf{z_{\rm best}}}}\xspace} 
\def \zp {\ensuremath {{\bf{z}_+}}\xspace} 
\def \zm {\ensuremath {{\bf{z}_-}}\xspace} 
\def \zpm {\ensuremath {{\bf{z}_\pm}}\xspace} 
\def \zmp {\ensuremath {{\bf{z}_\mp}}\xspace}

\def \pvec {\ensuremath{{\bf{p}}}\xspace} 
\def \pveczero {\ensuremath{{\bf{p_0}}}\xspace} 
\def \pvecbest {\ensuremath{{\bf{p_{\rm best}}}}\xspace} 
\def \qvec {\ensuremath{{\bf{q}}}\xspace} 
\def \qveczero {\ensuremath{{\bf{q_0}}}\xspace} 
\def \qvecbest {\ensuremath{{\bf{q_{\rm best}}}}\xspace} 
\def \qvecprimezero {\ensuremath{{\bf{q'_0}}}\xspace} 
\def \qvecprimebest {\ensuremath{{\bf{q'_{\rm best}}}}\xspace} 
\def\fisher    {\ensuremath{{\cal F}}\xspace} 
\def\mD    {\ensuremath{m_\D}\xspace} 
\def\deltam    {\ensuremath{\Delta m}\xspace} 
\newcommand{\kevc}{\ensuremath{{\mathrm{\,ke\kern -0.1em V\!/}c}}\xspace} 
\newcommand{\kevcc}{\ensuremath{{\mathrm{\,ke\kern -0.1em V\!/}c^2}}\xspace} 
\long\def\inst#1{\par\nobreak\kern 4pt\nobreak 
  {\it #1}\par\vskip 10pt plus 3pt minus 3pt} 
 
\begin{document} 
 
\begin{flushleft} 
\babar-PUB-\BABARPubYear/\BABARPubNumber \\ 
SLAC-PUB-\SLACPubNumber \\ 
\LANLNumber~[hep-ex]\\ 
\end{flushleft}

\title{\large \bf 
\boldmath 
\Large 
Improved measurement of the CKM angle $\gamma$  
in $\Bmp \to D^{(*)} K^{(*)\mp}$ decays  
with a Dalitz plot analysis of $D$ decays to $\kspipi$ and $\kskk$ 
} 
 
\input authors_feb2008.tex

\date{\today} 
 
\begin{abstract} 
\input abstract.tex 
\end{abstract} 
 
 
\maketitle 
 
\newpage

\setcounter{footnote}{0}

\section{ \boldmath Introduction and overview} 
\label{sec:intro} 
\input intro

\section{ \boldmath Event selection} 
\label{sec:selection} 
\input selection

\section{ \boldmath \Dztokspipi and \Dztokskk amplitudes} 
\label{sec:dalitzmodels} 
\input dalitzmodels

\section{ \boldmath Dalitz plot analysis of $\Bm \to D^{(*)} \Km$ and $\Bm \to D \Kstarm$ decays } 
\label{sec:cpanalysis} 
\input cpanalysis

\section{ \boldmath Interpretation of results} 
\label{sec:interpretation} 
\input interpretation

\section{ \boldmath Conclusion} 
\label{sec:conclusion} 
\input conclusion

\section{ \boldmath Acknowledgments} 
\label{sec:acknowl} 
\input acknowledgements

\appendix 
\section{\boldmath Correlation matrices} 
\label{sec:covariance} 
\input covariance.tex

\input biblio

\end{document}

%% file: authors_feb2008.tex
%
\author{B.~Aubert} 
\author{M.~Bona} 
\author{Y.~Karyotakis} 
\author{J.~P.~Lees} 
\author{V.~Poireau} 
\author{E.~Prencipe} 
\author{X.~Prudent} 
\author{V.~Tisserand} 
\affiliation{Laboratoire de Physique des Particules, IN2P3/CNRS et Universit\'e de Savoie, F-74941 Annecy-Le-Vieux, France } 
\author{J.~Garra~Tico} 
\author{E.~Grauges} 
\affiliation{Universitat de Barcelona, Facultat de Fisica, Departament ECM, E-08028 Barcelona, Spain } 
% ...
\collaboration{The \babar\ Collaboration} 
\noaffiliation

%% file: abstract.tex
 
\noindent 
We report on an improved measurement of the Cabibbo-Kobayashi-Maskawa \CP-violating phase $\gamma$ through a Dalitz plot analysis 
of neutral \D meson decays to \kspipi and \kskk produced in the processes $\Bmp \to \D \Kmp$, 
$\Bmp \to \D^{*} \Kmp$ with $\Dstar \to D\piz,D\g$, and $\Bmp \to \D\Kstarmp$ with $\Kstarmp \to \KS \pimp$. 
Using a sample of 383 million \BB pairs collected by the \babar\ detector, 
we measure 
$\gamma=(76 \pm 22 \pm 5 \pm 5)^\circ$ (\mbox{mod $180^\circ$}), 
where the first error is statistical, the second is the experimental 
systematic uncertainty and the third reflects the uncertainty on  
the description of the Dalitz plot distributions. 
The corresponding two standard deviation region is $29^\circ < \gamma < 122^\circ$. 
This result has a significance of direct \CP violation ($\gamma \ne 0$) of $3.0$ standard deviations. 

%% file: intro.tex
In the Standard Model (SM) the phase in the Cabibbo-Kobayashi-Maskawa (CKM) quark-mixing matrix~\cite{ref:CKM} 
is the sole source of \CP violation in  
the quark sector of the electroweak interactions. This phase can be directly determined using a variety of methods, involving 
either the interference between decays with and without mixing in time-dependent \CP asymmetries in neutral \B meson decays,  
or interference between neutral \B (self-tagged) or charged \B decays yielding the same final state (direct \CP violation). 
These multiple determinations in \CP-violating tree-level processes  
as well as  
in decays involving penguin diagrams 
test the CKM mechanism, thus probing the presence of physics beyond the SM~\cite{ref:globalCKMfits}. 
 
\begin{figure}[htb!] 
\includegraphics[width=0.23\textwidth]{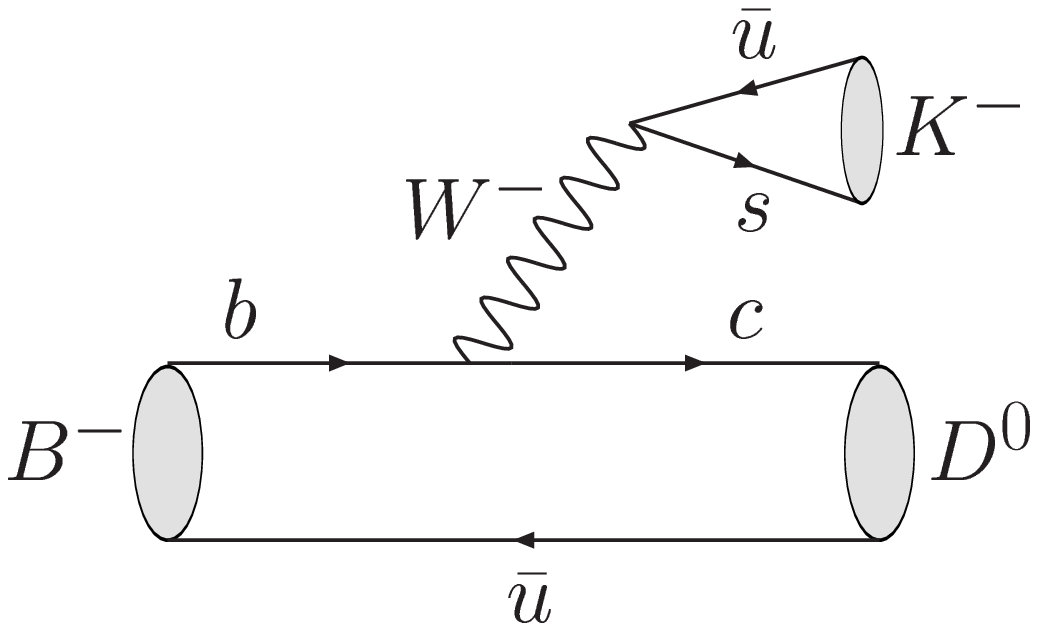} 
\includegraphics[width=0.23\textwidth]{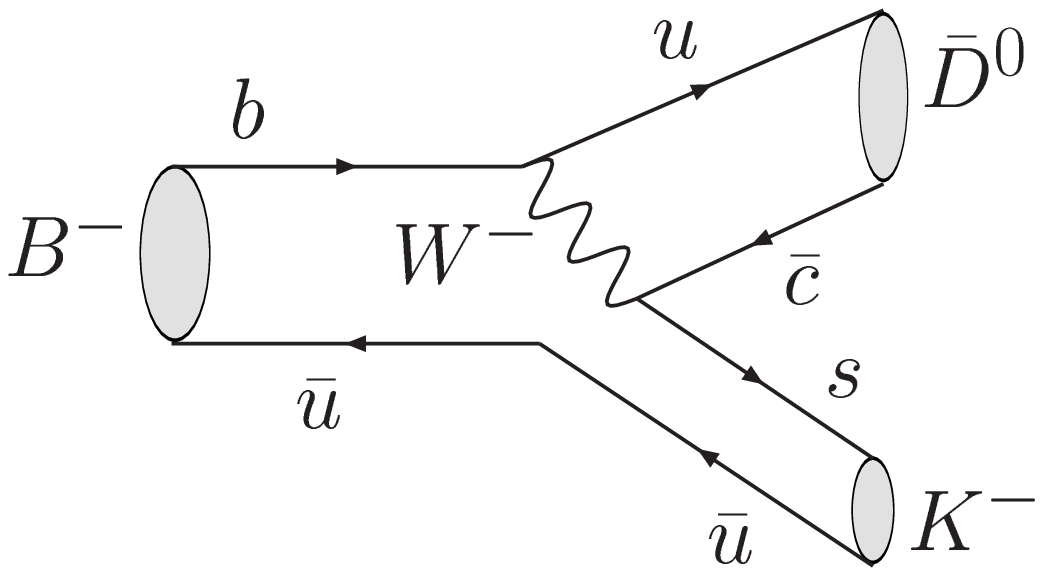} 
    \caption{\label{fig:DK_diag} Main Feynman diagrams contributing to the $\Bm \to \Dztilde \Km$ decay. The  
      left diagram proceeds via $\b \to \c\ubar\s$ transition, while the right diagram proceeds via  
       $\b \to \u \cbar \s$ transition and is color suppressed.} 
  \end{figure}

Among these determinations, the measurement of the angle $\gamma$, defined as $\arg{\left[-V_{ud}^{}V_{ub}^{*}/V_{cd}^{}V_{cb}^{*}\,\right]}$, 
where $V_{ij}$ are the elements of the CKM matrix, 
is one of the most difficult to achieve  
and constitutes an important goal of present and future \B physics experiments.  
Several methods have been proposed to extract $\gamma$. However, 
those using $B^\mp\to\DzDstarztilde K^{(*)\mp}$ 
decays~\cite{ref:symbol_star,ref:Kstar} (the symbol \DzDstarztilde indicates either a \DzDstarz or a \DzDstarzb meson)  
are theoretically clean 
and are unlikely to be affected by new physics because the main contributions to the amplitudes 
come from tree-level diagrams, as shown in Fig.~\ref{fig:DK_diag}. 
This is an important distinction from most of other direct measurements of phases of CKM elements. 
The decay amplitudes for the color allowed $\Bm \to \DzDstarz K^{(*)-}$ ($\b \to \c\ubar\s$)  
and the color suppressed $\Bm\to \DzDstarzb K^{(*)-}$ ($\b \to \u \cbar \s$)  
transitions~\cite{ref:chargeconj} differ by a factor $\rbrbst e^{i(\deltabdeltabst \mp \gamma)}$. 
Here, \rbrbst is the magnitude of the ratio of the amplitudes  
${\cal A}(\Bm \to \DzDstarzb \Km)$  
and 
${\cal A}(\Bm \to \DzDstarz \Km)$  
and \deltabdeltabst 
is their relative strong phase.  
The weak phase $\gamma$ leads to different \Bm and \Bp decay  
rates (direct \CP violation) and, when the \DzDstarz and \DzDstarzb decay to a common final  
state~\cite{ref:gronau,ref:soni,ref:ggsz_ads,ref:belle_dal04}, the phases become observable. 
The uncertainty in $\gamma$ scales roughly as $1/\rbrbst$. From the ratio of CKM matrix elements  
we expect $\rbrbst \approx c_F \mid V_{cs}^{}V_{ub}^{*} \mid / \mid V_{us}^{}V_{cb}^{*} \mid$ to be approximately in the range $0.1 - 0.2$,  
where $c_F\sim 0.2 - 0.4$ is the color suppression factor~\cite{ref:browder1996,ref:gronau2003}.

When the neutral \D meson is reconstructed in a three-body final state, like \kspipi, the distribution in the  
Dalitz plot~\cite{ref:dalitz} depends on the interference between  
Cabibbo allowed, doubly-Cabibbo suppressed, and \CP-eigenstate 
decay amplitudes of \Dz (from $\Bmp\to \DzDstarz K^{(*)\mp}$) and \Dzb (from $\Bmp \to \DzDstarzb K^{(*)\mp}$). 
The dominant interfering amplitudes in the \kspipi final state are 
$\Dz\to\Kstarm\pip$, $\Dz\to\Kstarp\pim$, and $\Dz\to\KS\rho(770)^0$~\cite{ref:chargeconj,ref:D0toKspipi-cleo}. 
Neglecting effects from $\Dz-\Dzb$ mixing and \CP asymmetries in neutral $D$ decays 
that are 1\% or less~\cite{ref:DmixingEvidence,ref:DmixingEffects,ref:CPVinDmixing}, 
the $\Bmp \to \DzDstarztilde \Kmp$, with $\Dstarztilde \rightarrow \Dztilde\piz,\Dztilde\gamma$, 
$\Dztilde \to \kspipi$ decay chain  
amplitude ${\cal A}^{(*)}_{\mp}(m^2_-,m^2_+)$ can be written as 
\bea 
{\cal A}^{(*)}_{\mp}(m^2_-,m^2_+) \propto {\cal A}_{D\mp} + \lambda \rbrbst e^{i(\deltabdeltabst \mp \gamma)}{\cal A}_{D\pm}~, 
\label{eq:ampB} 
\eea 
where $m^2_-$ and $m^2_+$ are the squared invariant masses of the $\KS\pim$ and $\KS\pip$ combinations, respectively, 
and ${\cal A}_{D\mp} \equiv {\cal A}_{D}(m^2_\mp,m^2_\pm)$, 
with ${\cal A}_{D-}$ (${\cal A}_{D+}$) the amplitude of the \Dztokspipi (\Dztokspipib) decay. 
For convenience, $m^2_0$ is defined analogously for the $\pip\pim$ combination. 
The factor $\lambda$ in Eq.~(\ref{eq:ampB}) takes the value $-1$ 
for the decay $\Bmp \to \Dstarztilde[\Dztilde\gamma] \Kmp$ and $+1$ for the remaining \B decays. 
This relative sign arises due to charge conjugation and angular momentum conservation in the \Dstarztilde decay~\cite{ref:bondar_gershon}. 
The corresponding decay rate $\Gamma^{(*)}_{\mp}(m^2_-,m^2_+)$ can therefore be written as 
\bea 
\Gamma_{\mp}^{(*)}(m^2_-,m^2_+) \propto |{\cal A}_{D\mp}|^2+{\rbrbst}^2 |{\cal A}_{D\pm}|^2 + \ \ \ \ \ \ \ \ \ \ \ \ \ \ \ \ \nonumber  \\  
   2 \lambda \left[ \xbxbstmp \real \{ {\cal A}_{D\mp} {\cal A}^*_{D_\pm} \} +  
                    \ybybstmp \imag \{ {\cal A}_{D\mp} {\cal A}^*_{D\pm}  \} \right],~ 
\label{eq:ampgen} 
\eea 
where we introduce the \CP parameters~\cite{ref:babar_dalitzpub} $\xbxbstmp = \rbrbst \cos(\deltabdeltabst \mp \gamma)$  
and  
$\ybybstmp = \rbrbst \sin(\deltabdeltabst \mp \gamma)$,  
where 
${\xbxbstmp}^2 + {\ybybstmp}^2 = {\rbrbst}^2$. 
 
Decays $\Bmp \to \Dztilde \Kstarmp$ with $\Kstarmp\to\KS\pi^\mp$ ~\cite{ref:Kstar} are also used 
in this analysis. For these, Eq.~(\ref{eq:ampgen}) requires the replacements  
$\Gamma_{\mp}^{(*)} \to \Gamma_{s\mp}$, 
$\rbrbst \to \rs$, $\deltabdeltabst \to \deltas$, 
$\xbxbstmp \to \xsmp = \kappa \rs \cos(\deltas \mp \gamma)$, and 
$\ybybstmp \to \ysmp = \kappa \rs \sin(\deltas \mp \gamma)$,  
where 
$\xsmptwo + \ysmptwo = \kappa^2 \rstwo$, where~\cite{ref:gronau2003} 
\bea 
\rstwo = \frac{\int A^2_u(p){\rm d}p}{\int A^2_c(p){\rm d}p} &,& 
\kappa e^{i\deltas} = \frac{\int A_c(p) A_u(p) e^{i\delta(p)}{\rm d}p}{\sqrt{\int A^2_c(p){\rm d}p \int A^2_u(p){\rm d}p}}.~~~~~~ 
\label{eq:kappa} 
\eea 
Here, $A_c(p)$ and $A_u(p)$ are the magnitudes of the $\b\to\c$ and $\b\to\u$ amplitudes as a function of  
the $\Bmp \to \Dztilde \KS \pimp$ phase space position $p$, and $\delta(p)$ is the relative strong phase. 
The parameter $\kappa$ accounts for the interference between $\Bmp \to\Dztilde \Kstarmp$ and other $\Bmp \to\Dztilde \KS\pi^\mp$ amplitudes 
with $0<\kappa<1$ in the most general case. 
This effective parameterization also accounts for efficiency variations as a function of the kinematics of the \B decay.

In this paper we present an improved measurement of $\gamma$ based on the analysis of the Dalitz plot distribution 
of \Dztildetokspipi and, for the first time, \Dztildetokskk,  
using a sample of 351 \invfb of integrated luminosity recorded at the \FourS resonance,  
corresponding to $383$ million \BB pairs. 
We analyze a total of seven signal samples (also referred to as \CP samples), 
$\Bm \to \DzDstarztilde \Km$ and $\Bm \to \Dztilde \Kstarm$, 
with $\Dstarztilde \rightarrow \Dztilde\piz,\Dztilde\gamma$, $\Dztildetokspipi,\kskk$, $\Kstarm \to \KS\pim$~\cite{ref:chargeconj}. 
Due to lack of statistics, the decay $\Bm \to \Dztilde \Kstarm$ with \Dztildetokskk has been excluded from the analysis. 
We also reconstruct high statistics control samples, one for each signal \B decay channel: 
$\Bm \to \DzDstarz \pim$ and (for $\Bm \to \Dztilde \Kstarm$) $\Bm \to \Dz a_1^-$ with $a_1^-\to\pi^-\pi^+\pi^-$~\cite{ref:chargeconj,ref:a1}. 
We exploit the same dataset to determine ${\cal A}_{D\mp}$ for \Dztokspipi and \Dztokskk decays 
from analyses of the respective Dalitz plots for high-statistics 
samples of flavor-tagged \Dz mesons from $\Dstarp \to \Dz \pip$ decays~\cite{ref:chargeconj},  
produced in $e^+ e^- \to \c\cbar$ events.   
Additional improvements compared to our previous publication~\cite{ref:babar_dalitzpub} include a higher  
reconstruction efficiency, an optimized treatment of the  
$e^+e^-\to \q\qbar$, $\q=\u,\d,\s,\c$ background and an improved \Dztokspipi description of the  
Dalitz plot distribution (referred to hereafter as Dalitz model), resulting in  
significant decrease of statistical, systematic and model uncertainties. 
This measurement supersedes our previous result based on 227 million \BB pairs~\cite{ref:babar_dalitzpub}.  
 
The paper is organized as follows. In Sec.~\ref{sec:selection} we describe the reconstruction and selection of the  
signal and control samples. Sec.~\ref{sec:dalitzmodels} is devoted to the determination 
of ${\cal A}_{D\mp}$ 
for \Dztokspipi and \Dztokskk decays. 
In Sec.~\ref{sec:cpanalysis} we describe the simultaneous maximum likelihood fit to the distributions  
$\Gamma_{-}^{(*)}$ and $\Gamma_{+}^{(*)}$ for the $\Bmp \to \DzDstarztilde \Kmp$ samples and to the analogous distributions  
for $\Bmp \to \Dztilde \Kstarmp$, to determine the \CP parameters \xbxbstmp, \ybybstmp, \xsmp, and \ysmp. 
In that section we also present the experimental results, including systematic uncertainties. 
We extract these \CP parameters since they have a good Gaussian behavior for small values of \rbrbst,\krs and relatively low statistics samples,  
independent of their values and precisions, in contrast to $\gamma$, \rbrbst, \deltabdeltabst, \krs, and \deltas. 
Finally, in Sec.~\ref{sec:interpretation}, we interpret the experimental results and extract the physically  
relevant quantities $\gamma$, \rbrbst, \deltabdeltabst, \krs and \deltas, using a statistical (frequentist) analysis.

%% file: selection.tex
\subsection{\babar~detector} 
 
This analysis is based on a data sample collected by the \babar\ 
detector at the Stanford Linear Accelerator Center \pep2 $e^+e^-$ asymmetric-energy storage ring.  
The \babar\ detector is described in detail elsewhere~\cite{ref:detector}.  
We summarize briefly the components that are crucial to this analysis. Charged-particle tracking is provided by  
a five-layer silicon vertex tracker (SVT) and a 40-layer drift chamber (DCH). In addition to providing precise  
space coordinates for tracking, the SVT and DCH also measure the specific ionization ($dE/dx$),  
which is used for particle identification of low-momentum charged particles. At higher momenta ($p>0.7$~\gevc) 
pions and kaons are identified by Cherenkov radiation detected in a ring-imaging 
device (DIRC). The typical separation between pions and kaons varies from 8$\sigma$ at 2~\gevc to 2.5$\sigma$ at 4~\gevc, 
where $\sigma$ denotes here the standard deviation. 
The position and energy of photons are 
measured with an electromagnetic calorimeter (EMC) consisting of 6580 thallium-doped CsI crystals. 
These systems are mounted inside a 1.5-T solenoidal super-conducting magnet. 
We use a GEANT4-based Monte Carlo (MC) simulation to model the response of the detector, taking into 
account the varying accelerator and detector conditions, and to generate large samples of signal and background  
for the \CP and control modes considered in the analysis.

\subsection{Event reconstruction and selection} 
 
The \Bm candidates are formed by combining a \DzDstarztilde candidate with a  
track identified as a kaon~\cite{ref:detector} or with a \Kstarm candidate formed as a  
combination of a \KS\ and a negatively charged pion, with an invariant mass within 55~\mevcc of the  
nominal \Kstarm mass. 
Here and in the following, nominal mass values are taken from~\cite{ref:pdg2006}. 
The \Dztilde candidates are selected by requiring  
the \kspipi or \kskk invariant mass to be within 12~\mevcc of the nominal \Dz mass, 
and the momentum in the center-of-mass (CM) frame to be greater than 1.3~\gevc. 
The \Kmp tracks in \Dztildetokskk are required to be positively identified as kaons in the DCH and DIRC. 
The \piz candidates from $\Dstarztilde\to \Dztilde\piz$ decays are formed from pairs of photons with invariant mass  
in the range $[115,150]$~\mevcc,  
and with photon energy greater than 30~\mev.  
Photon candidates from $\Dstarztilde\to \Dztilde\g$ decays are selected if their energy is greater than 100~\mev. 
The \Dztilde candidates are combined with a  
\piz (\g) to form the \Dstarztilde candidate,  
and are required to have a \Dstarztilde-\Dztilde mass difference within 2.5 (10) \mevcc of its nominal value. 
The \KS candidates are reconstructed from pairs of oppositely-charged pions constrained to originate from the same point and  
with an invariant mass within 9~\mevcc of the \KS nominal mass. 
The cosine of the collinearity angle  
between the \KS momentum and the line connecting its parent particle (the \Dztilde or the \Kstarm) 
and the \KS decay points in the plane transverse  
to the beam,  
is required to be larger than $0.990$ ($0.997$ for \KS from \Kstarm decays). 
This cut helps to significantly reduce background contributions from $\Dztilde\to\pi\pi\pi\pi$ decays and  
from $a_1^-$ misreconstructed as \Kstarm. 
The kinematic variables of the \DzDstarztilde, \KS, and \piz when forming the \Bm, \Dztilde, and \Dstarztilde, respectively,  
are fitted with masses constrained to nominal values. 
For $\Bm \to \Dztilde \Kstarm$ decays we also require $|\cos\theta_H|\ge 0.35$,  
where $\theta_H$ is the angle between the momentum of the \Kstarm daughter pion and the parent \Bm in the \Kstarm rest frame. 
The distribution of $\cos\theta_H$ is proportional to $\cos^2\theta_H$ for $\Bm\to \Dztilde \Kstarm$  
while it is approximately flat for $e^+e^-\to \q\qbar$, $\q=\u,\d,\s,\c$ (continuum) background. 
 
We characterize \B mesons using two almost independent variables, 
the beam-energy substituted mass,  
$\mes=\sqrt{(E^{*2}_0/2+{\bf p_0\cdot p_B})^2/E^2_0-p^2_B}$, 
and the energy difference $\Delta E=E^*_B-E^*_0/2$, with $p=(E,{\bf p})$, 
where the subscripts $0$ and $B$ refer to the initial $e^+e^-$ system and the \B candidate, 
respectively, and the asterisk denotes the CM frame.  
The signal events peak at the $B$ mass in \mes and at zero in \DeltaE.  
The \mes resolution is about 2.6~\mevcc and does not depend on the decay mode or on the nature  
of the prompt particle (\Km or \Kstarm candidate). 
In contrast, the \DeltaE resolution depends on the momentum resolution of the \DzDstarz meson and the prompt particle, 
and ranges between 15~\mev and 18~\mev, depending on the decay mode. 
We select events with $\mes>5.2\gevcc$ and $-80~\mev<\DeltaE<120~\mev$.  
We discriminate against the main background contribution coming from continuum events  
through the fit to the data, as described in Sec.~\ref{sec:step1fit}. 
 
For events in which multiple \B candidates satisfy the selection criteria, the one  
whose measured \Dztilde mass differs from the nominal value by the least number of standard deviations, 
is accepted as signal candidate. 
For $\Bm\to\Dztilde\Kstarm$ decays we select the candidate with the smallest value for the sum of the squares of the differences  
from nominal values, in standard deviations, of both \Kstarm and \Dztilde masses. 
The fraction of events in which we reconstruct more than one candidate is less than 1\% for 
$\Bm\to\DzDstarztilde\Km$ samples and about 6\% for $\Bm\to\Dztilde\Kstarm$. 
The cross-feed among the different samples is negligible except for $\Bm\to \Dstarztilde[\Dztilde\gamma]\Km$,  
where the background from $\Bm\to \Dstarztilde[\Dztilde\piz]\Km$ is below 5\% of the signal yield.  
If both $\Bm \to \Dstarztilde[\Dztilde\piz]\Km$ and $\Bm\to \Dstarztilde[\Dztilde\gamma]\Km$ candidates are  
selected in the same event, only the $\Bm\to \Dstarztilde[\Dztilde\piz]\Km$ is kept.  
This contamination has a negligible effect on the measurement of the \CP parameters. 
 
\begin{figure}[hbt!] 
\begin{tabular}{cc} 
\includegraphics[width=0.23\textwidth]{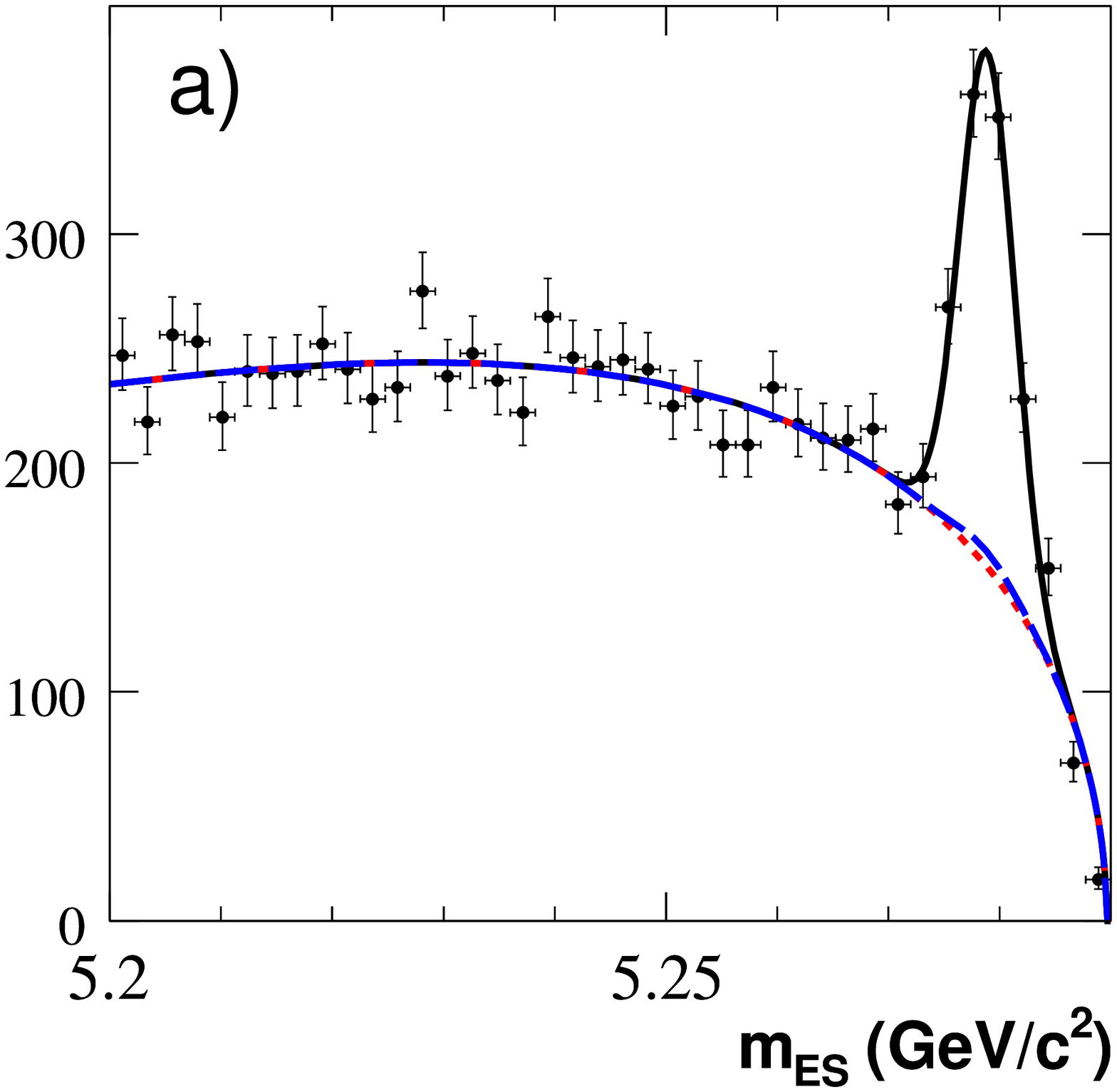} & 
\includegraphics[width=0.23\textwidth]{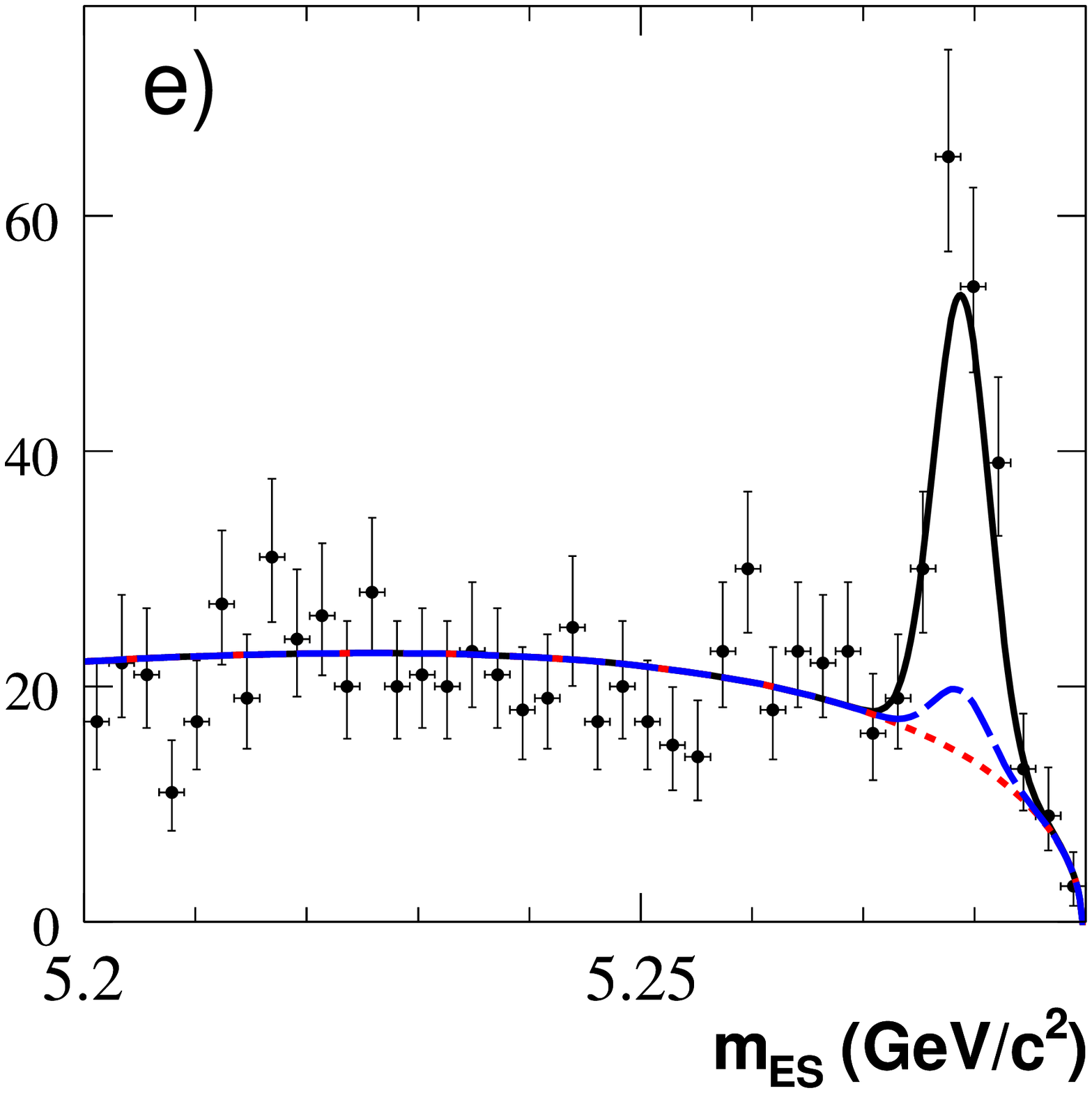} \\ 
\includegraphics[width=0.23\textwidth]{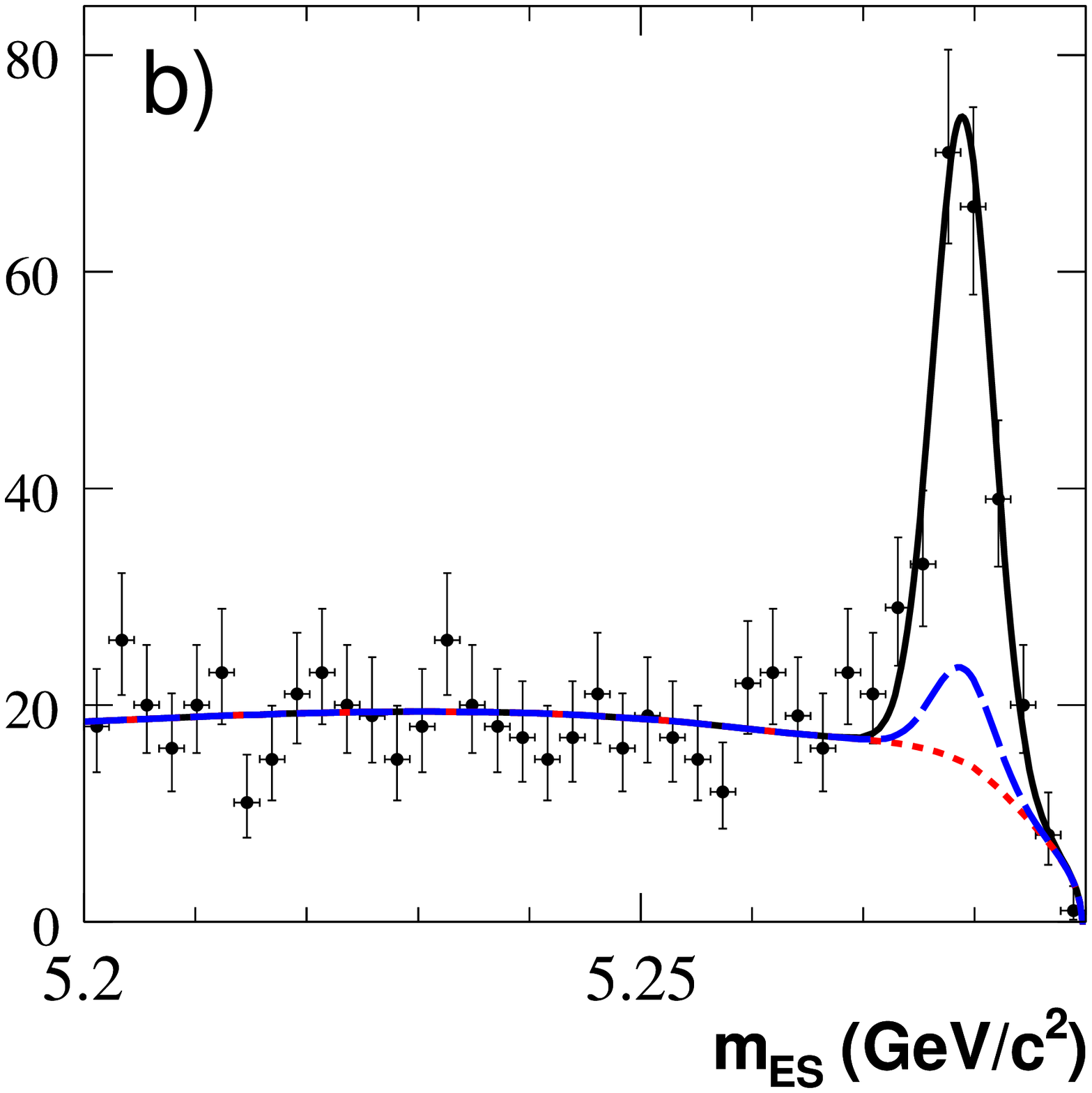} & 
\includegraphics[width=0.23\textwidth]{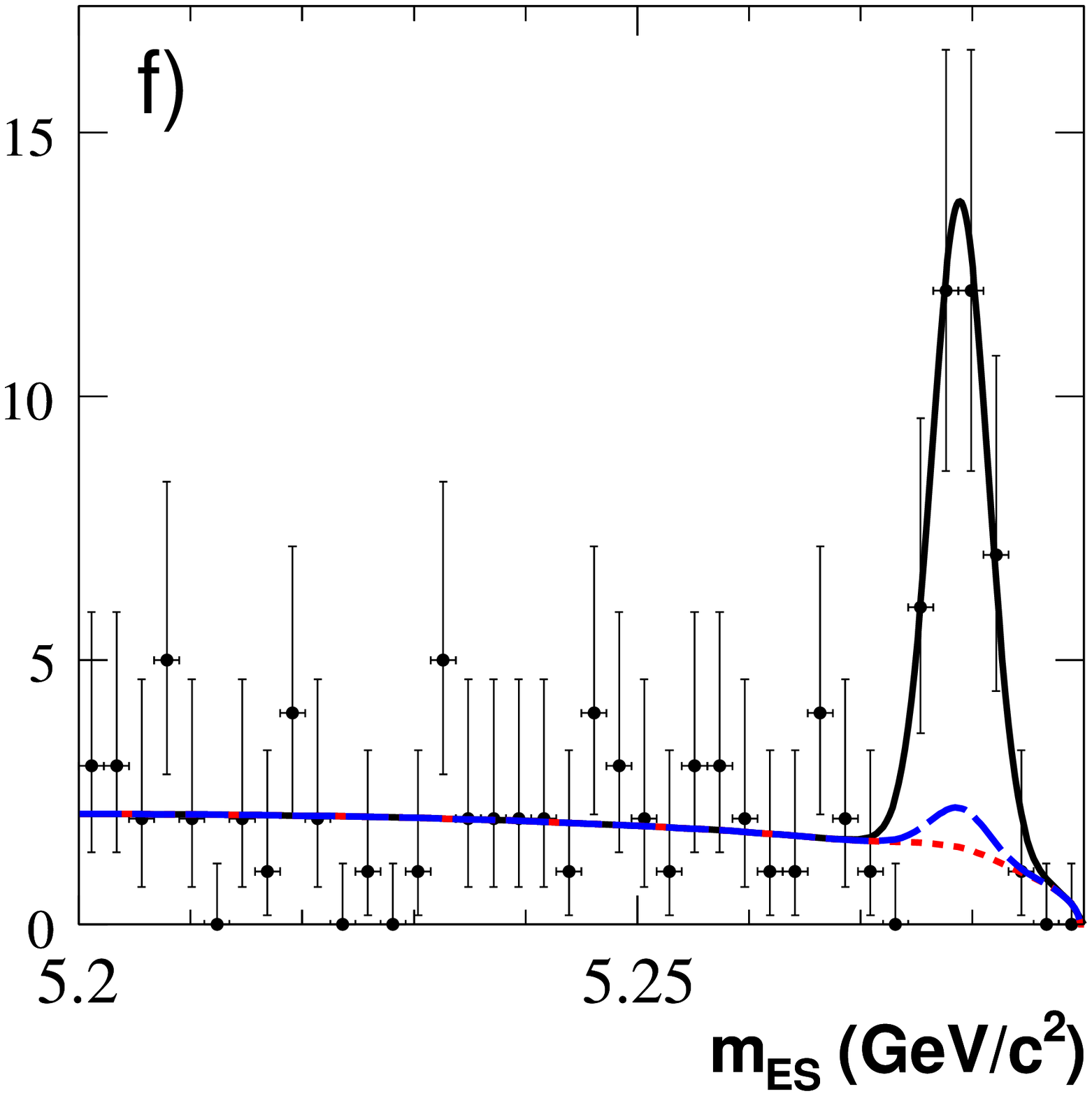} \\ 
\includegraphics[width=0.23\textwidth]{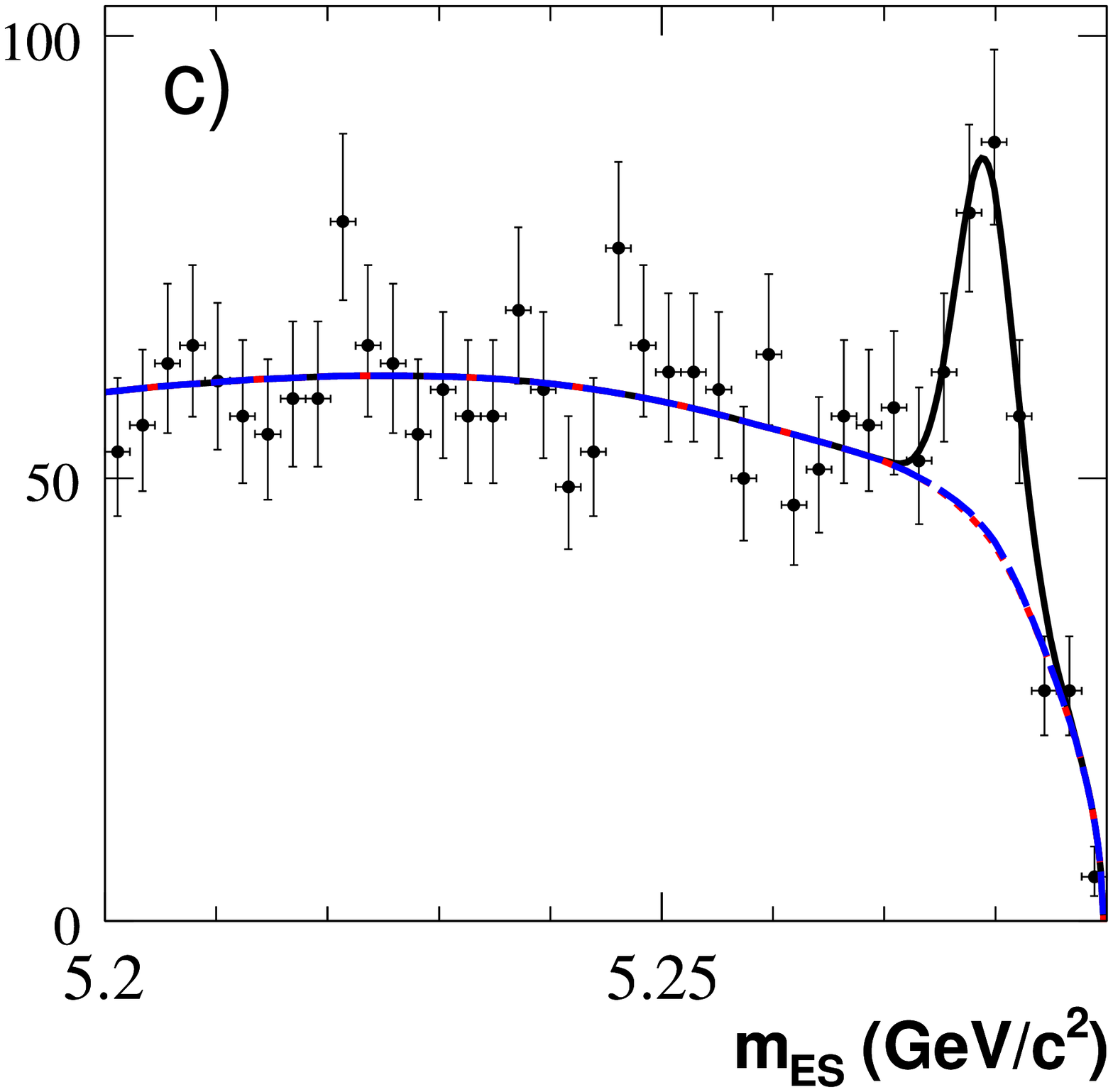} & 
\includegraphics[width=0.23\textwidth]{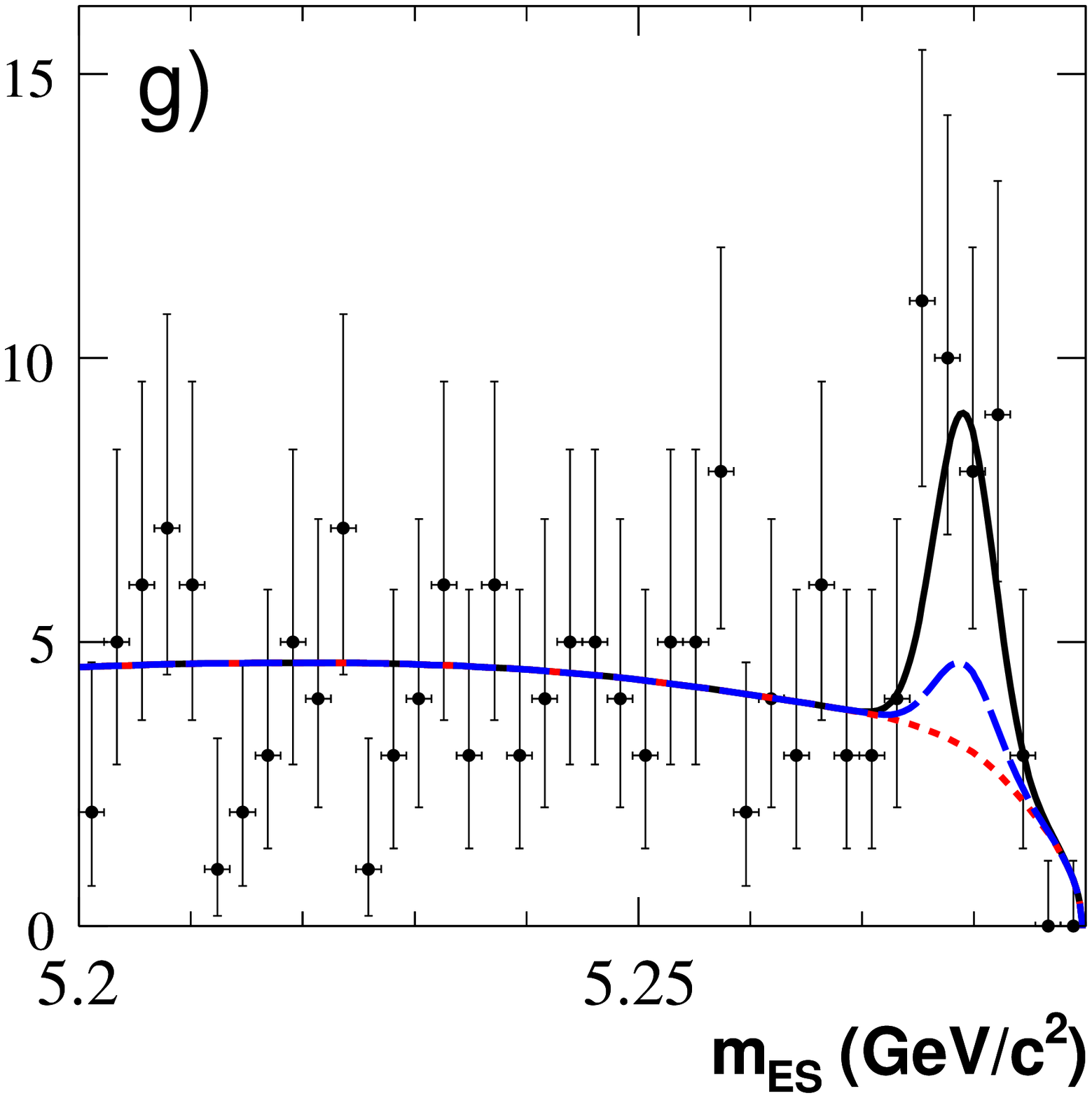} \\ 
\includegraphics[width=0.23\textwidth]{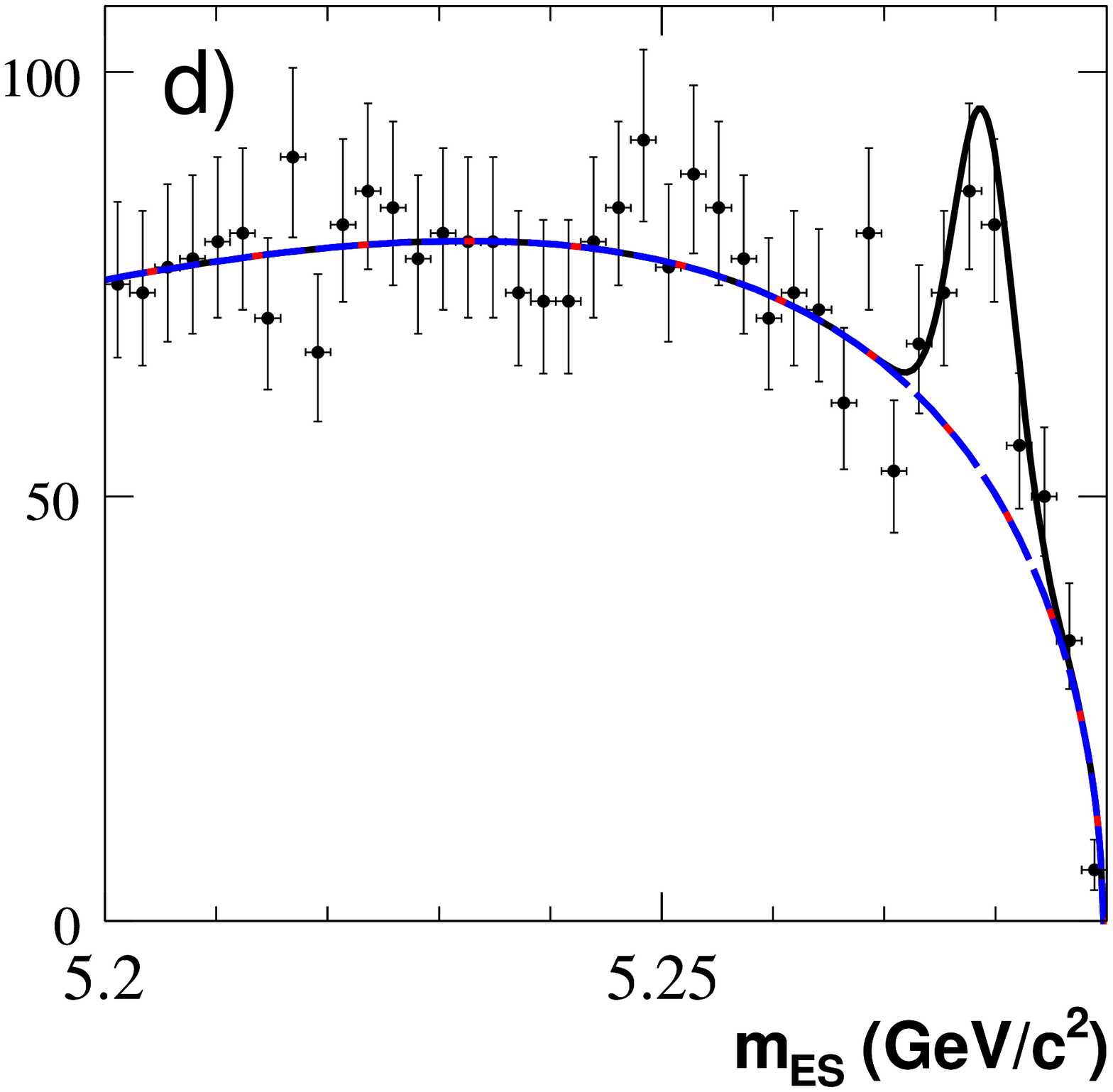} & 
\end{tabular} 
\caption{\label{fig:btodk_mes} (color online). Distributions of \mes\ in \DeltaE signal region for (a,e) $\Bm\to \Dztilde \Km$,  
(b,f) $\Bm\to \Dstarztilde[\Dztilde\piz] \Km$, (c,g) $\Bm\to \Dstarztilde[\Dztilde\gamma] \Km$, and (d) $\Bm\to \Dztilde \Kstarm$, 
with (a-d) \Dztildetokspipi and (e-g) \Dztildetokskk. 
The curves superimposed represent the projections of the fit described in Sec.~\ref{sec:step1fit}:  
signal plus background (solid black lines),  
the continuum plus \BB background contributions (dotted red lines),  
and the sum of the continuum, \BB, and $K/\pi$ misidentification background components (dashed blue lines).} 
\end{figure}

Figure~\ref{fig:btodk_mes} shows the \mes distributions  
in the \DeltaE signal region defined through the requirement $|\DeltaE|<30$~\mev, 
after all selection criteria are applied.  
The reconstruction efficiencies are $20\%$, $9\%$, $12\%$, and $12\%$, for  
$\Bm \to \Dztilde \Km$,  
$\Bm \to \Dstarztilde[\Dztilde\piz]\Km$,  
$\Bm \to \Dstarztilde[\Dztilde\gamma]\Km$,  
and $\Bm\to \Dztilde \Kstarm$ decay modes, respectively, with \Dztildetokspipi. Similarly, for \Dztildetokskk channels we 
obtain $19\%$, $8\%$, and $11\%$ ($\Bm\to \Dztilde \Kstarm$ is not reconstructed).  
 
\begin{figure}[hbt!] 
\begin{tabular}{cc} 
\includegraphics[width=0.23\textwidth]{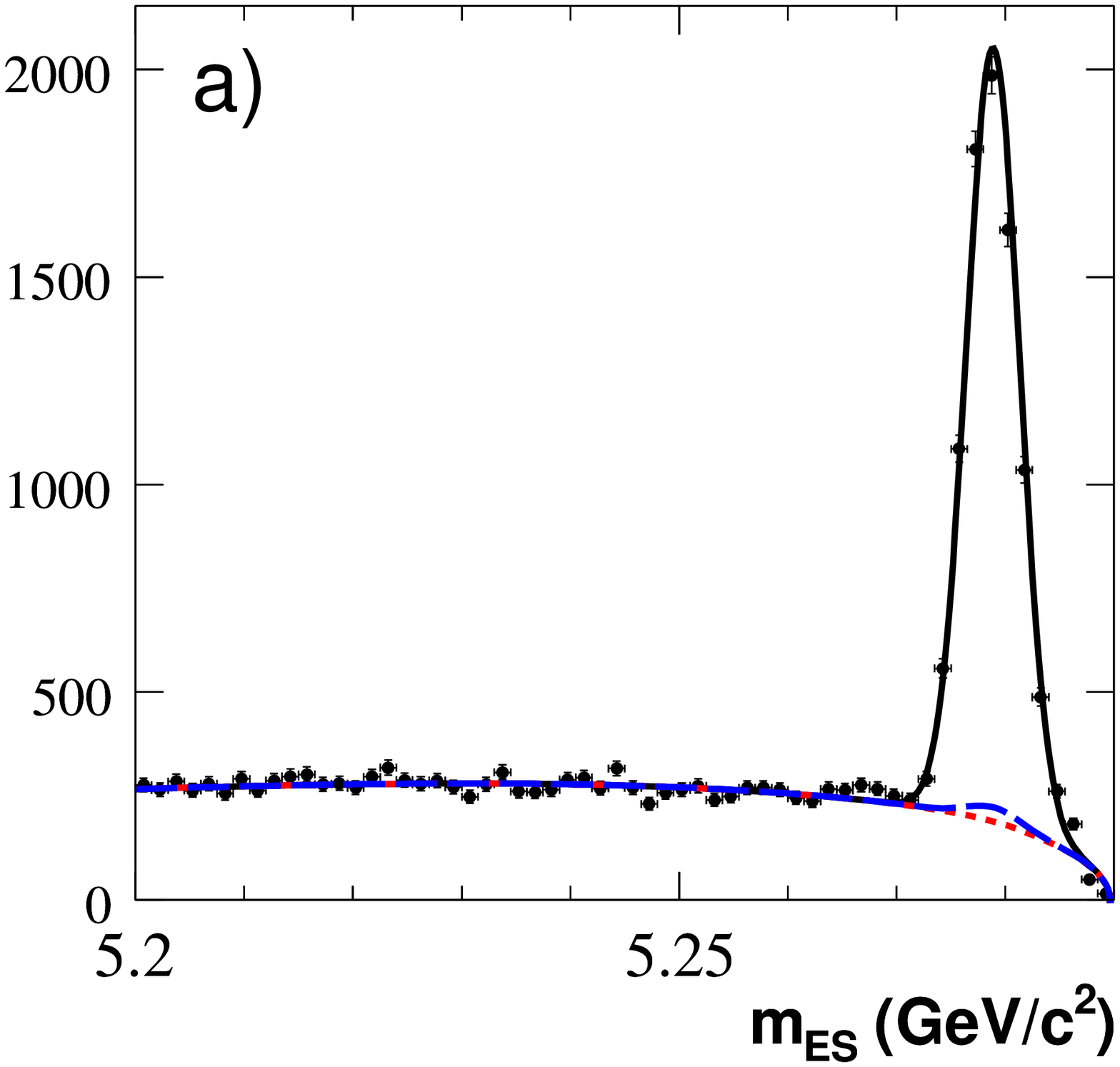} & 
\includegraphics[width=0.23\textwidth]{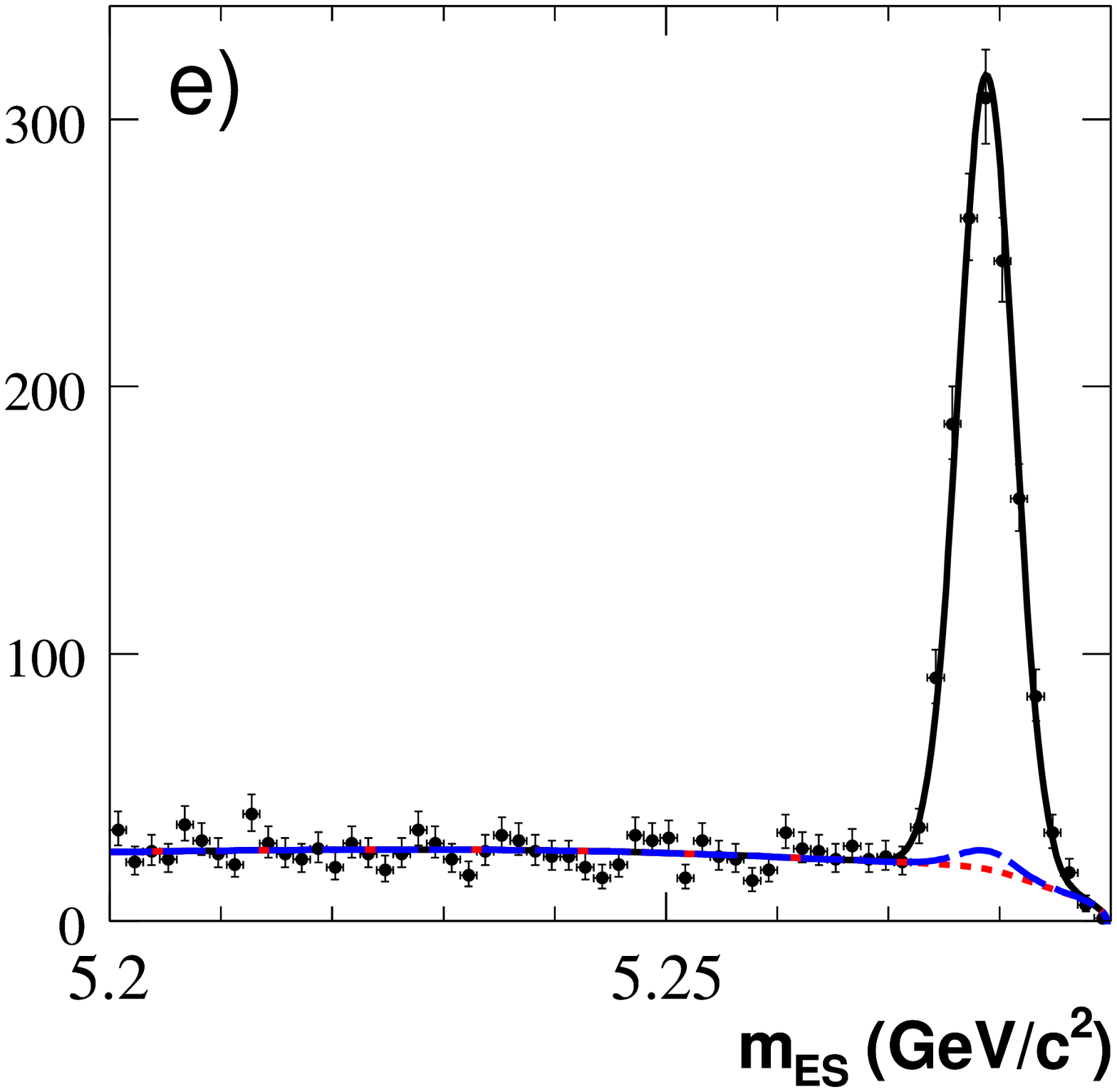} \\ 
\includegraphics[width=0.23\textwidth]{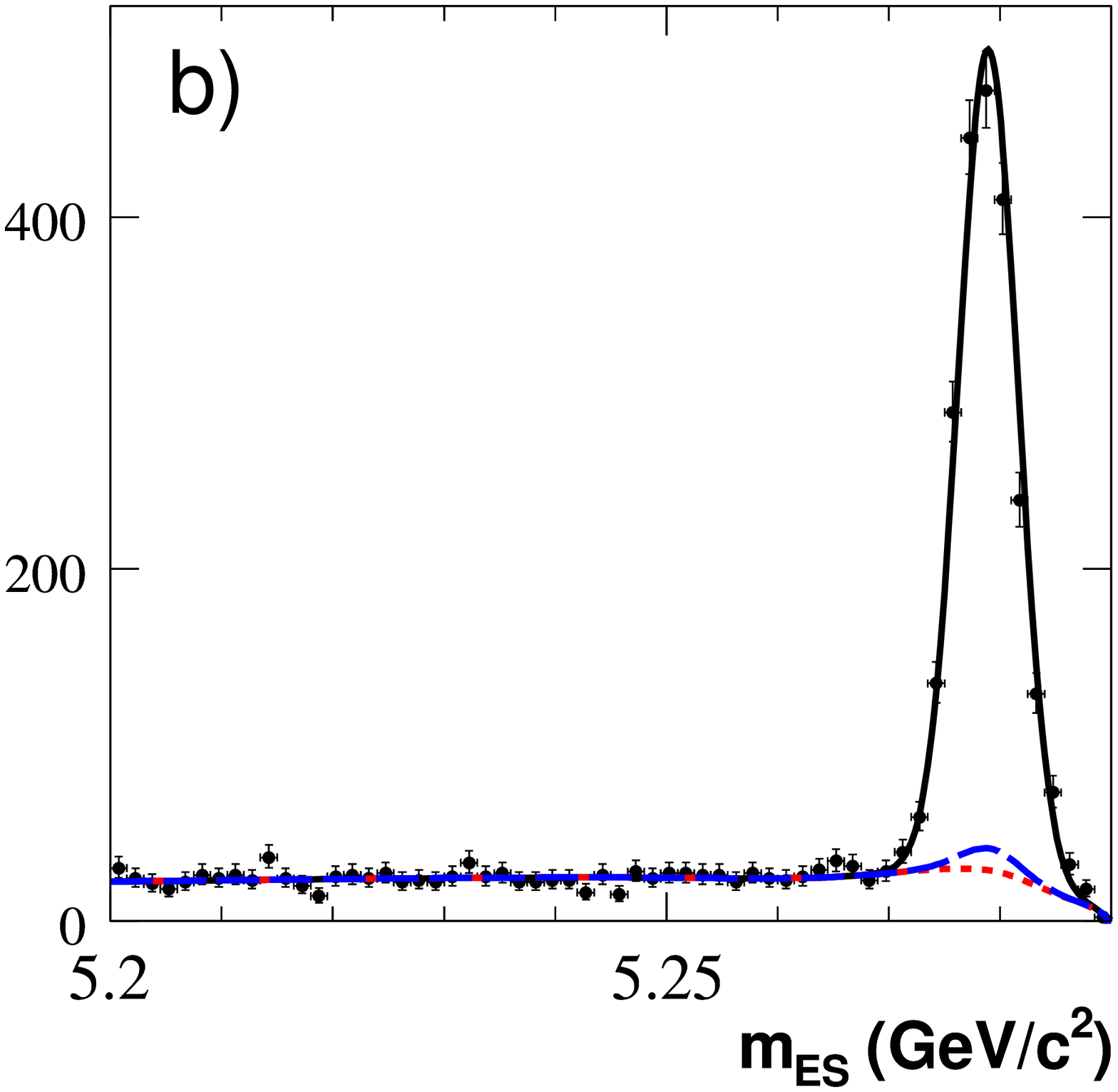} & 
\includegraphics[width=0.23\textwidth]{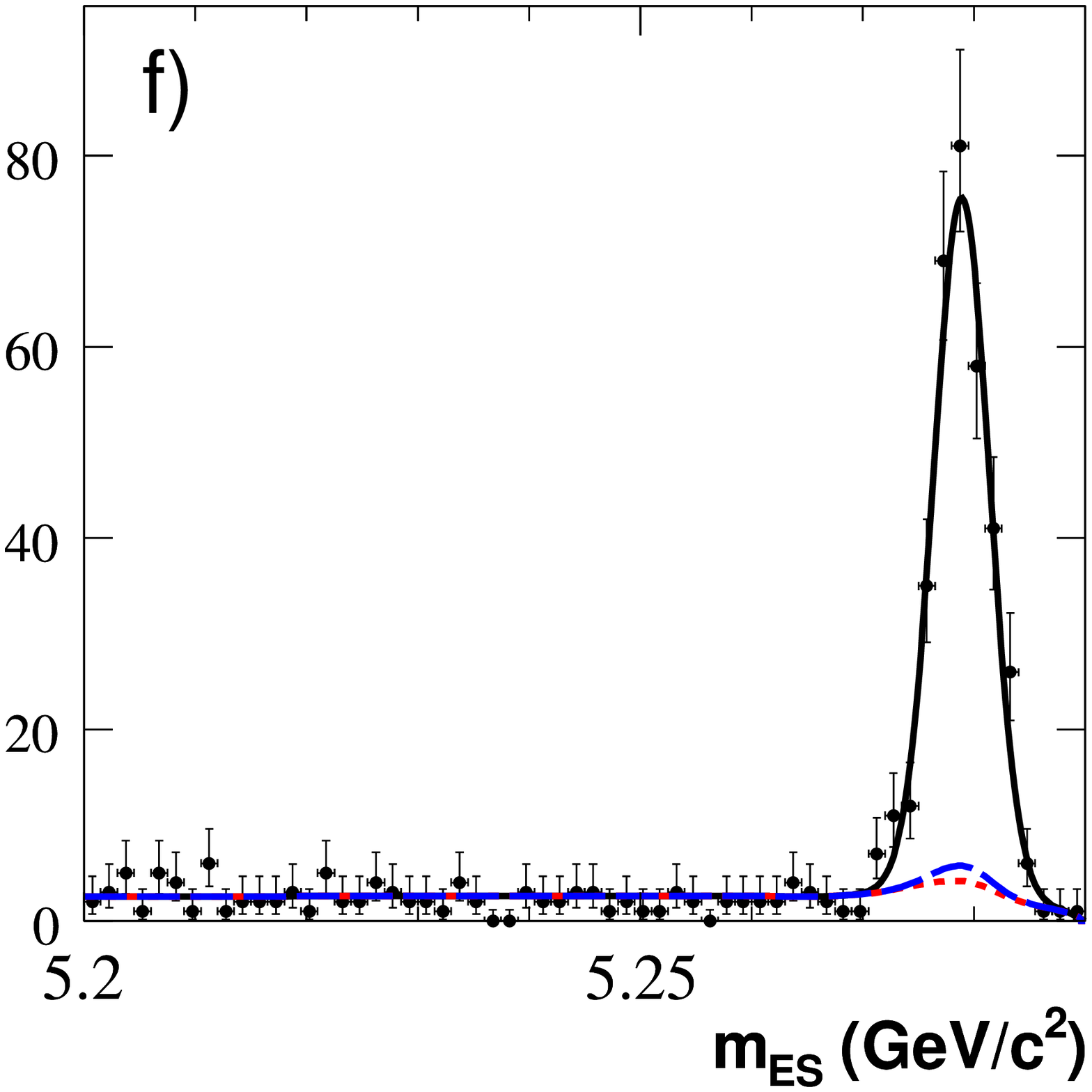} \\ 
\includegraphics[width=0.23\textwidth]{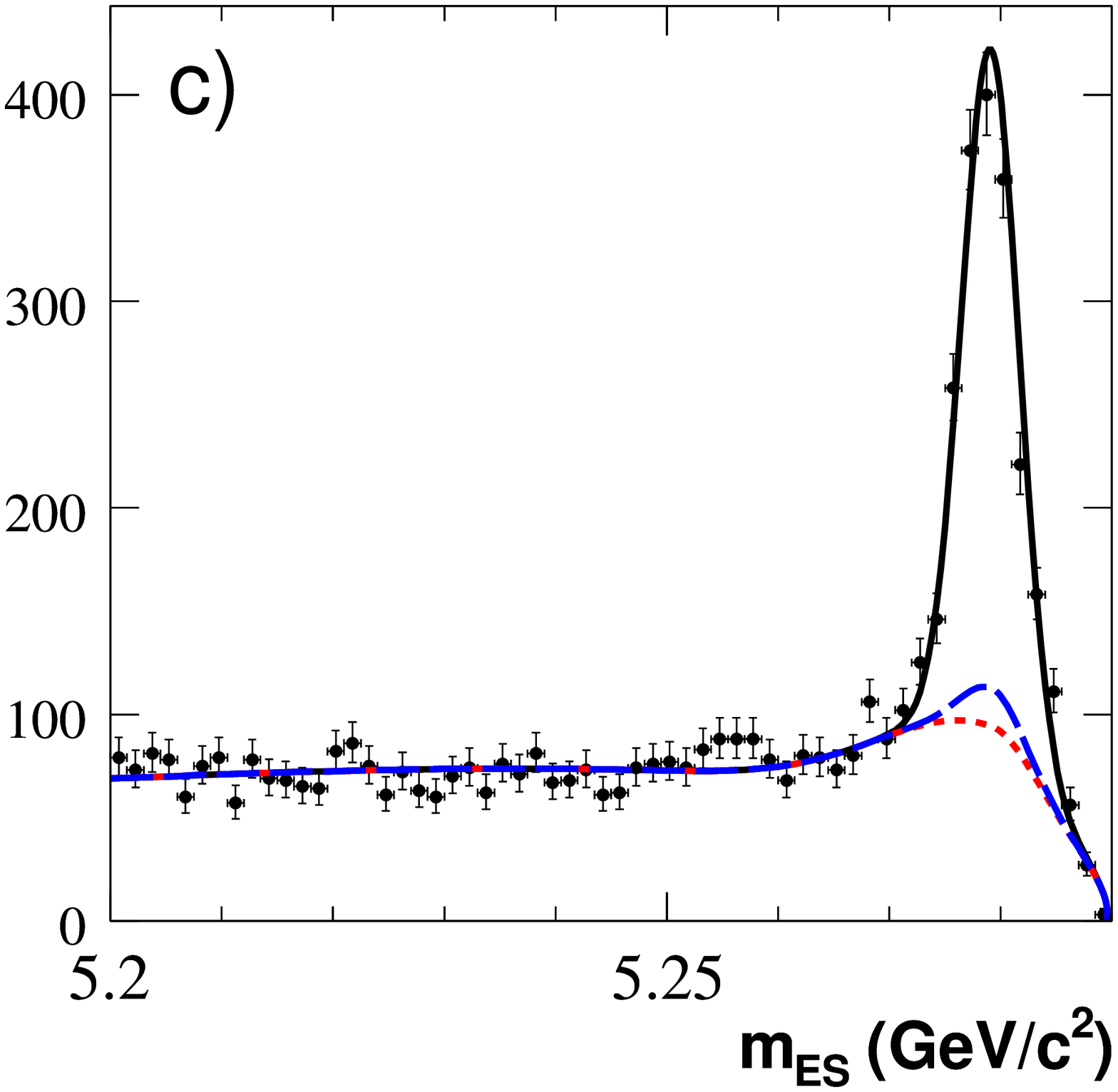} & 
\includegraphics[width=0.23\textwidth]{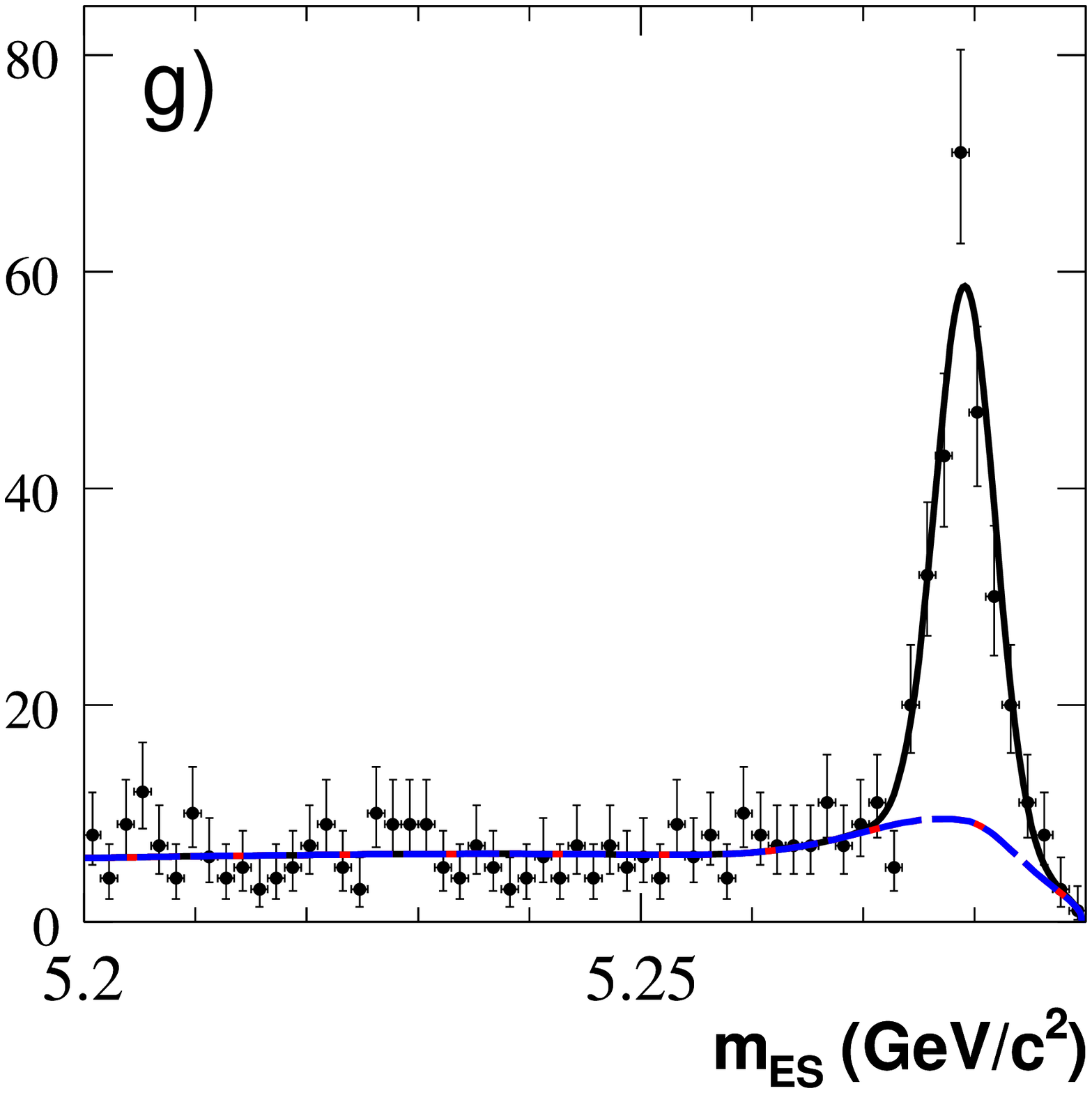} \\ 
\includegraphics[width=0.23\textwidth]{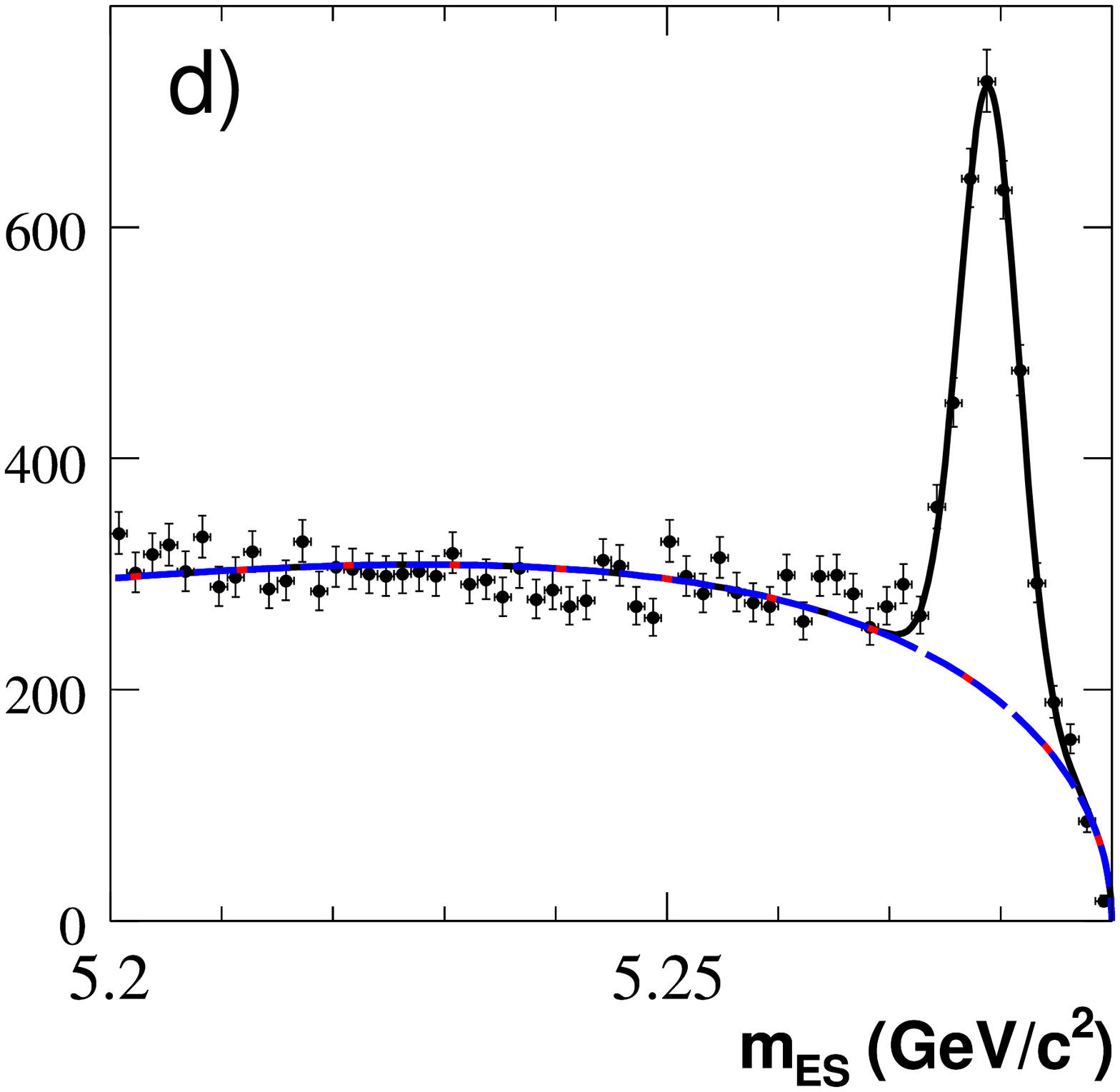} & 
\end{tabular} 
\caption{\label{fig:btodpi_mes} (color online). Distributions of \mes\ in \DeltaE signal region  
for (a,e) $\Bm\to \Dztilde \pim$,  (b,f) $\Bm\to \Dstarztilde[\Dztilde\piz] \pim$,  
(c,g) $\Bm\to \Dstarztilde[\Dztilde\gamma] \pim$, and (d) $\Bm\to \Dztilde a_1^-$, 
with (a-d) \Dztildetokspipi and (e-g) \Dztildetokskk. 
The curves superimposed represent the projections of the fit described in Sec.~\ref{sec:step1fit}:  
signal plus background (solid black lines),  
the continuum plus \BB background contributions (dotted red lines),  
and the sum of the continuum, \BB, and $K/\pi$ misidentification background components (dashed blue lines).} 
\end{figure}

The same selection criteria are applied to select $\Bm \to \DzDstarz \pim$ control samples, apart from the 
particle identification requirement on the prompt track, which is  
replaced by a kaon identification veto. 
Since we evaluate \DeltaE\ with the kaon mass hypothesis, the \DeltaE distributions are shifted by approximately $+50$~\mev, 
as given by Eq.~(\ref{eq:DeltaEshift}). 
$\Bm \to \Dz a_1^-$ candidates are reconstructed similarly to $\Bm \to \Dztilde \Kstarm$, with $a_1^- \to \rho(770)^0 \pim$ candidates 
made using combinations of three charged tracks with the requirements that the $a_1^-$ invariant mass must be 
in the range $[1.0,1.6]$~\gevcc, and that of the $\rho(770)^0$ within 150~\mevcc of its nominal mass. 
As for signal samples, we select events with $-80~\mev<\DeltaE<120~\mev$. 
Figure~\ref{fig:btodpi_mes} shows the \mes distributions for all control samples  
in the \DeltaE signal region defined through the requirement $20~\mev<\DeltaE<80~\mev$, 
after all selection criteria.  
The corresponding reconstruction efficiencies are similar to those estimated for 
the \CP samples.

\subsection{Background composition and signal yields} 
\label{sec:step1fit} 
 
The largest background contribution is from continuum events, where a fake or true \DzDstarztilde is combined with a random  
track ($\Bm \to \DzDstarztilde \Km$ samples), or a fake or true \Dztilde is combined with a random or fake \Kstar ($\Bm \to \Dztilde \Kstarm$ sample). 
To separate continuum from \BB events in a likelihood fit (discussed below and in Sec.~\ref{sec:cpanalysis}),  
variables that characterize the event shape are used. We construct a Fisher discriminant \fisher~\cite{ref:fisher} from  
a linear combination of four topological variables:  
the monomials  
$L_0=\sum_{i} p_i^*$ and $L_2 =\sum_{i} p_i^* |\cos \theta^*_i|^2$, $|\cos\theta_T^*|$, and $|\cos\theta_B^*|$. 
Here, $p_i^*$ and $\theta_i^*$ are the 
CM momentum and the angle of the remaining tracks and clusters in the event, with respect to the \B candidate thrust axis~\cite{ref:thrust}. 
$\theta_T^*$ is the angle between the thrust axis of the \B candidate and that of the rest of the event, 
and $\theta_B^*$ is the polar angle of the \B candidate momentum, in the CM frame. 
The first three variables account for the jet-like shape of continuum events, in comparison to the spherical 
topology of \BB events. In particular, the variable $|\cos\theta_T^*|$ peaks close to one for continuum while for 
\BB it is essentially uniformly distributed.  
The angular distribution of the variable $|\cos\theta_B^*|$ follows  
$1-\cos^2\theta_B^*$ for \BB events and $1+\cos^2\theta_B^*$ for $e^+e^-\to \q\qbar$~\cite{ref:epemtomupmumxsection}. 
The strategy of using the Fisher discriminant in a likelihood fit enhances significantly, typically about 25\%, the signal reconstruction  
efficiency compared to our previous analysis~\cite{ref:babar_dalitzpub},  
where we required $|\cos\theta_T^*|< 0.8$. At the same time it provides a larger sample of continuum events in the \mes sidebands,  
thus allowing the determination of the background properties directly from data (see Sec.~\ref{sec:cpanalysis}). 
 
Another source of background is related to \BB decays where a fake or true \DzDstarztilde is combined with a random or misidentified track or \Kstar.  
The main single contribution for $\Bm\to \DzDstarztilde \Km$ signal comes from  
$\Bm\to \DzDstarz \pim$ decays when the pion is misidentified as a kaon.  
This source is accounted for separately from other \BB backgrounds.  
This contribution can be discriminated from the signal due to the shift in the \DeltaE distribution relative to that of signal events. 
Since \DeltaE is computed by assigning the kaon mass hypothesis to the prompt track, it 
is shifted by a quantity 
\bea 
\DeltaE_{\rm shift} & = & \gamma_{\rm \pep2} \left( \sqrt{m_\K^2+|\bf{p}|^2} -\sqrt{m_\pi^2+|\bf{p}|^2} \right),~~~ 
\label{eq:DeltaEshift} 
\eea 
which depends on the momentum ${\bf p}$ of the prompt track in the laboratory frame and the Lorentz  
parameter $\gamma_{\rm \pep2}$ characterizing the boost of the CM relative to the laboratory frame,  
estimated from the \pep2 beam energies. 
For $\Bm\to \Dztilde \Kstarm$ signal the main \BB background source comes from $\Bm\to \Dz a_1^-$ decays. Since 
this contribution is highly suppressed by the cut on the cosine of the collinearity angle at $0.997$,  
it is not treated separately from other \BB backgrounds. 
Non-\Kstar decays contributing to the $\Bm\to \Dztilde \Kstarm$ sample are  
considered as signal, and their effect is accounted for by the factor $\kappa$ defined in Eq.~(\ref{eq:kappa}). 
 
We fit  
the seven signal samples $\Bm \to \DzDstarztilde \Km$ and $\Bm \to \Dztilde \Kstarm$, 
and their control samples $\Bm \to \DzDstarz \pim$ and $\Bm\to \Dz a_1^-$,  
using an unbinned extended maximum likelihood method to extract signal and background yields, and probability density functions (PDFs) 
for the variables \mes, \DeltaE, and the \fisher discriminant, in the \DeltaE selection region. 
Three different background components are considered: 
continuum events, $\K/\pi$ misidentification (for $\Bm \to \DzDstarztilde \Km$ and $\Bm \to \DzDstarz \pim$ samples only), 
and other $\FourS \to \BB$ decays.  
The log-likelihood for each of the \CP and control samples is 
\bea 
\ln {\cal L} = -\eta + \sum_j \ln \left[ \sum_c N_c{\cal P}_c({\bf u}_j) \right], 
\label{eq:likehood_sel} 
\eea 
where ${\bf u}_j = \{\mes,\DeltaE,\fisher\}_j$ characterizes the event $j$. 
Here, ${\cal P}_{c}({\bf u}) = {\cal P}_{c}(\mes) {\cal P}_{c}(\DeltaE) {\cal P}_{c}(\fisher)$ is the combined selection PDF, 
verifying the normalization condition $\int {\cal P}_{c}({\bf u}) {\rm d{\bf u}} = 1$,  
$N_c$ the event yield for signal or background component $c$, and $\eta = \sum_c N_c$. 
 
The signal \mes distributions for each \CP sample and its corresponding control sample  
are parameterized using a common single Gaussian. Similarly, the \fisher PDF makes use of a 
double Gaussian with different widths for the left and right parts of the curve (bifurcated Gaussian), 
and is assumed common for all \CP and control samples. 
The signal \DeltaE distribution for $\Bm \to \DzDstarztilde \Km$ events  
is parameterized with a double Gaussian function, while 
for $\Bm \to \DzDstarz \pim$ events we use the same function, shifted event-by-event using Eq.~(\ref{eq:DeltaEshift}). 
For $\Bm \to \Dztilde \Kstarm$ and $\Bm \to \Dztilde a_1^-$ signal events a common double Gaussian is used instead. 
 
The continuum background in the \mes distribution is described by a 
threshold function~\cite{ref:argus} while  
the continuum \DeltaE distribution is described using a first order  
polynomial parameterization. The free parameters are different for each \CP sample 
but common to the corresponding control sample. 
The \fisher distribution for continuum background is parameterized with the sum  
of two Gaussian functions and assumed common for all samples. 
 
The shape of the \mes distribution for $\FourS \to \BB$ background 
(excluding the $K/\pi$ misidentification contribution, as indicated previously) 
is taken from generic \BB simulated events for each \CP and control sample independently, 
and uses a threshold function~\cite{ref:argus} to describe the 
combinatorial component plus a bifurcated Gaussian to parameterize the contribution  
peaking at the $B$ mass. The fraction of the peaking contribution  
is extracted directly from the fit to the data, except for the \Dztildetokskk \CP samples,  
where it is taken from the generic \BB MC due to lack of statistics. 
The \DeltaE distribution for \BB background is taken similarly from simulation and is 
parameterized with the sum of a second order polynomial and an exponential function that takes into account the  
increase of combinatorial  
background at negative \DeltaE values. A Gaussian function is also included to  
account for potential \DeltaE peaking background,  
although we find no significant peaking structure in any of our samples. 
The \fisher distributions  
for signal and control samples, and for  
the generic \BB\ background, are assumed to be the same as found in the simulation. 
 
The selection fit yields, respectively, $610 \pm 34$, $156 \pm 17$,  $114 \pm 16$, and $110 \pm 15$ signal candidates, for  
$\Bm \to \Dztilde \Km$,  
$\Bm \to \Dstarztilde[\Dztilde\piz]\Km$,  
$\Bm \to \Dstarztilde[\Dztilde\gamma]\Km$,  
and $\Bm\to \Dztilde \Kstarm$ reconstructed in the \Dztildetokspipi\ mode, 
in agreement with expectations based on measured branching fractions and efficiencies estimated from Monte Carlo simulation.  
Similarly, for \Dztildetokskk channels we obtain, respectively, $132 \pm 14$, $35 \pm 7$, and $16 \pm 6$.  
The corresponding signal yields for control samples are 
$8262 \pm 105$, $2227 \pm 55$,  $1446 \pm 53$, and $2321 \pm 75$, for \Dztildetokspipi decay modes,  
and  
$1402 \pm 41$, $350 \pm 20$, and $236 \pm 20$, for \Dztildetokskk. All errors are statistical only. 
The curves in Fig.~\ref{fig:btodk_mes} and Fig.~\ref{fig:btodpi_mes}  
represent the fit projections on the \mes variable,  
for the \DeltaE signal region.

%% file: dalitzmodels.tex
\subsection{Selection of flavor-tagged \Dz mesons} 
\label{sec:dalitzmodels-selection} 
 
The \Dztokspipi and \Dztokskk (referred to hereafter collectively as \Dztokshh) decay amplitudes  
are determined from Dalitz plot analyses of \Dz mesons from $\Dstarp \to \Dz \pip$ decays  
produced in $e^+ e^- \to \c\cbar$ events.  
The charge of the low momentum $\pip$ from the \Dstarp decay identifies (``tags'')  
the flavor of the \Dz. Reconstruction and selection of \Dztokshh candidates from $\Dstarp \to \Dz \pip$ decays are similar 
to those from $\Bm \to \DzDstarztilde\Km$ decays,  
the only exception being the kaon identification of only one charged kaon for \Dztokskk. 
The \Dstarp candidates are formed by combining the \Dz with the low momentum charged track.  
The two \Dstarp decaying daughters are constrained to originate from  
the same point inside the \pep2 luminous region.  
To reduce combinatorial background and contamination from \BB decays, the \Dz candidates are  
required to have a CM momentum greater than 2.2~\gevc. 
 
\begin{figure}[htb!] 
\begin{tabular} {cc}   
{\includegraphics[width=0.23\textwidth]{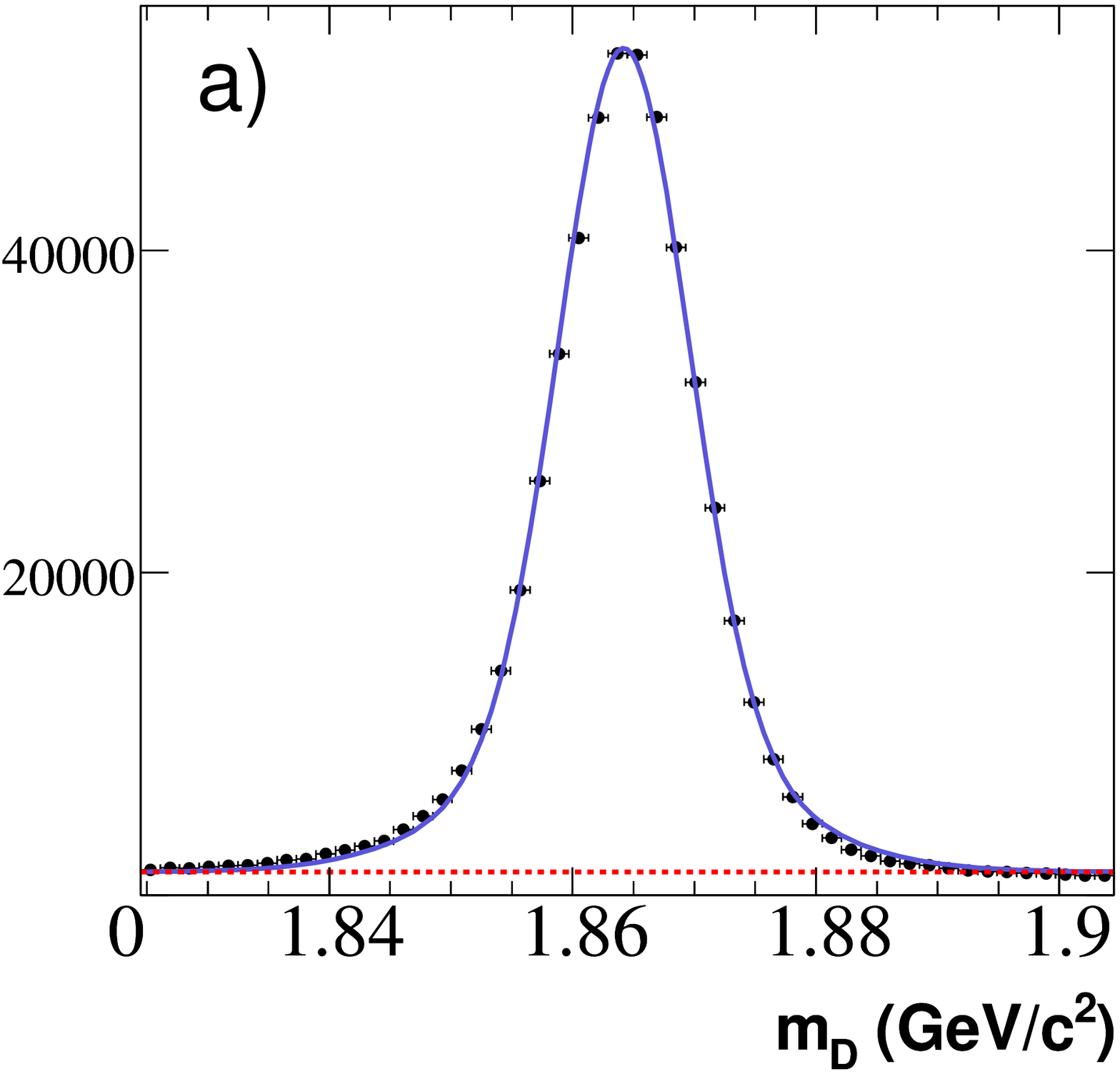}} & 
{\includegraphics[width=0.23\textwidth]{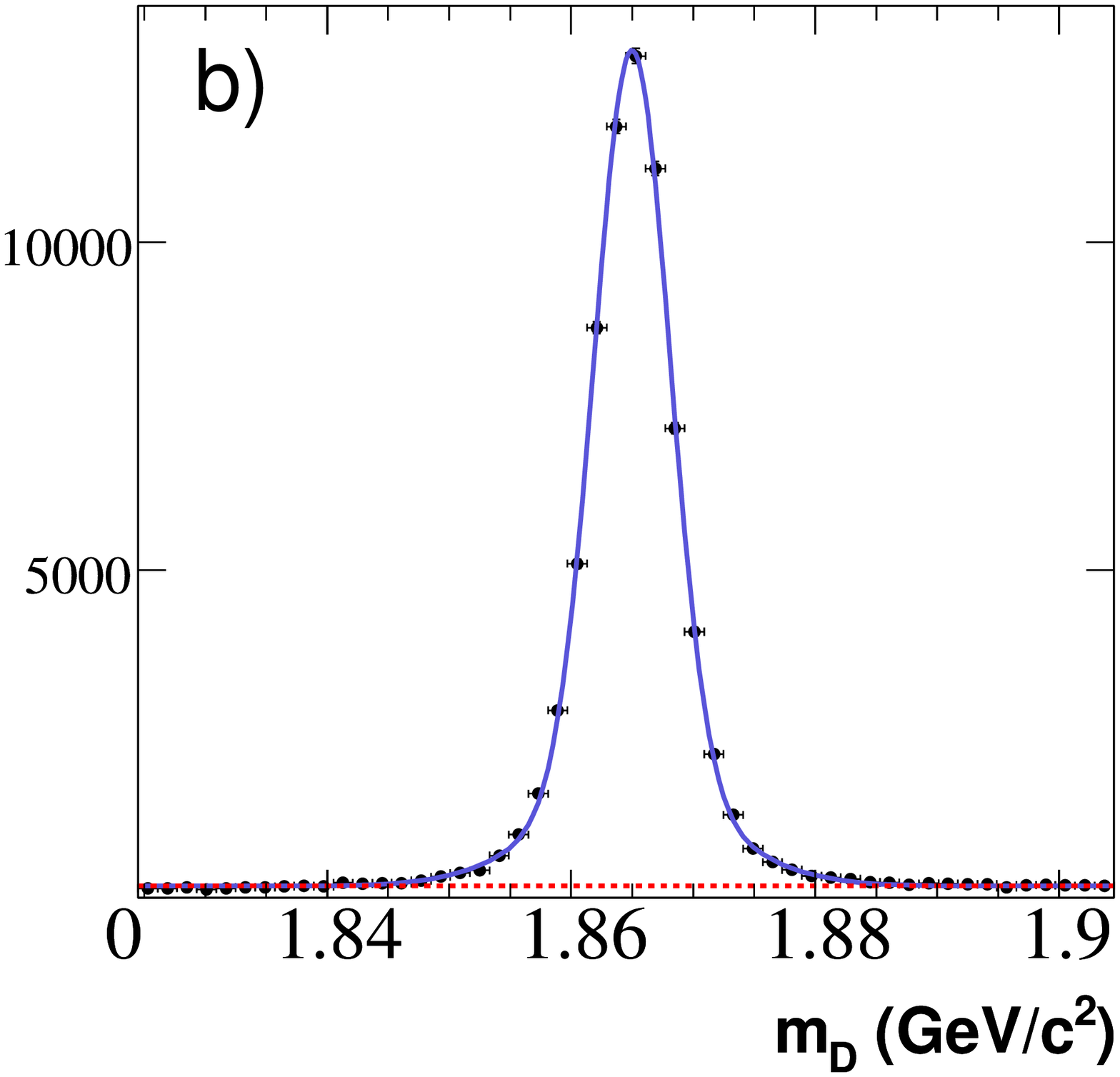}} \\ 
\end{tabular}    
\caption{\label{fig:mD} (color online). \Dz mass distributions after all selection criteria, for 
(a) $\Dstarp \to \Dz \pip$, \Dztokspipi and (b) $\Dstarp \to \Dz \pip$, \Dztokskk. The curves superimposed represent the result from the \mD fit  
(solid blue lines) and the linear background contribution (dotted red lines). 
} 
\end{figure}

\begin{figure}[hbt!] 
\begin{tabular} {cc}   
{\includegraphics[width=0.23\textwidth]{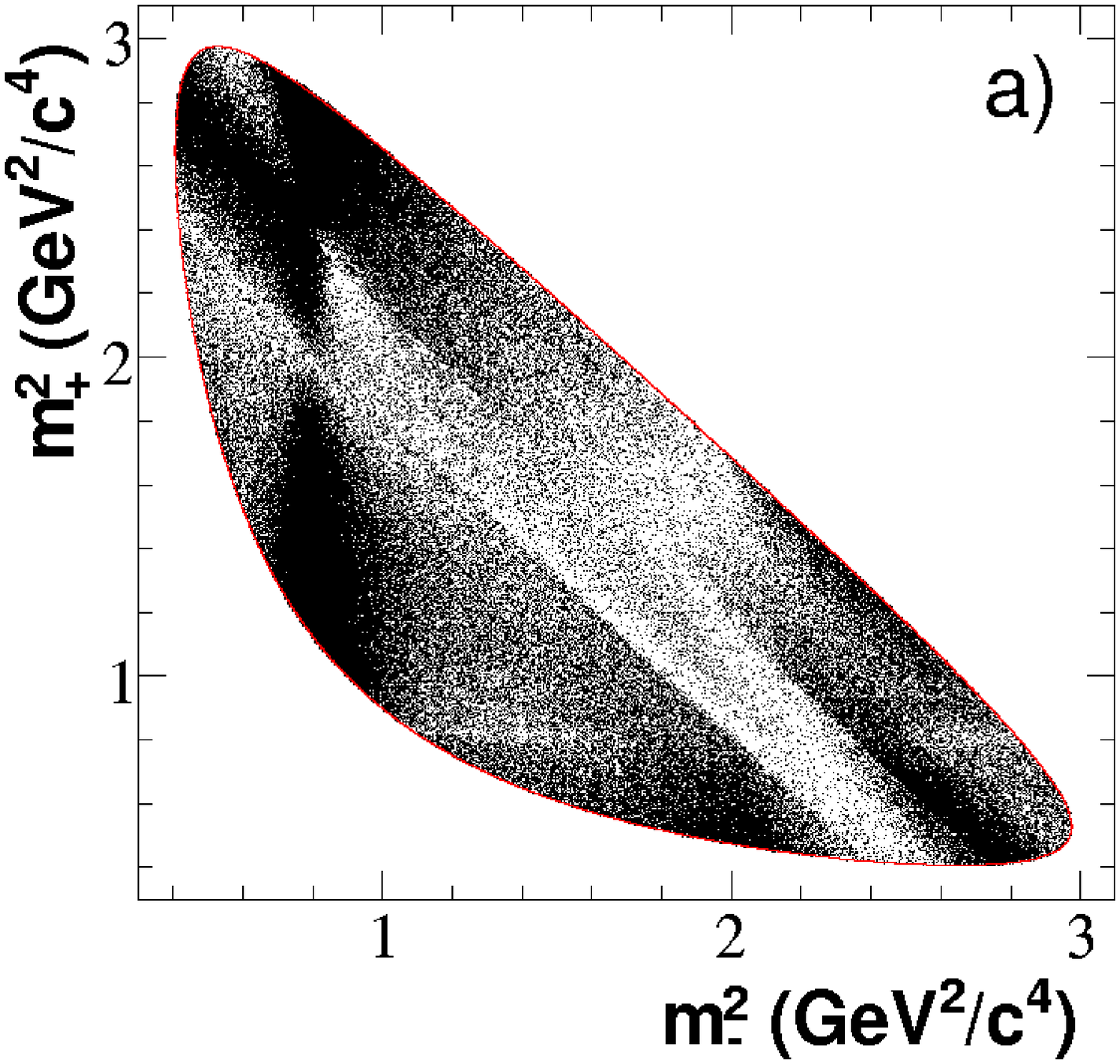}} & 
{\includegraphics[width=0.23\textwidth]{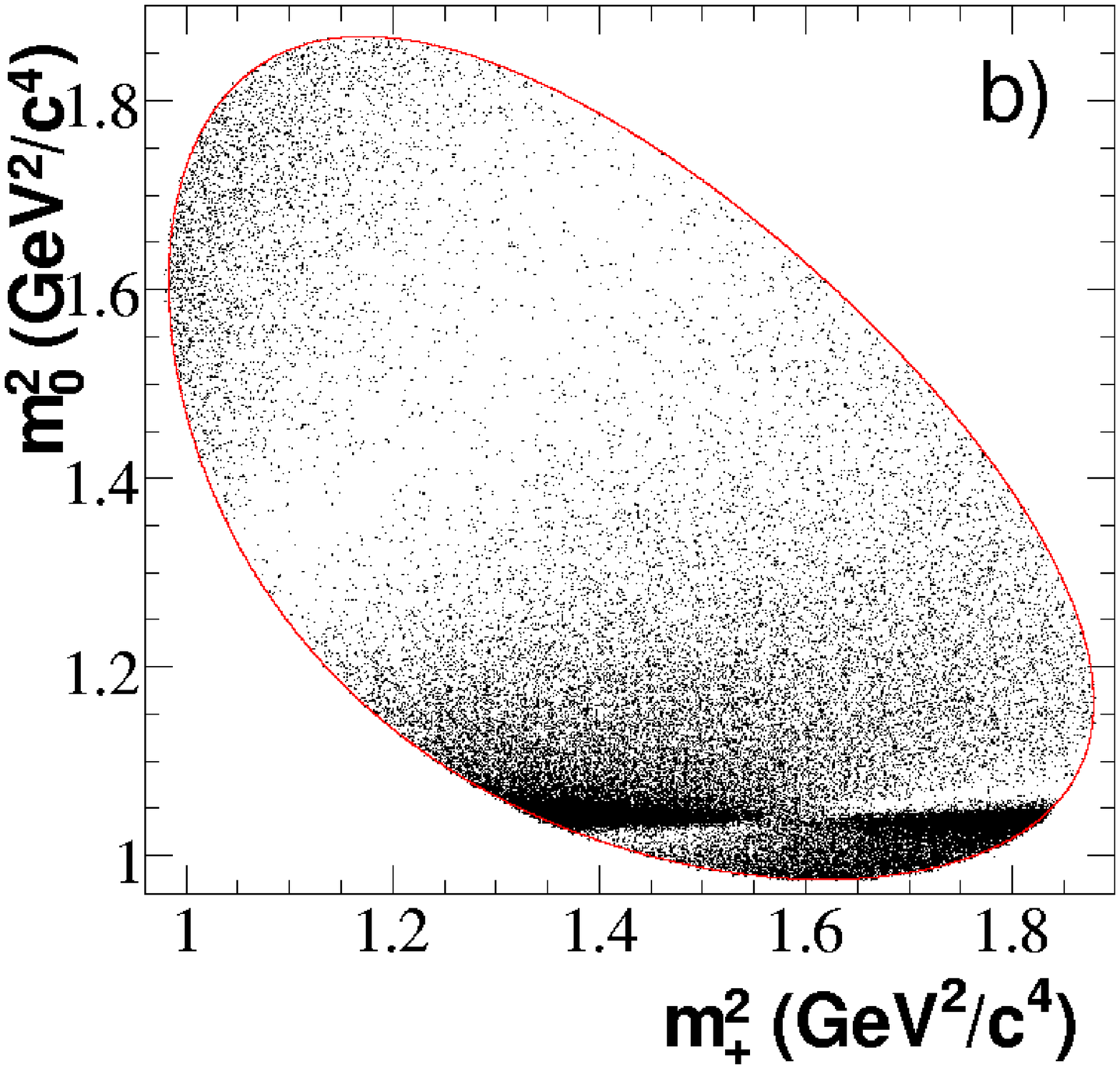}} \\ 
\end{tabular}    
\caption{\label{fig:Dalitz} (color online). Dalitz plot distributions for (a) \Dztokspipi and (b) \Dztokskk from $\Dstarp \to \Dz \pip$ events 
after all selection criteria, in the \Dz mass signal signal region.  
The contours (solid red lines) represent the kinematical limits of the \Dztokspipi and \Dztokskk decays.  
} 
\end{figure}

Each \Dz sample is characterized by the distributions of two variables, the invariant mass of the \Dz candidate \mD and 
the $\deltam = \Dstarp-\Dz$ mass difference. We select \Dz candidates within $\pm 0.64$~\mevcc and $\pm 0.61$~\mevcc,  
corresponding to $\pm2$ standard deviations, around the nominal \deltam~\cite{ref:pdg2006},  
for \Dztokspipi and \Dztokskk, respectively. 
Figure~\ref{fig:mD} shows the resulting \Dztokshh mass distributions. 
The \mD lineshape is described 
using a two Gaussian function for the signal and a linear background, as also shown in Fig.~\ref{fig:mD}. 
The \mD resolutions are $6.7$~\mevcc and $3.9$~\mevcc, for \Dztokspipi and \Dztokskk. 
The mass resolution for the latter is better than that of the former because of the much smaller $Q$-value involved. 
The signal purity in the signal box ($\pm 2\sigma$ cutoff on \mD, where $\sigma$ stands for the \mD resolution) is 97.7\% and 99.3\%,  
with  
about 487000 and 69000 candidates, for \Dztokspipi and \Dztokskk. The Dalitz plot distributions for these events 
are shown in Fig.~\ref{fig:Dalitz}, with $m_\mp^2 = m^2_{\KS h^\mp}$ and $m_0^2 = m^2_{h^+ h^-}$.


\subsection{Dalitz plot analysis} 
\label{sec:dalitzmodels-analysis} 
 
Three-body charm decays are expected to proceed through intermediate quasi-two body  
modes~\cite{ref:bauer} and this is the observed pattern. We therefore use,  
as a baseline model to describe ${\cal A} _{D}(m^2_\mp,m^2_\pm)$,  
an isobar approach consisting of a coherent sum  
of two-body amplitudes (subscript $r$) and a ``non-resonant'' (subscript NR) contribution~\cite{ref:RevDalitzPlotFormalism}, 
\bea 
{\cal A}_D(\mtwo) = \sum_r a_r e^{i \phi_r} {\cal A}_r(\mtwo) + a_{\rm NR} e^{i \phi_{\rm NR}}, 
\label{eq:genAmp} 
\eea 
where we have introduced the notation $\mtwo \equiv (m^2_-,m^2_+)$. 
The parameters $a_r$ ($a_{\rm NR}$) and $\phi_r$ ($\phi_{\rm NR}$) are the magnitude and phase of the  
amplitude for component $r$ (NR).  
The function ${\cal A}_r = F_\D \times F_r \times T_r \times W_r$ is a Lorentz-invariant expression that describes the 
dynamic properties of the \Dz meson decaying into \kshh through an intermediate resonance $r$, as a function  
of position in the Dalitz plane.  
Here, $F_\D$ ($F_r$) is the Blatt-Weisskopf centrifugal barrier factor for the \D (resonance)  
decay vertex~\cite{ref:blatt-weisskopf} with  
radius $R=1.5~\gev^{-1}\hbar c \equiv0.3~\fm$,  
$T_r$ is the resonance propagator, and $W_r$ describes the angular distribution in the decay. 
For $T_r$ we use a relativistic Breit-Wigner (BW) parameterization with mass-dependent width~\cite{ref:RevDalitzPlotFormalism},  
except for $r=\rho(770)^0$ and $\rho(1450)^0$ resonances where 
we use the Gounaris-Sakurai functional form~\cite{ref:gounarissakurai}. 
The angular dependence $W_r$ is described using either Zemach tensors~\cite{ref:zemach,ref:zemach-covariant} where transversality is enforced 
or the helicity formalism~\cite{ref:jacob-wick,ref:cleomodel,ref:cleo-bug} when we allow for a longitudinal component in the resonance  
propagator (see Ref.~\cite{ref:RevDalitzPlotFormalism} for a comprehensive summary). 
Mass and width values are taken from~\cite{ref:pdg2006}, unless otherwise specified. 
 
The complex $\pi\pi$ S-wave dynamics in the \Dztokspipi reaction~\cite{ref:RevScalarMesons},  
with the presence of several broad and overlapping scalar resonances, 
is more adequately described through the use of a K-matrix formalism~\cite{ref:Kmatrix} with the P-vector approximation~\cite{ref:Kmatrix-Pvector}. 
This approach offers a direct way of imposing the unitarity constraint of the scattering matrix,  
not guaranteed in the case of the isobar model. 
The Dalitz plot amplitude ${\cal A}_D(\mtwo)$ given by Eq.~(\ref{eq:genAmp}) is then modified as 
 
\bea 
{\cal A}_D(\mtwo) = F_1(s) + \sum_{r\ne(\pi\pi)_{L=0}} a_r e^{i \phi_r} {\cal A}_r(\mtwo) + a_{\rm NR} e^{i \phi_{\rm NR}},~\label{eq:k-matrix} 
\label{eq:genAmp-kmatrix} 
\eea 
where $F_1(s)$ is the contribution of $\pi\pi$ S-wave states written in terms of the K-matrix formalism, 
 
\bea 
F_u(s) = \sum_l\left[I-iK(s)\rho(s)\right]^{-1}_{uv} P_{v}(s). 
\label{eq:F1} 
\eea 
Here, $s = m_0^2$ is the squared invariant mass of the $\pip\pim$ system, 
$I$ is the identity matrix, $K$ is the matrix describing the S-wave scattering process, $\rho$ is the phase-space matrix,  
and $P$ is the initial production vector (P-vector).  
The index $u$ (and similarly $v$) represents the $u^{\rm th}$ channel ($1=\pi\pi$, $2=\K\Kbar$, $3=\pi\pi\pi\pi$, $4=\eta\eta$, $5=\eta\eta'$). 
In this framework, the production process can be viewed as the initial preparation of several states, 
which are then propagated by the $\left[I-iK(s)\rho(s)\right]^{-1}$ term into the final one. 
The propagator can be described using scattering data, provided that the two-body system in the final state is  
isolated and does not interact with the rest of the final state in the production process. 
The P-vector has to be determined from the data themselves since it depends on the production mechanism.  
Only the $F_1$ amplitude appears in Eq.~(\ref{eq:genAmp-kmatrix}) since we are describing the $\pi\pi$ channel.  
See Sec.~\ref{sec:dalitzmodels-kspipi} for more details.

The decay amplitude ${\cal A}_D(\mtwo)$ is then determined from a maximum likelihood fit to the \Dztokshh Dalitz plot  
distribution \mtwo in a $\pm 2\sigma$ cutoff region of the \Dz mass, with log-likelihood function 
\bea 
\ln {\cal L} = \sum_j \ln \left[ f_{\rm sig} {\cal D}_{{\rm sig},\mp}(\mtwoj) + 
                                 (1-f_{\rm sig}) {\cal D}_{{\rm bkg},\mp}(\mtwoj) 
                          \right], 
\eea 
where $f_{\rm sig}$ represents the fraction of signal obtained from the fit to the mass spectrum, and 
${\cal D}_{\rm sig (bkg),\mp}(\mtwoj)$ is the signal (background) Dalitz plot PDF for event $j$,  
satisfying the condition $\int {\cal D}_{\rm sig (bkg),\mp}(\mtwo){\rm d}\mtwo = 1$.  
For \Dz signal events, ${\cal D}_{\rm sig,+}(\mtwo) = |{\cal A}_{D}(\mtwo)|^2 \epsilon (\mtwo)$,  
while for \Dzb, ${\cal D}_{\rm sig,-}(\mtwo) = |{\cal A}_{D}(\mtwob)|^2 \epsilon (\mtwo)$, 
with $\mtwob \equiv (m^2_+,m^2_-)$. Here $\epsilon (\mtwo)$ represents the efficiency variations on the Dalitz plot, evaluated 
using high statistics signal MC samples. These are generated according to a uniform distribution and parameterized using  
third-order polynomial functions in two dimensions, symmetric for \Dztokspipi and asymmetric for \Dztokskk to account for 
possible charge asymmetries in the \Km and \Kp detection efficiencies. 
The Dalitz plot distributions for the background, ${\cal D}_{\rm bkg,\mp}(\mtwo)$, are determined using \Dz mass sideband data. 

For each contribution $r$  
we evaluate the fit fraction as the normalized integral of $a_r^2 |{\cal A}_r(\mtwo)|^2$ over the Dalitz  
plane~\cite{ref:RevDalitzPlotFormalism}, 
\bea 
f_r & = & \frac{a_r^2 \int |{\cal A}_r(\mtwo)|^2 {\rm d\mtwo}}{ \sum_r \sum_{r'} a_r a_{r'}^* \int {\cal A}_r(\mtwo) {\cal A}_{r'}^*(\mtwo) {\rm d\mtwo} }. 
\label{eq:fitfraction} 
\eea 
The sum of fit fractions does not necessarily add up to unity because of interference effects among the amplitudes.

\subsection{\Dztokspipi Dalitz model} 
\label{sec:dalitzmodels-kspipi}

The P- and D-waves of the \Dztokspipi decay amplitude are described using a total of  
6 resonances leading to 8 two-body decay amplitudes:  
the Cabibbo allowed (CA) $K^{*}(892)^-$, $K^{*}(1680)^-$, $K^{*}_2(1430)^-$,  
the doubly-Cabibbo suppressed (DCS) $K^{*}(892)^+$, $K^{*}_2(1430)^+$,  
and the \CP eigenstates $\rho(770)^0$, $\omega(782)$, and $f_2(1270)$. 
Since the $K\pi$ P-wave is largely dominated by the $K^*(892)^\mp$,  
the mass and width of this resonance are simultaneously determined from our fit to the tagged \Dz sample, 
$M_{K^{*}(892)^\mp}=893.61\pm0.08$~\mevcc and $\Gamma_{K^{*}(892)^\mp}=46.34\pm0.16$~\mevcc (errors are statistical only). 
The mass and width values of the $K^{*}(1680)^-$ are taken from~\cite{ref:lass}, 
where the interference between the $K\pi$ S- and P-waves is properly accounted for.

We adopt the same parameterizations for $K$, $\rho$, and $P$ in Eq.~(\ref{eq:F1}) as in Refs.~\cite{ref:RevDalitzPlotFormalism,ref:AS,ref:FOCUS}.  
For the $K$ matrix we have 
\begin{eqnarray} 
K_{uv}(s) = \left( \sum_{\alpha} \frac{g^{\alpha}_u g^{\alpha}_v}{m^2_{\alpha}-s} +  
            f^{\rm scatt}_{uv}\frac{1-s^{\rm scatt}_0}{s-s^{\rm scatt}_0} \right) f_{A0}(s), 
\label{eq:kmatrix} 
\end{eqnarray} 
where $g^{\alpha}_u$ is the coupling constant of the K-matrix pole $m_{\alpha}$ to the $u^{\rm th}$ channel.  
The parameters $f^{\rm scatt}_{uv}$ and $s^{\rm scatt}_0$ describe the slowly-varying part of the K-matrix. The factor 
 
\bea 
f_{A0} (s) & = & \frac{1-s_{A0}}{s-s_{A0}} \left(s-s_A \frac{m^2_{\pi}}{2} \right), 
\eea 
suppresses the false kinematical singularity at $s=0$ in the physical region near the $\pi\pi$ threshold (the Adler zero~\cite{ref:Adler}). 
The parameter values used in this analysis are listed in Table~\ref{tab:AS}, and are obtained 
from a global analysis of the available $\pi\pi$ scattering data from threshold up  
to $1900$~\mevcc~\cite{ref:AS}.  
The parameters $f^{\rm scatt}_{uv}$, for $u \ne 1$, are all set to zero since they are not related to the $\pi\pi$ scattering process.  
Similarly, for the $P$ vector we have 
 
\begin{eqnarray} 
P_{v}(s)= \sum_{\alpha} \frac{\beta_{\alpha} g^{\alpha}_v}{m^2_{\alpha}-s} + f^{\rm prod}_{1v}\frac{1-s^{\rm prod}_0}{s-s^{\rm prod}_0}. 
\label{eq:pvector} 
\end{eqnarray} 
Note that the P-vector has the same poles as the K-matrix, otherwise the $F_1$ vector would vanish (diverge) 
at the K-matrix (P-vector) poles. 
The parameters $\beta_{\alpha}$, $f_{1v}^{\rm prod}$ and $s^{\rm prod}_0$ of the initial P-vector are obtained from our fit to the  
tagged \Dztokspipi data sample.

\begin{table}[hbt!] 
\caption{\label{tab:AS} K-matrix parameters from a global analysis of the available $\pi\pi$ scattering data  
from threshold up to  
$1900$~\mevcc~\cite{ref:AS}.  
Masses and coupling constants are given in~\gevcc.} 
\begin{ruledtabular} 
\begin{tabular}{cccccc} 
$m_{\alpha}$ & $g_{\pipi}^\alpha$ & $g_{\K\Kb}^\alpha$ & $g_{4\pi}^\alpha$ & $g_{\eta\eta}^\alpha$ & $g_{\eta\eta^{\prime}}^\alpha$  \\ 
\hline 
     $\phm0.65100$  &   $\phm0.22889$  &  $-0.55377$      &    $\phm0.00000$   &  $-0.39899$      &  $-0.34639$ \\ 
     $\phm1.20360$  &   $\phm0.94128$  &  $\phm0.55095$   &    $\phm0.00000$   &  $\phm0.39065$   &  $\phm0.31503$ \\ 
     $\phm1.55817$  &   $\phm0.36856$  &  $\phm0.23888$   &    $\phm0.55639$   &  $\phm0.18340$   &  $\phm0.18681$ \\ 
     $\phm1.21000$  &   $\phm0.33650$  &  $\phm0.40907$   &    $\phm0.85679$   &  $\phm0.19906$   &  $-0.00984$ \\ 
     $\phm1.82206$  &   $\phm0.18171$  &  $-0.17558$      &    $-0.79658$      &  $-0.00355$      &  $\phm0.22358$ \\ 
\hline 
     $s^{\rm scatt}_0$    &    $f^{\rm scatt}_{11}$  &   $f^{\rm scatt}_{12}$  &   $f^{\rm scatt}_{13}$  &   $f^{\rm scatt}_{14}$  &   $f^{\rm scatt}_{15}$ \\ 
    $-3.92637$            &    $\phm0.23399$         &   $\phm0.15044$         &   $-0.20545$            &     $\phm0.32825$       &   $\phm0.35412$   \\	 
\hline 
     $s_{A0}$      &     $s_{A}$ \\ 
     $-0.15$       &     $1$ \\ 
\end{tabular} 
\end{ruledtabular} 
\end{table} 
 
For the $\K\pi$ S-wave contribution to Eq.~(\ref{eq:k-matrix}) we use a parameterization extracted from  
scattering data~\cite{ref:lass} which consists of a $K^{*}_0(1430)^-$ or $K^{*}_0(1430)^+$ BW (for CA or DCS contribution, respectively)  
together with an effective range non-resonant component with a phase shift, 
 
\bea 
{\cal A}_{\K\pi\ L=0}(\mtwo) & = & F \sin\delta_F e^{i\delta_F} + R \sin\delta_R e^{i\delta_R}e^{i2\delta_F}~,~~~~~ 
\label{eq:lass} 
\eea 
with 
\bea 
 \delta_R & = & \phi_R + \tan^{-1}\left[\frac{M \Gamma(m^2_{\K\pi})}{M^2-m^2_{\K\pi}}\right]~,\nn \\ 
 \delta_F & = & \phi_F + \cot^{-1}\left[ \frac{1}{aq}+\frac{rq}{2} \right]. 
\eea 
The parameters $a$ and $r$ play the role of a scattering length and effective interaction length, respectively, 
$F$ ($\phi_F$) and $R$ ($\phi_R$) are the amplitudes (phases) for the non-resonant and resonant terms, and 
$q$ is the momentum of the spectator particle in the $K\pi$ system rest frame. 
Note that the phases $\delta_F$ and $\delta_R$ depend on $m^2_{\K\pi}$. 
$M$ and $\Gamma(m^2_{\K\pi})$ are the mass and running width of the resonant term. 
This parameterization corresponds to a K-matrix approach describing a rapid phase shift coming from the resonant term and a slow rising  
phase shift governed by the non-resonant term, with relative strengths $R$ and $F$~\cite{ref:BenLau-PhDThesis}.  
The parameters $M$, $\Gamma$, $F$, $\phi_F$, $R$, $\phi_R$, $a$ and $r$ are determined from our fit to the tagged  
\Dz sample, along with the other parameters of the model. 
Other recent experimental efforts to improve the description of the $\K\pi$ S-wave using K-matrix and model independent 
parameterizations from high-statistics samples of $\Dp \to \Km\pip\pip$ decays are described in Ref.~\cite{ref:KpiS-wave}.  
 
\begin{figure*}[hbt!] 
\begin{tabular} {ccc}   
{\includegraphics[width=0.3\textwidth]{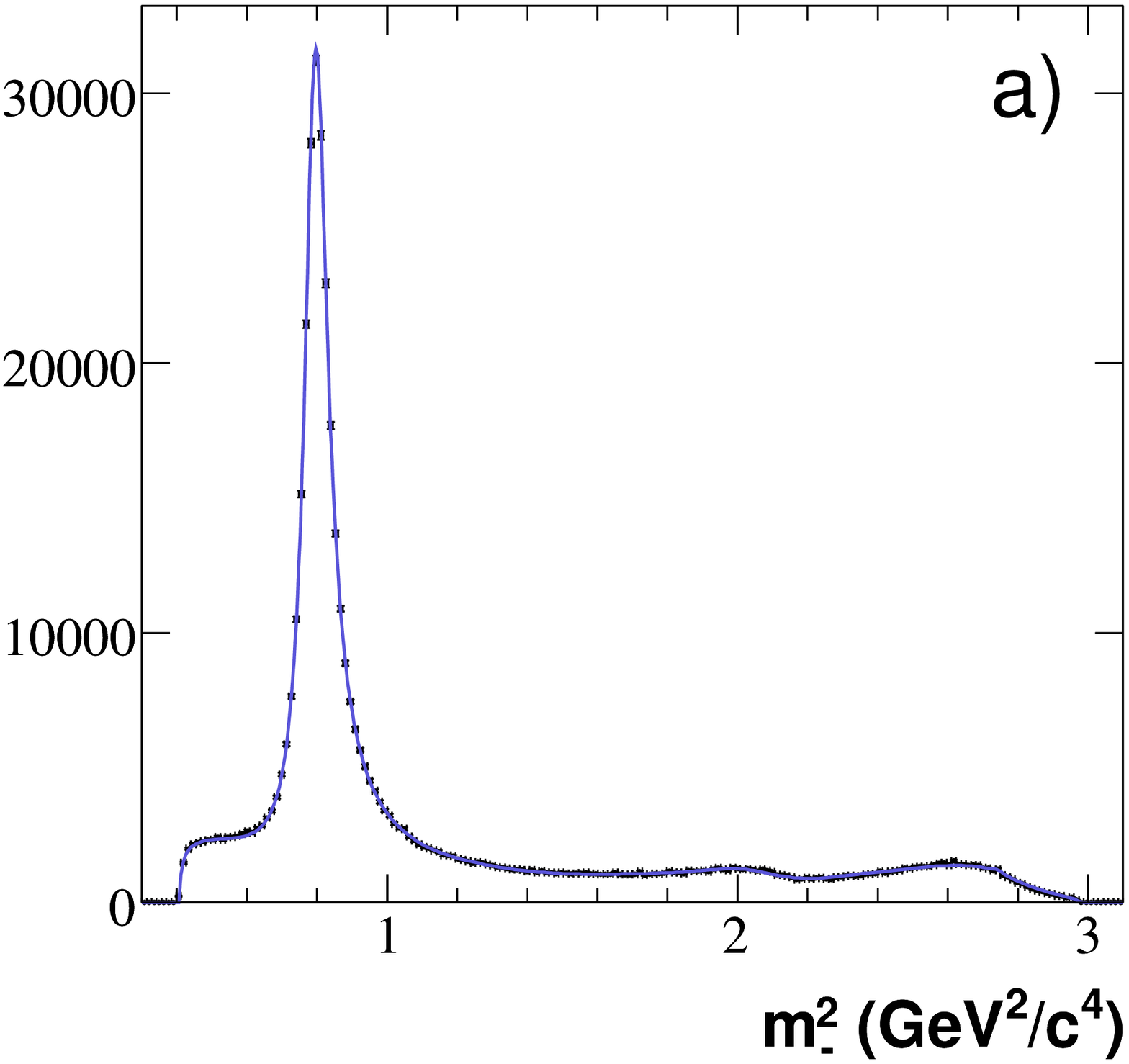}} & 
{\includegraphics[width=0.3\textwidth]{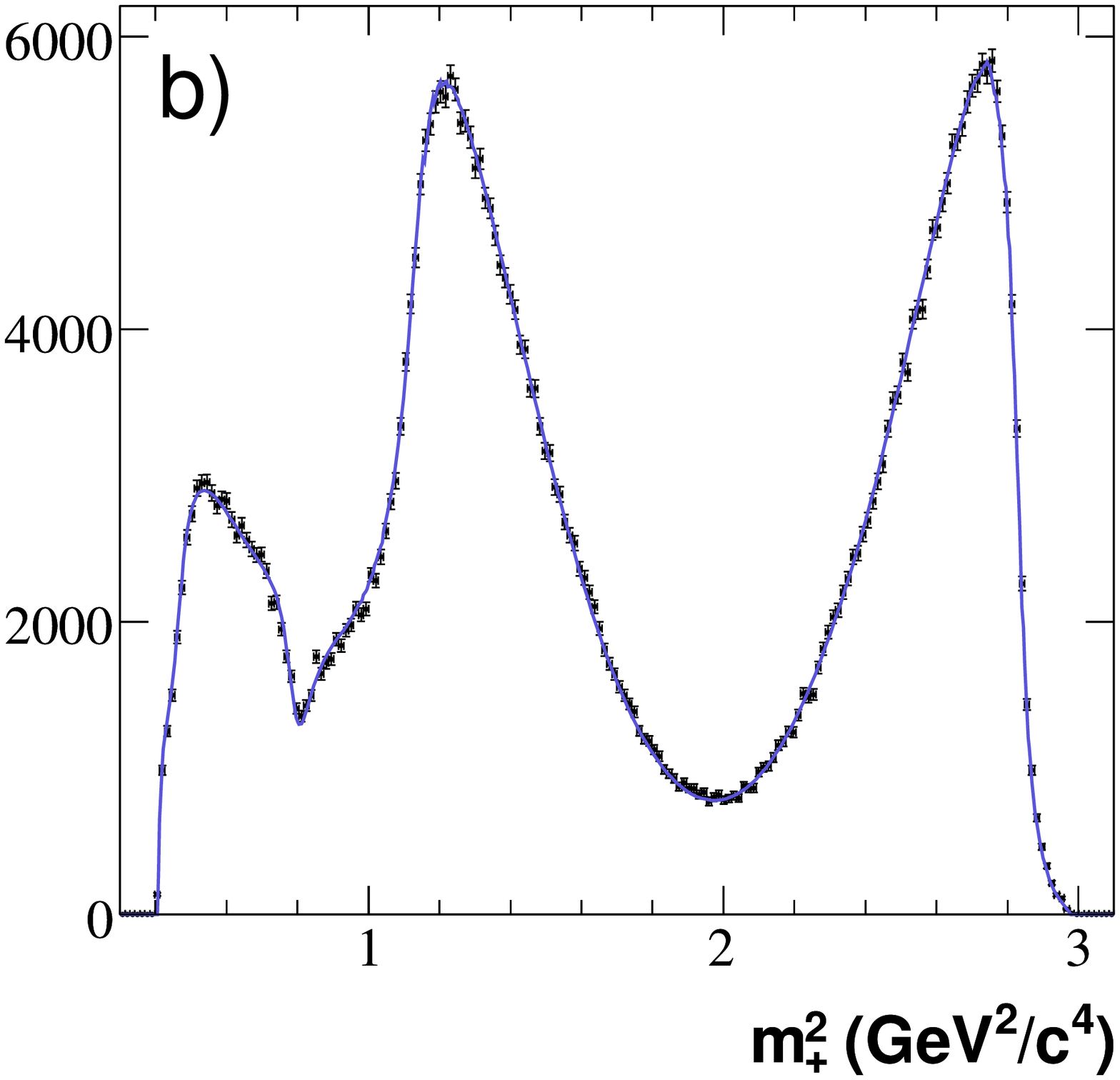}} & 
{\includegraphics[width=0.3\textwidth]{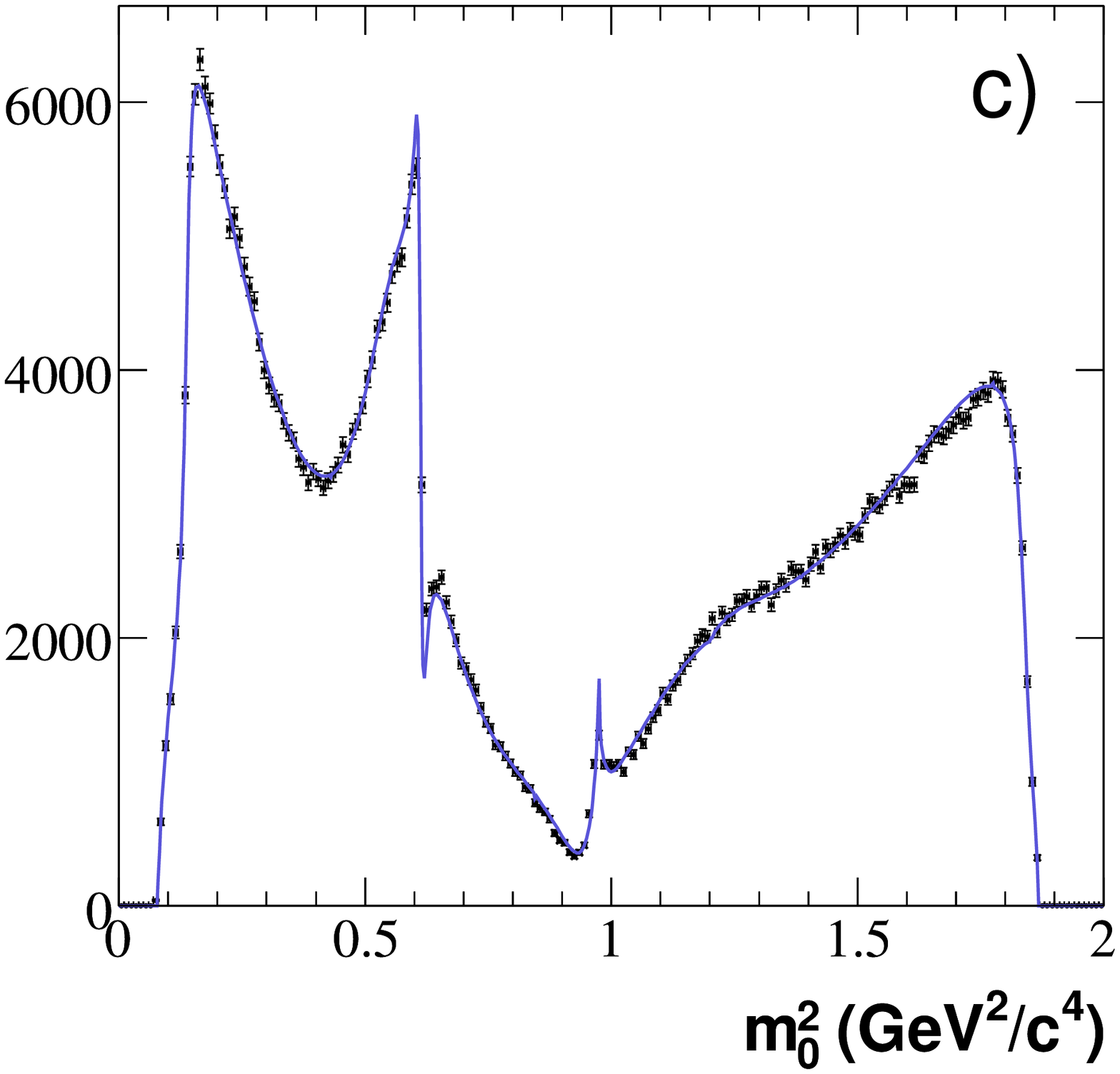}} \\ 
{\includegraphics[width=0.3\textwidth]{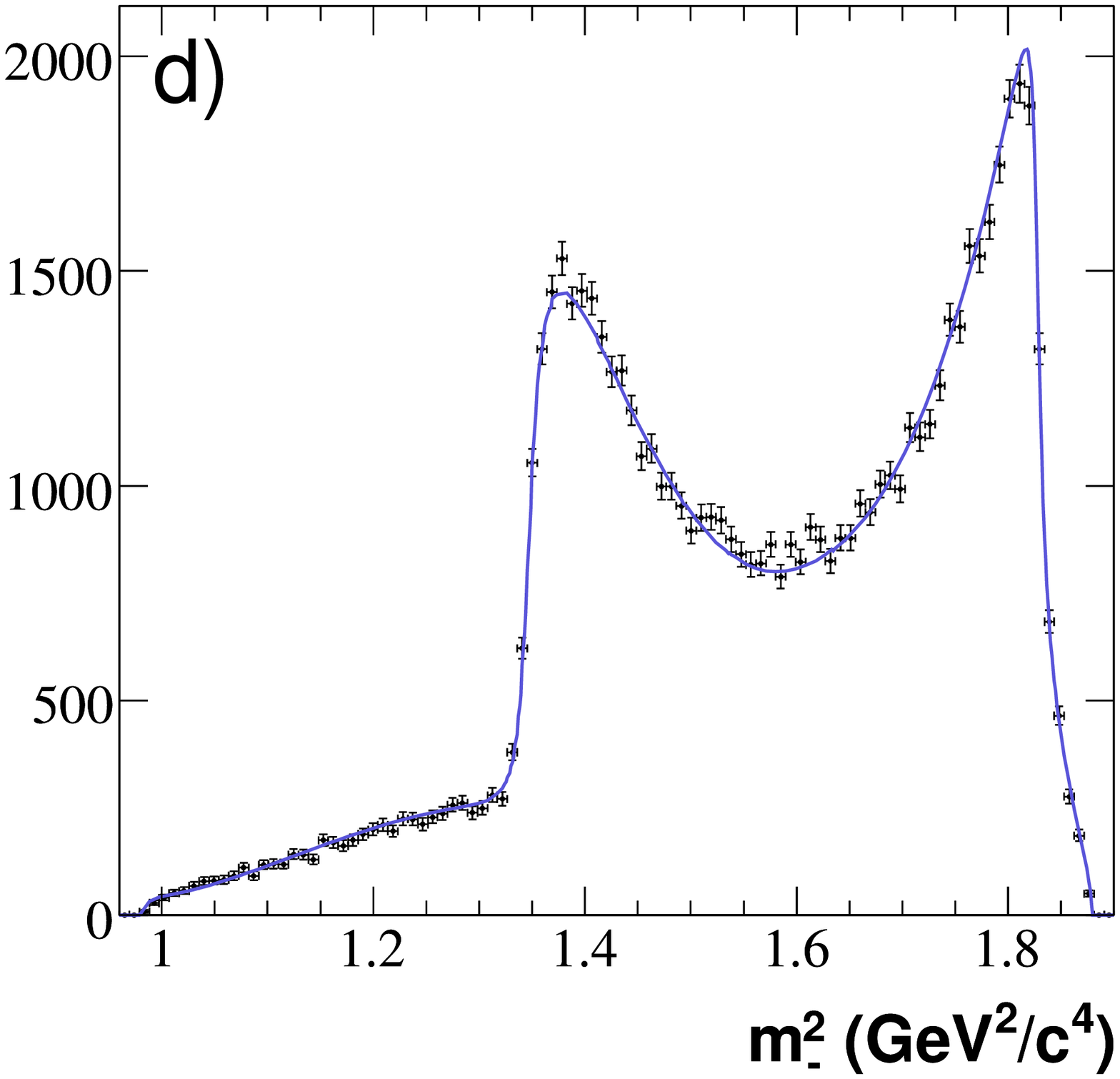}} & 
{\includegraphics[width=0.3\textwidth]{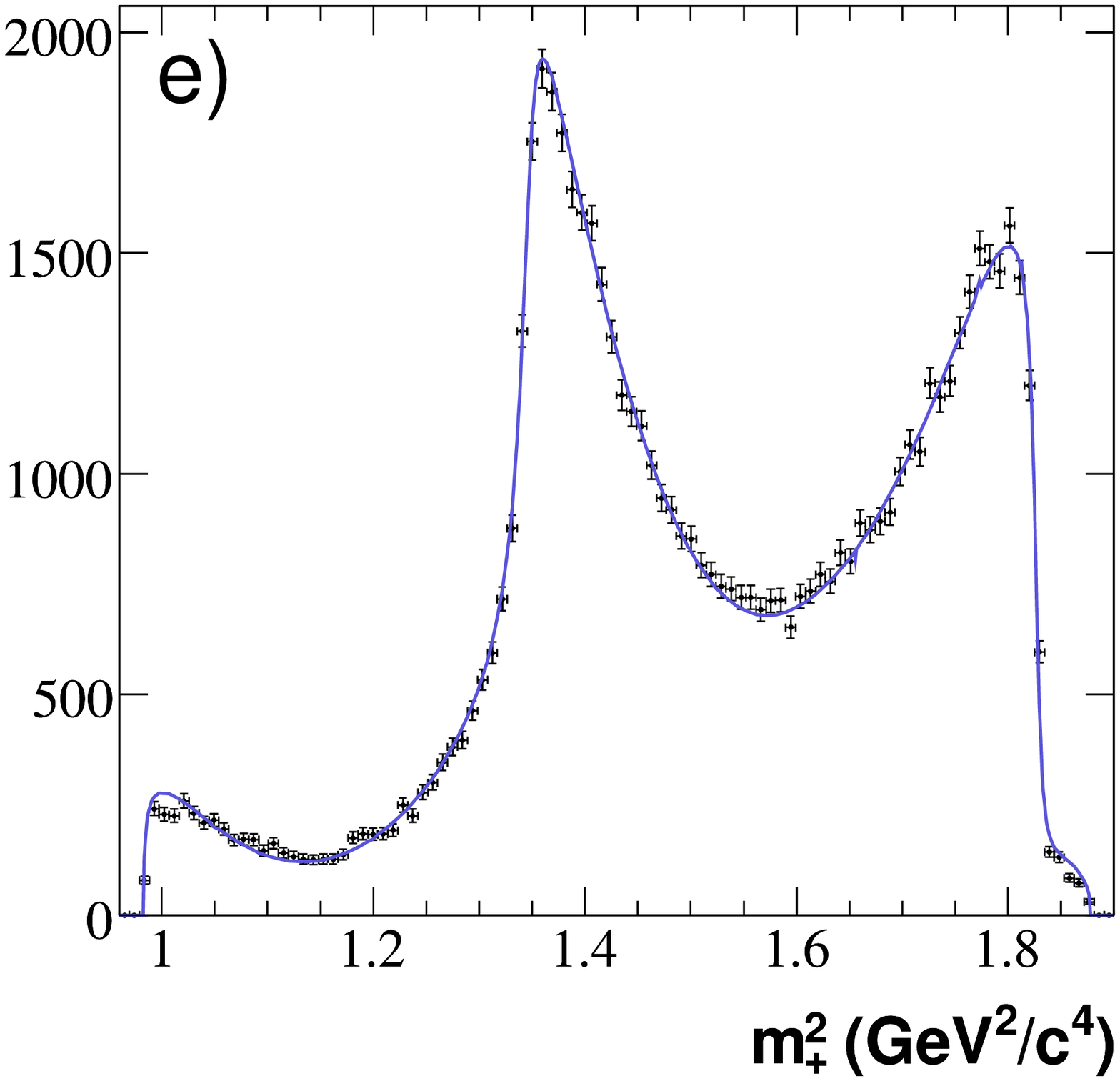}} & 
{\includegraphics[width=0.3\textwidth]{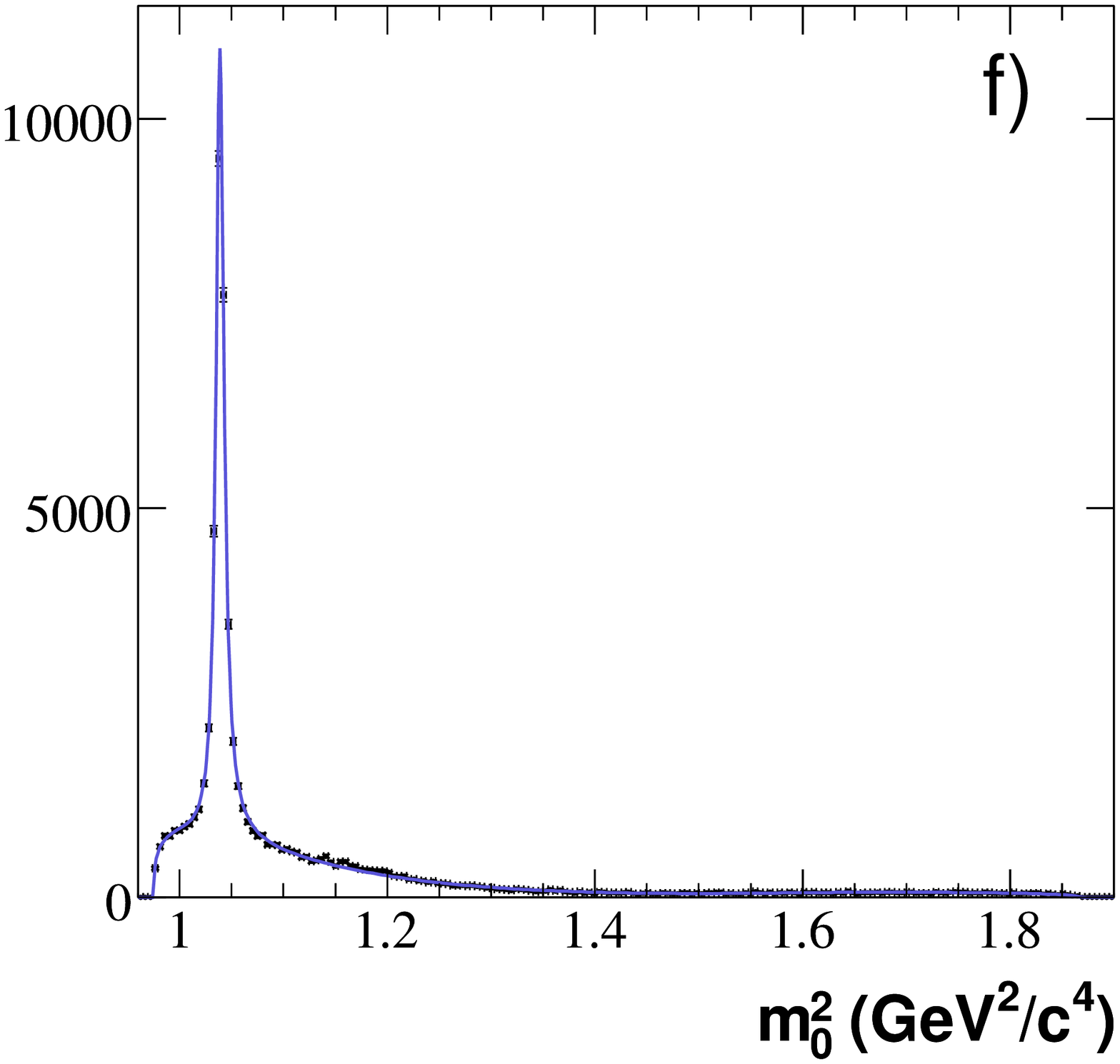}} \\ 
\end{tabular}   
\caption{\label{fig:DalitzModels} (color online). \Dztokshh Dalitz plot projections from $\Dstarp \to \Dz \pip$ events on  
(a,d) $m^2_-$, (b,e) $m^2_+$, and (c,f) $m^2_0$, for  
(a,b,c) \Dztokspipi and (d,e,f) \Dztokskk. 
The curves are the reference model fit projections.  
} 
\end{figure*}

Table~\ref{tab:kspipimodel} summarizes the values obtained for all free parameters of the \Dztokspipi Dalitz model: 
CA, DCS, and \CP eigenstates complex amplitudes $a_r e^{i\phi_r}$, $\pipi$ S-wave P-vector parameters, and $\K\pi$ S-wave parameters,  
along with the fit fractions.  
The non-resonant term of Eq.~(\ref{eq:k-matrix}) has not been included since the $\pi\pi$ and $K\pi$ S-wave  
parameterizations naturally account for their respective non-resonant contributions. 
The fifth P-vector channel and pole have also been excluded since the $\eta\eta'$ threshold and the pole mass $m_5$  
are both far beyond our $\pi\pi$ kinematic range, and thus there is little sensitivity to the associated parameters, 
$f_{15}^{\rm prod}$ and $\beta_5$, respectively. 
The amplitudes are measured with respect to $\Dz \to \KS \rho(770)^0$ which gives the second largest contribution. 
We report statistical errors only for the amplitudes, but for the fit fractions we also include systematic  
uncertainties (see Sec.~\ref{sec:dalitzmodels-systematics}), which largely dominate.  
The $\K\pi$ and $\pi\pi$ P-waves dominate the decay, but significant contributions from the corresponding S-waves 
are also observed (above 6 and 4 standard deviations, respectively). 
We obtain a sum of fit fractions of $(103.6 \pm 5.2)\%$, and the goodness of fit is estimated through a two-dimensional $\chi^2$ test 
performed binning the Dalitz plot into square regions of size $0.015~\gev^2/c^4$,  
yielding a reduced $\chi^2$ of $1.11$ (including statistical errors only) for $19274$ degrees of freedom.  
The variation of the contribution to the $\chi^2$ as a function of the Dalitz plot position is approximately uniform. 
Figure~\ref{fig:DalitzModels}(a,b,c) shows the Dalitz fit projections overlaid with the data distributions.  
The Dalitz plot distributions are well reproduced, with some small discrepancies in low and high mass regions  
of the $m^2_0$ projection, and in the $\rho(770)^0-\omega(782)$ interference region. 
 
%
 
\begin{table}[hbt!] 
\caption{\label{tab:kspipimodel} CA, DCS, and \CP eigenstates complex amplitudes $a_r e^{i\phi_r}$, $\pi\pi$ S-wave P-vector parameters, 
$\K\pi$ S-wave parameters, and fit fractions, as obtained from the fit of the \Dztokspipi Dalitz plot distribution from $\Dstarp \to \Dz \pip$.  
P-vector parameters $f_{1v}^{'\rm prod}$, for $v\ne1$, are defined as $f_{1v}^{\rm prod}/f_{11}^{\rm prod}$.  
Errors for amplitudes are statistical only, while for fit fractions include statistical and systematic uncertainties, 
largely dominated by the latter. Upper limits on fit fractions are quoted at 95\% confidence level.} 
\begin{ruledtabular} 
\begin{tabular}{lccc} 
\\[-0.15in] 
    Component  &  $a_r$   &  $\phz\phz\phz\phi_r~({\rm deg})$ & Fraction (\%) \\ [0.01in] 
\hline 
$K^{*}(892)^-$        &    $1.740\pm0.010$                  &  $\phm139.0\pm0.3$             & $55.7\pm2.8$ \\   
$K^{*}_0(1430)^-$     &    $\phz\phz8.2\pm0.7\phz\phz$      &  $\phm153\pm8$                 & $10.2\pm1.5$ \\   
$K^{*}_2(1430)^-$     &    $1.410\pm0.022$               &  $\phm138.4\pm1.0$                 &  $\phz2.2\pm1.6$ \\   
$K^{*}(1680)^-$       &    $\phz1.46\pm0.10\phz$            &  $-174\pm4$   &  $\phz0.7\pm1.9$ \\  
 
\hline 
$K^{*}(892)^+$        &    $0.158\pm0.003$                  & $\phz\phz-42.7\pm1.2\phz\phz$    & $\phz0.46\pm0.23$ \\  
$K^{*}_0(1430)^+$     &    $\phz0.32\pm0.06\phz$            & $\phm\phz143\pm11$      & $<0.05$ \\ 
$K^{*}_2(1430)^+$     &    $0.091\pm0.016$                  & $\phm\phz\phz85\pm11$    & $<0.12$ \\ 
\hline 
$\rho(770)^0$         &    $1$                              & $\phm\phz\phz0$             & $21.0\pm1.6$ \\  
$\omega(782)$         &    $0.0527\pm0.0007$               & $\phm126.5\pm0.9$    & $\phz0.9\pm1.0$ \\  
$f_2(1270)$           &    $0.606\pm0.026$                  & $\phm157.4\pm2.2$    & $\phz0.6\pm0.7$ \\  
\hline 
$\beta_1$             &    $\phz\phz9.3\pm0.4\phz\phz$      & $\phz-78.7\pm1.6\phz$   &  \\ 
$\beta_2$             &    $10.89\pm0.26\phm$         & $-159.1\pm2.6$ &   \\  
$\beta_3$             &    $\phz24.2\pm2.0\phz\phz$             & $\phm168\pm4$ & \\  
$\beta_4$             &    $\phz\phz9.16\pm0.24\phz\phz$      & $\phm\phz90.5\pm2.6$ &\\  
$f_{11}^{\rm prod}$   &    $\phz7.94\pm0.26\phz$            & $\phm\phz73.9\pm1.1$  &\\  
$f_{12}^{'\rm prod}$  &    $\phz\phz2.0\pm0.3\phz\phz$      & $\phz-18\pm9\phz$   & \\ 
$f_{13}^{'\rm prod}$  &    $\phz\phz5.1\pm0.3\phz\phz$      & $\phm\phz33\pm3$    & \\ 
$f_{14}^{'\rm prod}$  &    $\phz3.23\pm0.18\phz$            & $\phm\phz\phz4.8\pm2.5$   & \\ 
$s_0^{\rm prod}$      & \multicolumn{2}{c}{$-0.07\pm0.03$} & \\ 
$\pi\pi$ S-wave       &                    &               & $11.9\pm2.6$\\  
\hline 
$M$~(\gevcc)          &  \multicolumn{2}{c}{$\phm1.463\pm0.002$}   &  \\ 
$\Gamma$~(\gevcc)     &  \multicolumn{2}{c}{$\phm0.233\pm0.005$}   &  \\ 
$F$                   &  \multicolumn{2}{c}{$\phm\phz0.80\pm0.09\phz$}    &\\  
$\phi_F$              &  \multicolumn{2}{c}{$\phm\phz2.33\pm0.13\phz$}   &\\  
$R$                   &  \multicolumn{2}{c}{$\phm1$}   &\\  
$\phi_R$              &  \multicolumn{2}{c}{$-5.31\pm0.04$}  &\\  
$a$                   &  \multicolumn{2}{c}{$\phm1.07\pm0.11$}    &\\  
$r$                   &  \multicolumn{2}{c}{$-1.8\pm0.3$}    &\\  
\end{tabular} 
\end{ruledtabular} 
\end{table}

As a cross-check,  
we alternatively parameterize the $\pi\pi$ and $K\pi$ S-waves using the isobar approximation with the following BW amplitudes 
(plus the non-resonant contribution):  
the CA $K^{*}_0(1430)^-$, 
the DCS $K^{*}_0(1430)^+$, and the \CP eigenstates $f_0(980)$, $f_0(1370)$, $\sigma$ and an ad hoc $\sigma'$. 
This model is very similar to that used in our previous measurement of $\gamma$~\cite{ref:babar_dalitzpub}, except that 
the $K^{*}(1410)^-$ and $\rho(1450)^0$ resonances have been removed because of their negligible  
fit fractions. 
Masses and widths of the $\sigma$ and $\sigma'$ scalars are obtained from the fit, 
$M_{\sigma}=528\pm5$, $\Gamma_{\sigma}=512\pm9$, $M_{\sigma'}=1033\pm4$, and $\Gamma_{\sigma'}=99\pm6$, 
given in~\mevcc. 
Mass and width values for the $K^{*}_0(1430)^\mp$, $f_0(980)$, and $f_0(1370)$ are taken  
from~\cite{ref:Kst1430-E791,ref:f01370-E791}. 
We obtain a sum of fit fractions of $122.5\%$, and a reduced $\chi^2$ of $1.20$ (with statistical errors only) for $19274$ degrees of freedom,  
which strongly disfavors the isobar approach in comparison to the K-matrix formalism.

\subsection{\Dztokskk Dalitz model} 
\label{sec:dalitzmodels-kskk}

The description of the \Dztokskk decay amplitude uses Eq.~(\ref{eq:genAmp}) and consists of five distinct resonances 
leading to 8 two-body decays: $\KS a_0(980)^0$, $\KS \phi(1020)$, $K^- a_0(980)^+$, $\KS f_0(1370)$,  
$K^+ a_0(980)^-$, $\KS f_2(1270)^0$, $\KS a_0(1450)^0$, and $K^- a_0(1450)^+$. This isobar model is essentially identical 
to that used in our previous analysis of the same reaction~\cite{ref:babar_D0toKsKK},  
but for 
the addition of the $a_0(1450)$ scalar, whose contribution is strongly supported by the much larger data sample,  
as well as of a D-wave contribution parameterized with the $f_2(1270)$ tensor.  
Attempts to improve the model quality by adding 
other contributions (including the non-resonant term) did not give better results.  
 
The $\phi(1020)$ resonance is described using a relativistic BW, with mass and width left free in our fit to the 
\Dz tagged sample in order to account for mass resolution effects. The $a_0(980)$ resonance has a mass very close to 
the $\K\Kb$ threshold and decays mostly to $\eta\pi$. Therefore it is described using a coupled  
channel BW~\cite{ref:RevDalitzPlotFormalism,ref:babar_D0toKsKK}, where the mass pole and coupling constant 
to $\eta\pi$ are taken from~\cite{ref:crystallbarrel}, while the coupling constant to 
$\K\Kb$, $g_{\K\Kb}$, is determined from our fit.  
 
Table~\ref{tab:kskkmodel} summarizes the values obtained for all free parameters of the \Dztokskk Dalitz model, 
the complex amplitudes $a_r e^{i\phi_r}$, the mass and width of the $\phi(1020)$ and the coupling constant $g_{\K\Kb}$, 
together with the fit fractions.  
The value of $g_{\K\Kb}$ is consistent with our previous result~\cite{ref:babar_D0toKsKK}, and differs significantly  
from the measurement reported in~\cite{ref:crystallbarrel}. 
All amplitudes are measured with respect to $\Dz \to \KS a_0(980)^0$, which gives the largest contribution. 
The sum of fit fractions is $152.3\%$, and the reduced $\chi^2$ is $1.09$ (with statistical errors only)  
for $6856$ degrees of freedom, estimated from a binning of the Dalitz plot into square regions of size $0.045~\gev^2/c^4$. 
The variation of the contribution to the $\chi^2$ as a function of the Dalitz plot position is approximately uniform in the 
regions where most of the decay dynamics occurs. 
Figure~\ref{fig:DalitzModels}(d,e,f) shows the fit projections overlaid with the data distributions.  
The Dalitz plot distributions are well reproduced, with some small discrepancies at the peaks of the 
$m^2_-$ and $m^2_+$ projections.


\begin{table}[hbt!] 
\caption{\label{tab:kskkmodel} \CP eigenstates, CA, and DCS complex amplitudes $a_r e^{i\phi_r}$ and fit fractions,  
obtained from the fit of the \Dztokskk Dalitz plot distribution from $\Dstarp \to \Dz \pip$.  
The mass and width of the $\phi(1020)$, and the $g_{\K\Kb}$ coupling constant are simultaneously determined in the fit, 
yielding $M_{\phi(1020)} = 1.01943 \pm 0.00002$~\gevcc, $\Gamma_{\phi(1020)} = 4.59319 \pm 0.00004$~\mevcc,  
and $g_{\K\Kb} = 0.550 \pm 0.010$~\gevcc.  
Errors for amplitudes are statistical only.  
Uncertainties (largely dominated by systematic contributions) are not estimated for the fit fractions. 
} 
\begin{ruledtabular} 
\begin{tabular}{lccc} 
\\[-0.15in] 
    Component  &  $a_r$   &  $\phz\phz\phz\phi_r~({\rm deg})$ & {\hskip-0.2cm}Fraction (\%) \\ [0.01in] 
\hline 
$\KS a_0(980)^0$      &  $1$                      &  $\phm\phz\phz0$                & $55.8$  \\ 
$\KS \phi(1020)$      &  $0.227\pm0.005$          &  $\phz\phz-56.2\pm1.0\phz\phz$  & $44.9$   \\ 
$\KS f_0(1370)$       &  $0.04\pm0.06$            &  $\phz\phz\phz-2\pm80\phz$      & $\phz0.1$  \\ 
$\KS f_2(1270)$       &  $\phz0.261\pm0.020\phz$  &  $\phz\phz\phz-9\pm6\phz\phz$   & $\phz0.3$      \\ 
$\KS a_0(1450)^0$     &  $\phz0.65\pm0.09\phz$    &  $\phz\phz-95\pm10\phz$         & $12.6$       \\ 
\hline 
$K^- a_0(980)^+$      &  $0.562\pm0.015$          &  $\phm179\pm3$                  & $16.0$  \\ 
$K^- a_0(1450)^+$     &  $\phz0.84\pm0.04\phz$    &  $\phm\phz97\pm4$               & $21.8$  \\ 
\hline 
$K^+ a_0(980)^-$      &  $0.118\pm0.015$          &  $\phm138\pm7$                  & $\phz0.7$       \\ 
\end{tabular} 
\end{ruledtabular} 
\end{table}

\subsection{Systematic uncertainties} 
\label{sec:dalitzmodels-systematics} 
 
Systematic uncertainties on ${\cal A}_D(\mtwo)$ are evaluated by repeating the fit 
to the tagged \Dz samples with alternative assumptions to those adopted in the reference \Dztokspipi and \Dztokskk amplitude analyses. 
These uncertainties can then be directly propagated to the measurement of the \CP parameters, 
as discussed in Sec.~\ref{sec:systematics-cp}, and the total systematic error can be obtained from the sum square of the  
individual contributions.  
In this paper we have also propagated these systematic uncertainties to the measurement  
of the \Dztokspipi fit fractions, as reported in Table~\ref{tab:kspipimodel}. 
In general, each of the considered alternative models has a reduced $\chi^2$ poorer than that of the reference. Therefore,  
our systematic uncertainties do not include potential contributions due to the residual poor quality of the reference model fit,  
as reported in Secs.~\ref{sec:dalitzmodels-kspipi} and \ref{sec:dalitzmodels-kskk}.

\subsubsection{Model contributions} 
\label{sec:dalitzmodels-systematics-model} 
 
Dalitz model systematic uncertainties on ${\cal A}_D(\mtwo)$ are related to the model dependence of the  
strong charm decay phase as a function of the Dalitz plot position when it is determined from the  
Dalitz plot density, which only depends on decay rates.  
 
We use alternative models where the BW parameters are varied according to their uncertainties or changed 
by values measured by other experiments. This is the case of the $f_0(1370)$, where 
the reference values~\cite{ref:Kst1430-E791} are replaced by alternative measurements~\cite{ref:BES}, 
and the $K^{*}(1680)^-$, where the reference parameters~\cite{ref:lass} are replaced by those from~\cite{ref:pdg2006}. 
We also build models using alternative parameterizations, as in the 
case of the $\rho(770)^0$ where the reference Gounaris-Sakurai form is replaced by the standard relativistic BW.

To estimate the $\pi\pi$ S-wave systematic error we replace the reference K-matrix solution (Table~\ref{tab:AS}) 
by all alternative solutions analyzed in Ref.~\cite{ref:AS}. Analogously, the uncertainty on the  
parameterization of the $K\pi$ S-wave is estimated using a standard relativistic BW describing the  
$K^*_0(1430)^\mp$ with parameters taken either from~\cite{ref:Kst1430-E791} 
or simultaneously determined from our fit to the \Dz sample. Additionally, the isobar model is used as a cross-check of  
the combined $\pi\pi$ and $\K\pi$ S-wave effect. 
  
Uncertainties due to our choice of the angular dependence are estimated by replacing the reference Zemach tensors by the  
helicity formalism. The effect is negligible for S-waves, very small for P-waves, but larger for D-waves~\cite{ref:zemach-covariant}.   
Other alternative models are built by changing the Blatt-Weisskopf radius between $0$ and $3~\gev^{-1}\hbar c$, 
and removing and adding resonances with small or negligible fit fractions.  
For \Dztokspipi, we added $\rho(1450)^0$ and CA $K^*(1410)^-$.  
Similarly, for \Dztokskk, we removed all the $a_0(1450)$ charged states and the $f_2(1270)$, and added the $f_0(980)$ and the charged DCS $a_0(1450)$. 
The $f_0(980)$ resonance is described using a coupled channel BW with parameters taken from a variety of  
experiments~\cite{ref:BES,ref:f0980-W76,ref:f01370-E791}.

\subsubsection{Experimental contributions} 
\label{sec:dalitzmodels-systematics-exp}

Experimental systematic errors come from uncertainties in the knowledge of variations of the reconstruction 
efficiency on the Dalitz plot, background Dalitz plot shapes, mass resolution, mistag rate, and binning. 
 
The uncertainty from the efficiency variations on the Dalitz plot $\epsilon (\mtwo)$ has been evaluated assuming 
the efficiency to be flat. Tracking efficiency studies in data and MC show that this method gives a conservative estimate  
of the imperfections of the detector simulation, which appear mainly at the boundaries of the phase space because of 
the presence of very low momentum tracks.

Systematic errors related to the background Dalitz plot profile ${\cal D}_{\rm bkg,\mp}(\mtwo)$ are determined  
assuming a flat shape, which gives the largest effect among other alternative profiles obtained  
using either the \mD sideband from continuum MC, or the \mD signal region from continuum MC after  
removal of true \Dz mesons.  
 
All the resonances, except for the $\omega(782)$ and $\phi(1020)$, have intrinsic width significantly larger than possible bias  
on invariant mass measurement and resolution. We estimate the systematic uncertainty  
associated with $\omega(782)$ by repeating the model fit using an overall width resulting from  
adding in quadrature its natural width and the mass resolution in the 782~\mevcc $\pi\pi$ mass region. No systematic 
error is assigned for the $\phi(1020)$, since the reference model has been extracted  
with its mass and width as free parameters. 
 
The uncertainty due to a wrong identification of the flavor of the \Dz (\Dzb) meson from the $\Dstarp \to \Dz \pip$ ($\Dstarm \to \Dzb \pim$) decay, 
due to the association of the \D meson with a random soft pion of incorrect charge has been evaluated taking into account 
explicitly the rate of mistags observed in the MC, at 0.7\% level. 
 
Effects from limited numerical precision in the computation of normalization integrals and binning  
in the Dalitz plane have been evaluated using coarser and thinner  
bins.

%% file: cpanalysis.tex
\begin{figure}[hbt!] 
\begin{tabular}{cc} 
\includegraphics[width=0.23\textwidth]{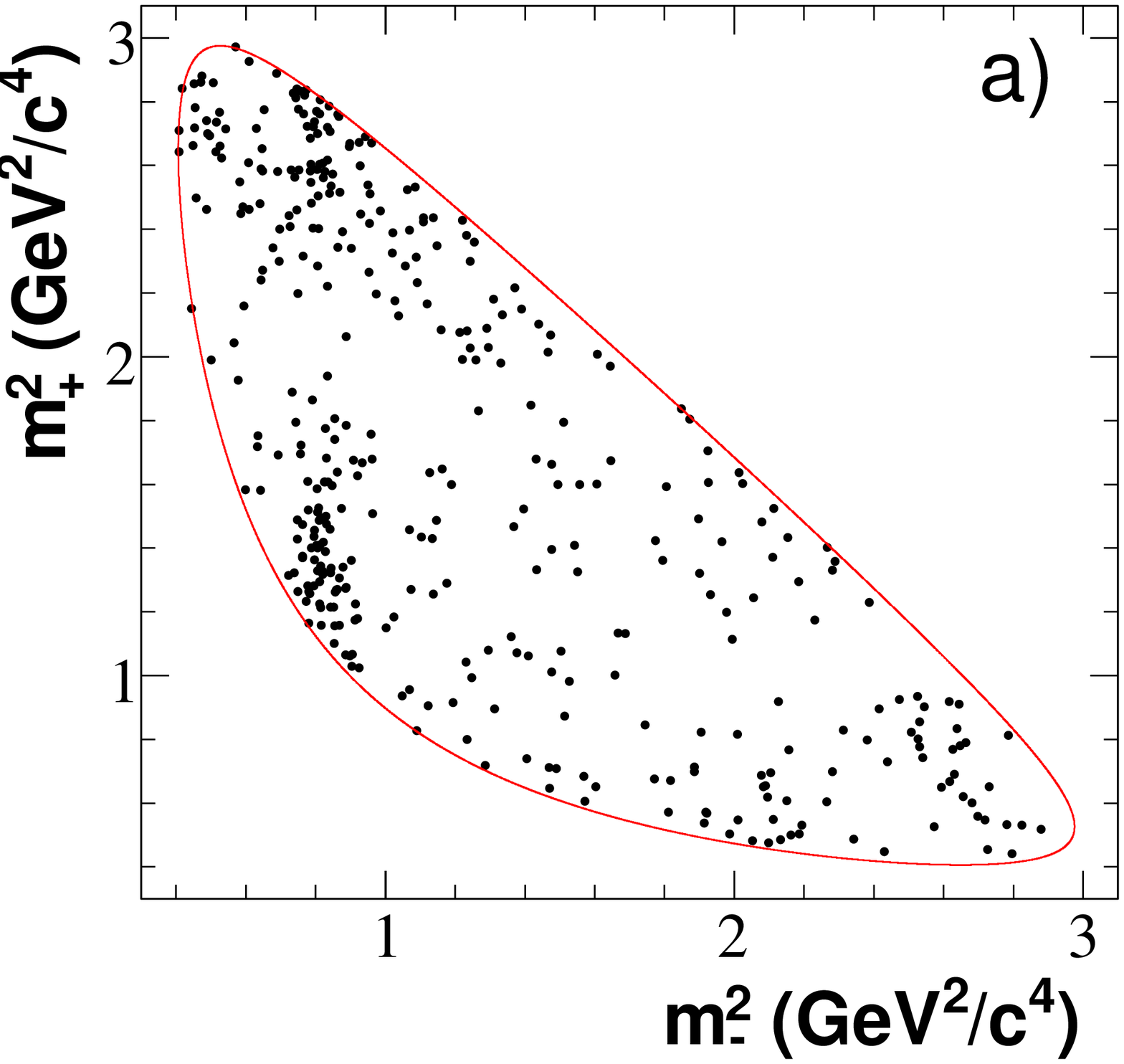}& 
\includegraphics[width=0.23\textwidth]{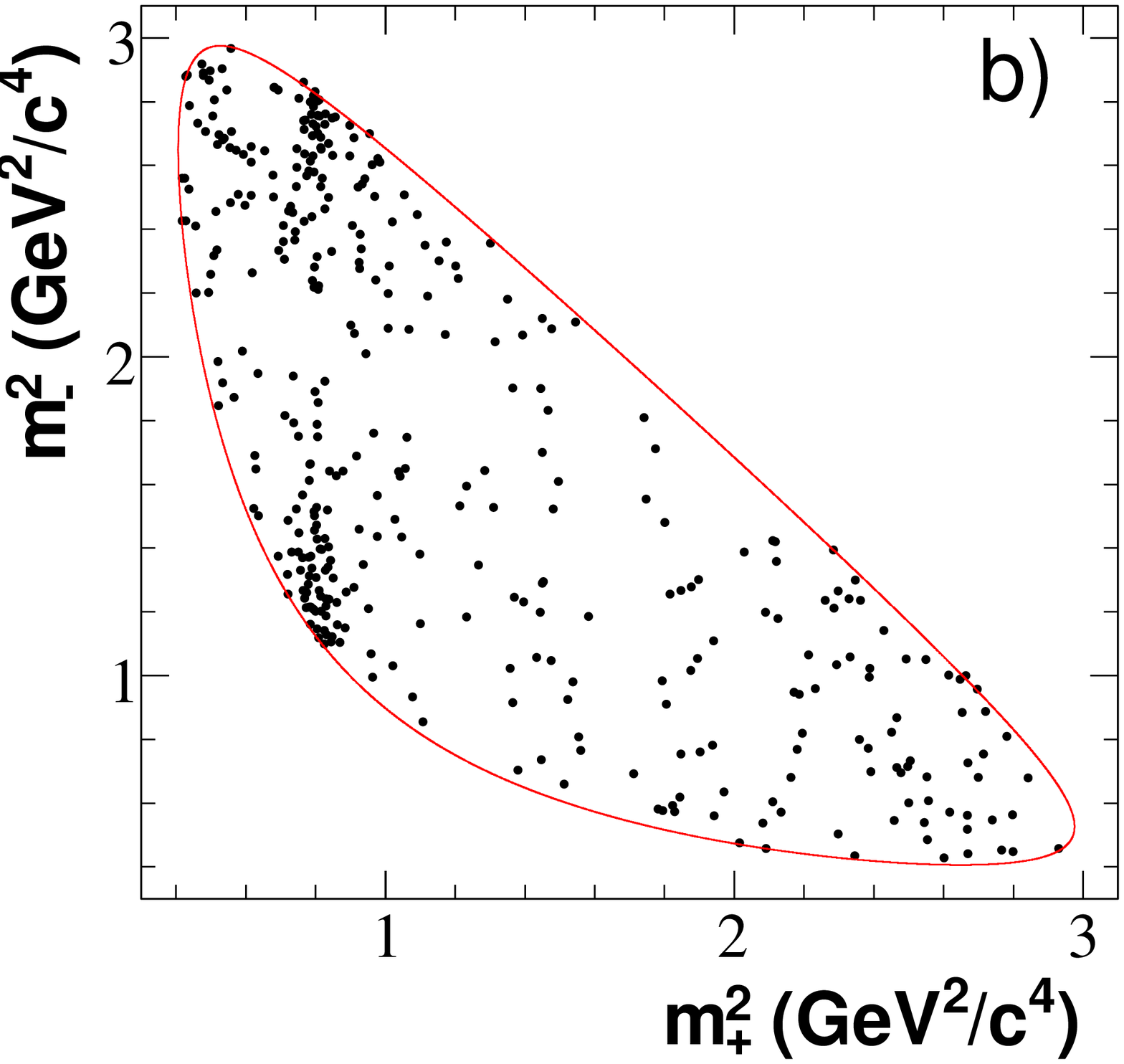} \\ 
\includegraphics[width=0.23\textwidth]{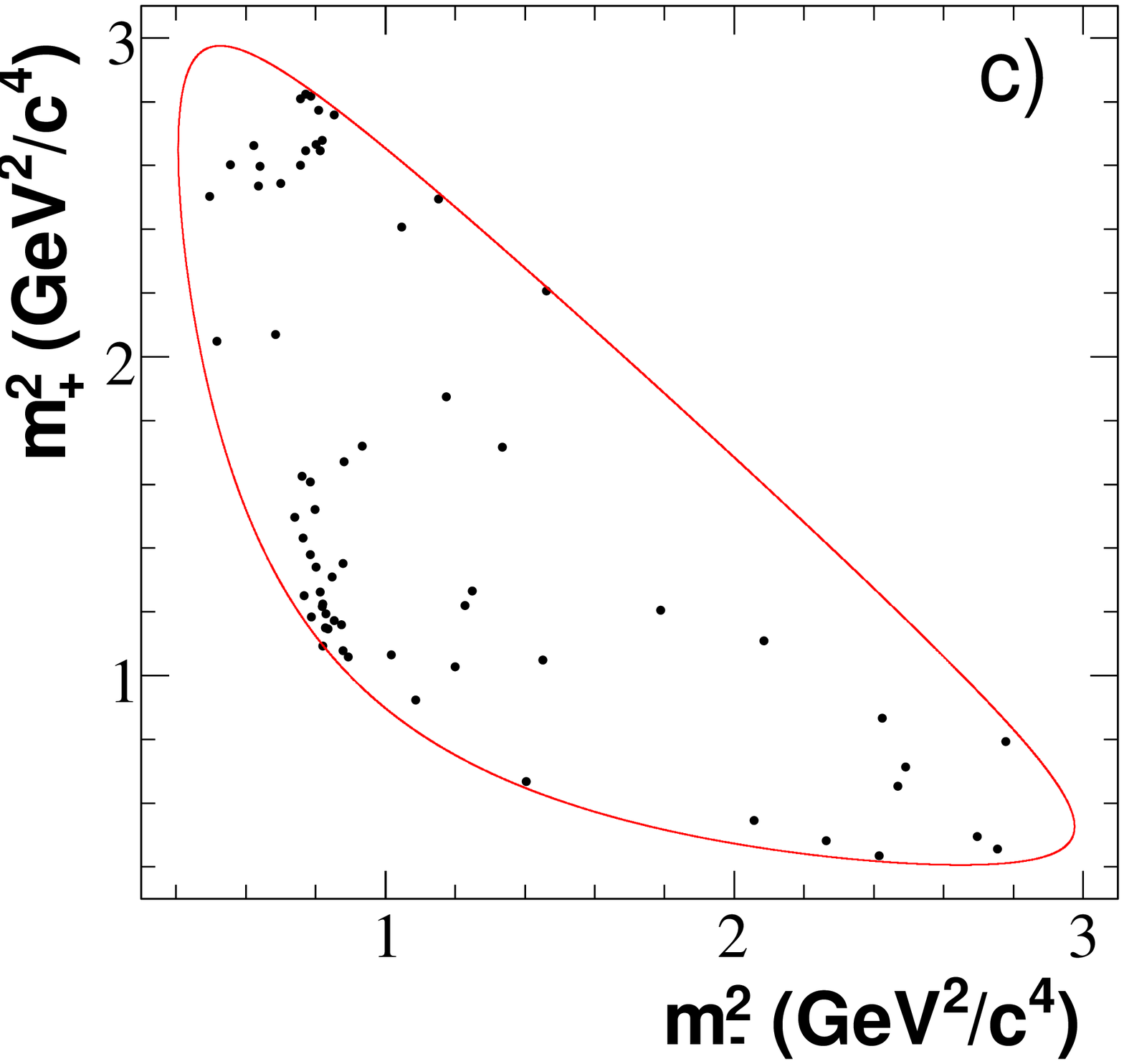}& 
\includegraphics[width=0.23\textwidth]{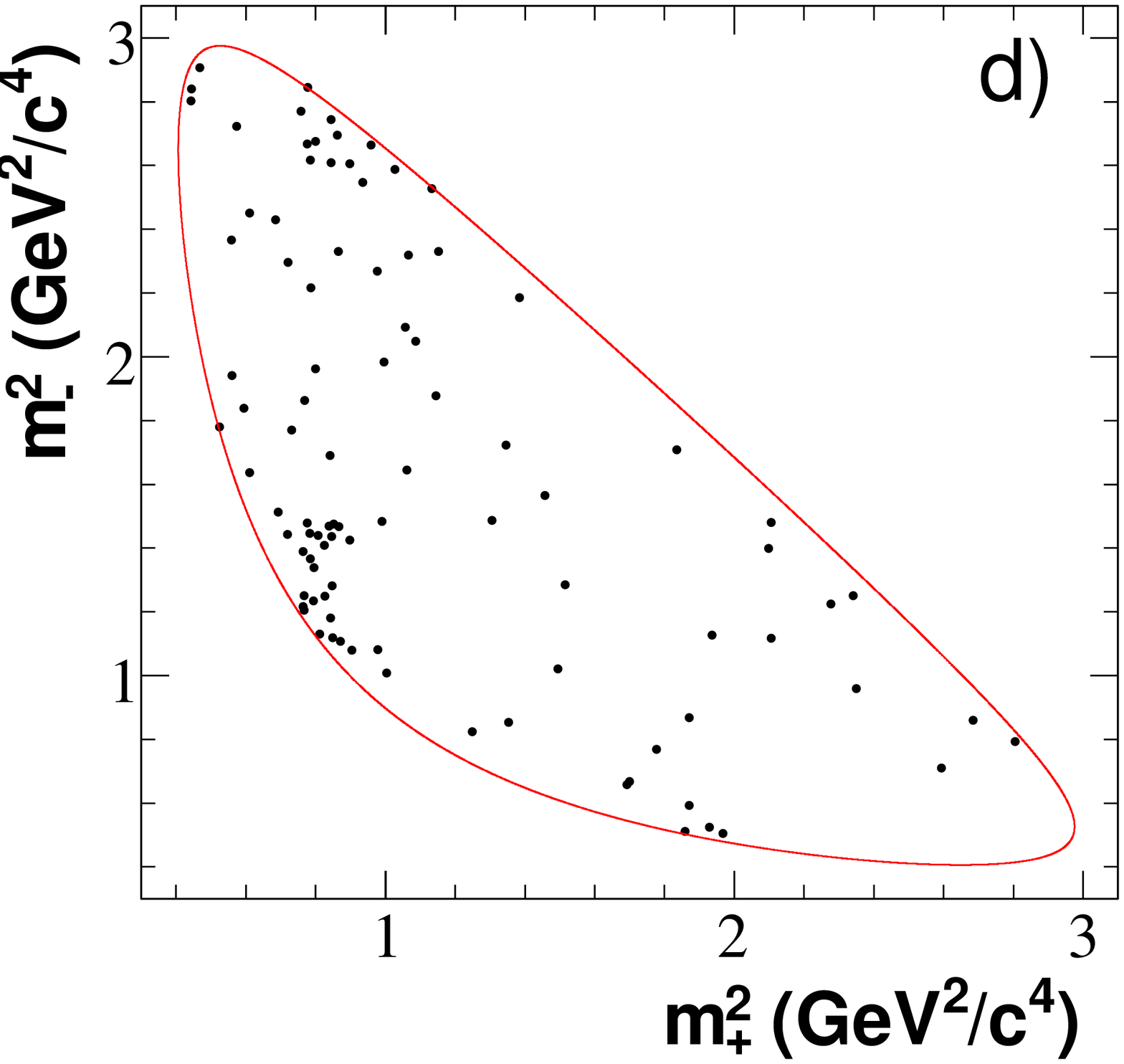} \\ 
\includegraphics[width=0.23\textwidth]{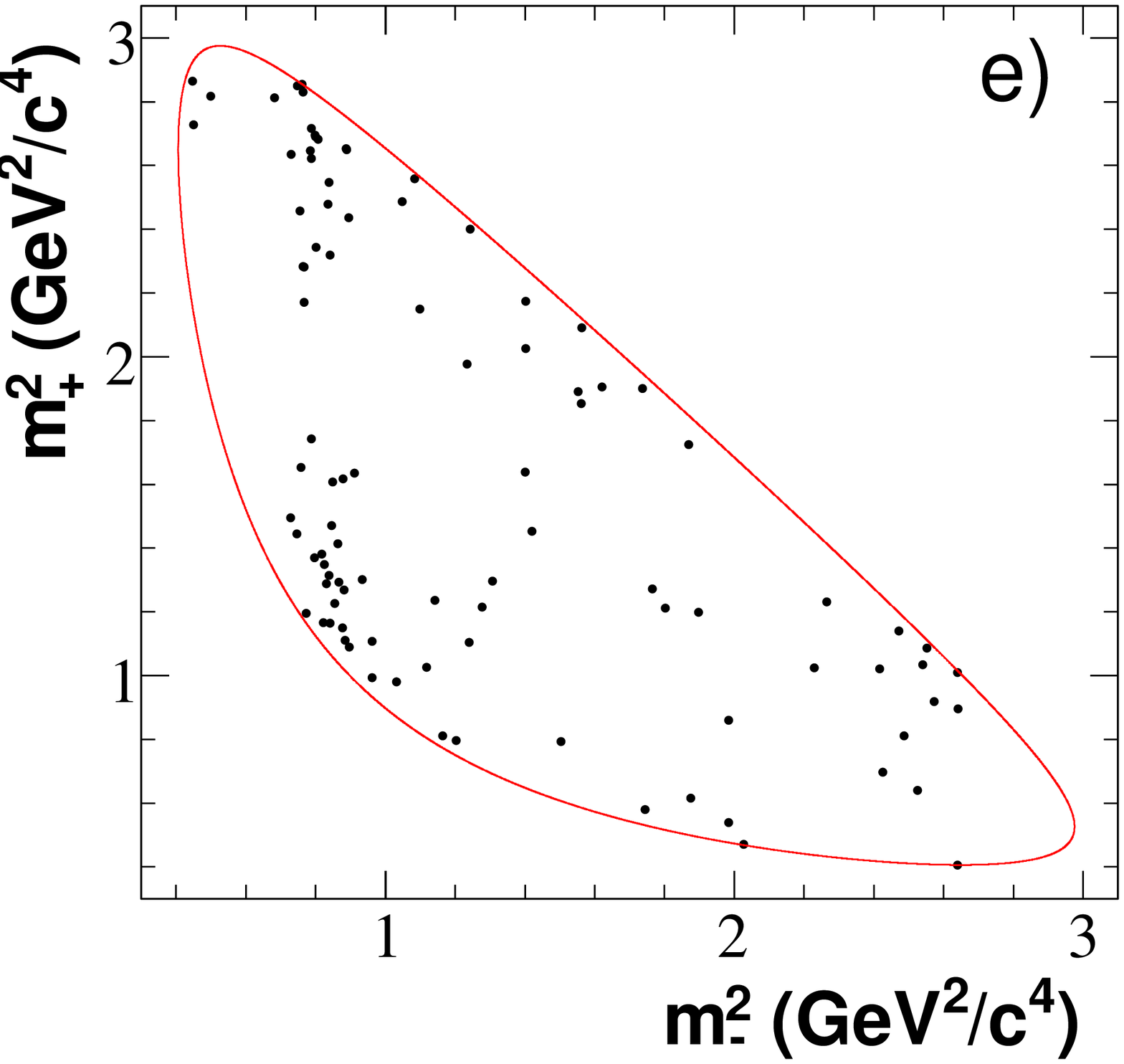}& 
\includegraphics[width=0.23\textwidth]{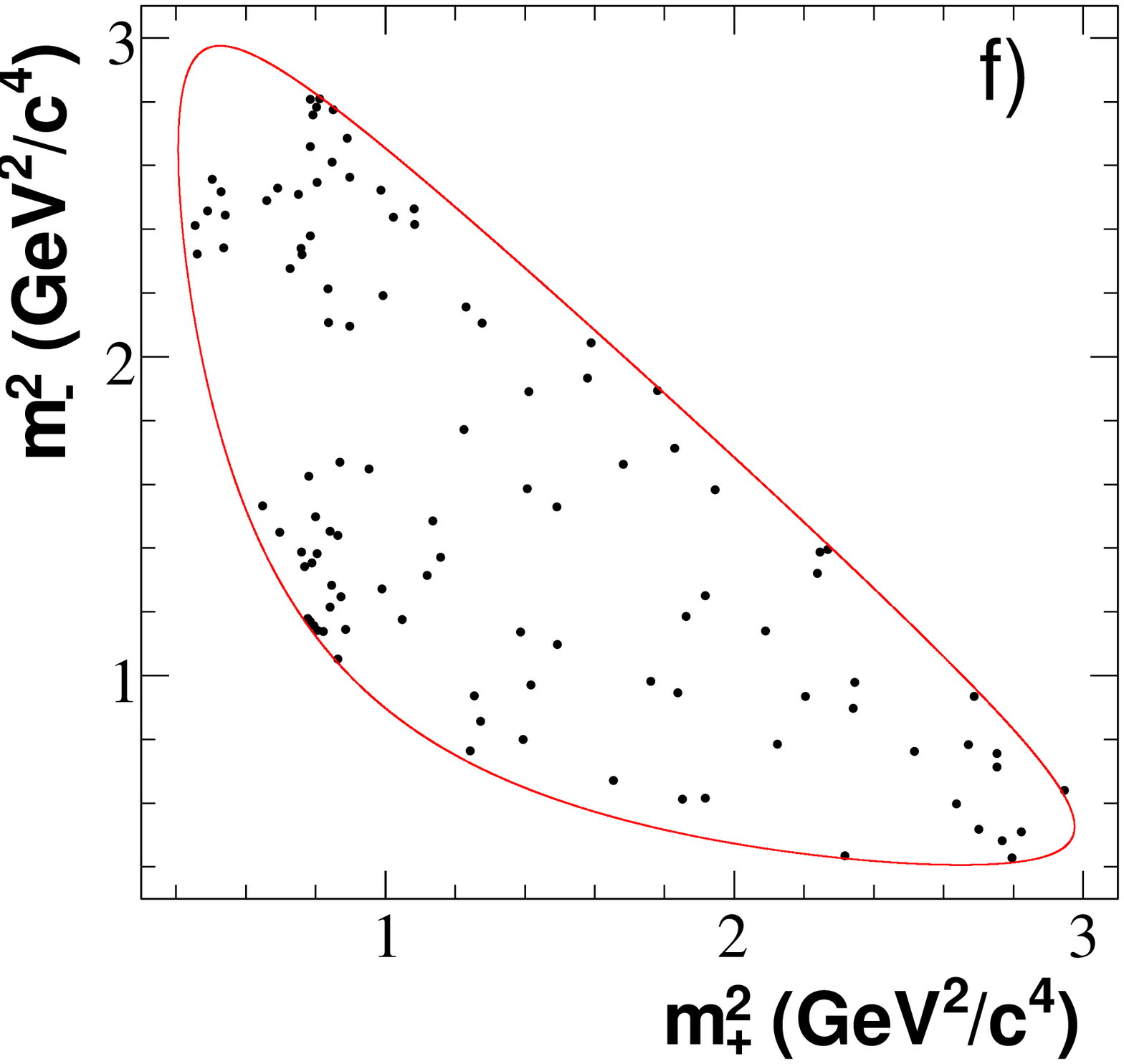} \\ 
\includegraphics[width=0.23\textwidth]{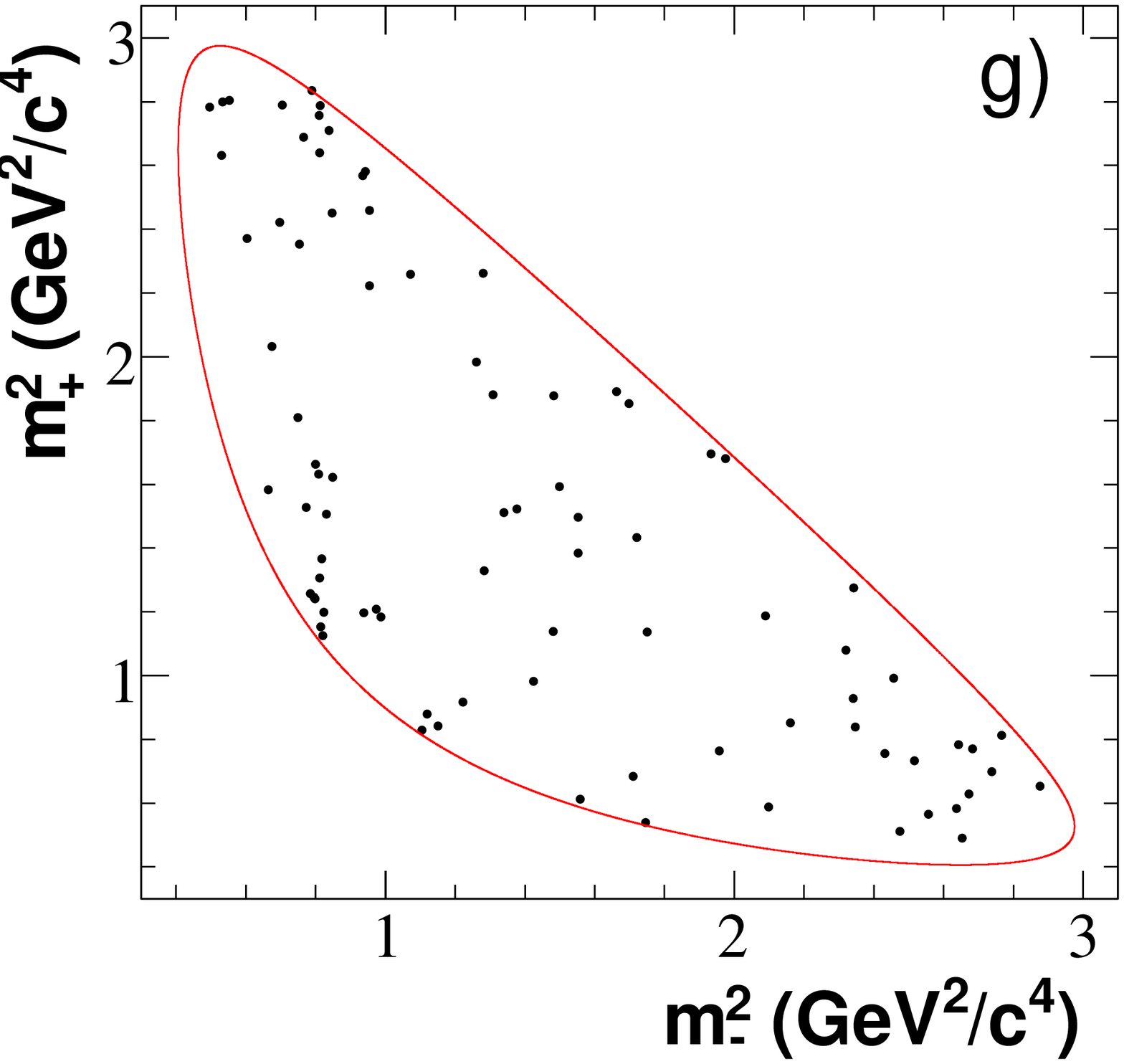}& 
\includegraphics[width=0.23\textwidth]{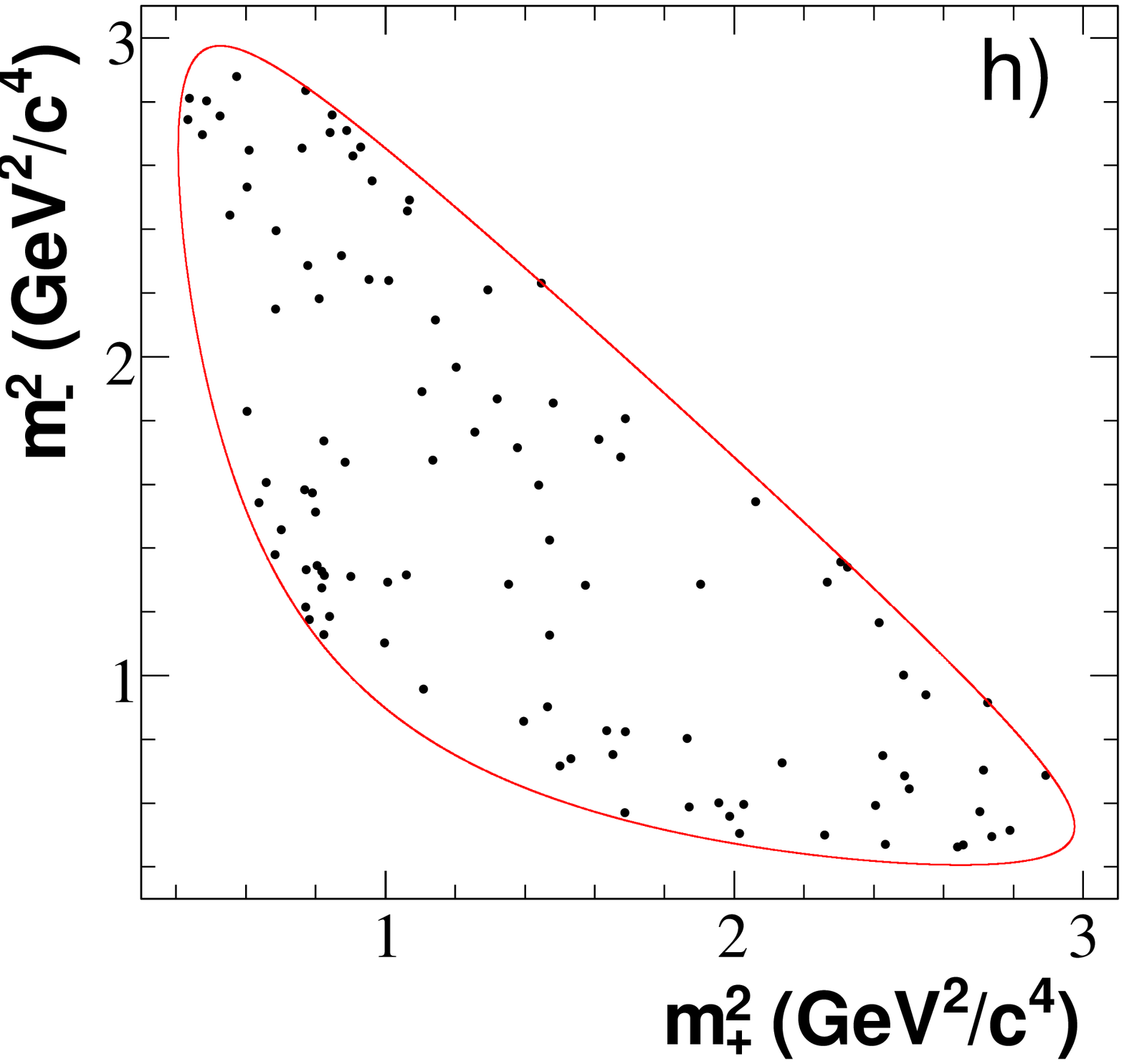} \\ 
\end{tabular} 
\caption{\label{fig:dalitz-kspipi} (color online). \Dztildetokspipi Dalitz plot distributions for  
(a) $\Bm \to \Dztilde \Km$, (b) $\Bp \to \Dztilde \Kp$,  
(c) $\Bm \to \Dstarztilde[\Dztilde\piz] \Km$, (d) $\Bp \to \Dstarztilde[\Dztilde\piz] \Kp$,  
(e) $\Bm \to \Dstarztilde[\Dztilde\gamma] \Km$, (f) $\Bp \to \Dstarztilde[\Dztilde\gamma] \Kp$,  
(g) $\Bm \to \Dztilde \Kstarm$, and (h) $\Bp \to \Dztilde \Kstarp$, for the \DeltaE signal region. 
The requirements $\mes>5.272$~\gevcc and $\fisher>-0.1$ have been applied to reduce the background contamination,  
mainly from continuum events. 
The contours (solid red lines) represent the kinematical limits of the \Dztildetokspipi decay.  
} 
\end{figure}
\begin{figure}[hbt!] 
\begin{tabular}{cc} 
\includegraphics[width=0.23\textwidth]{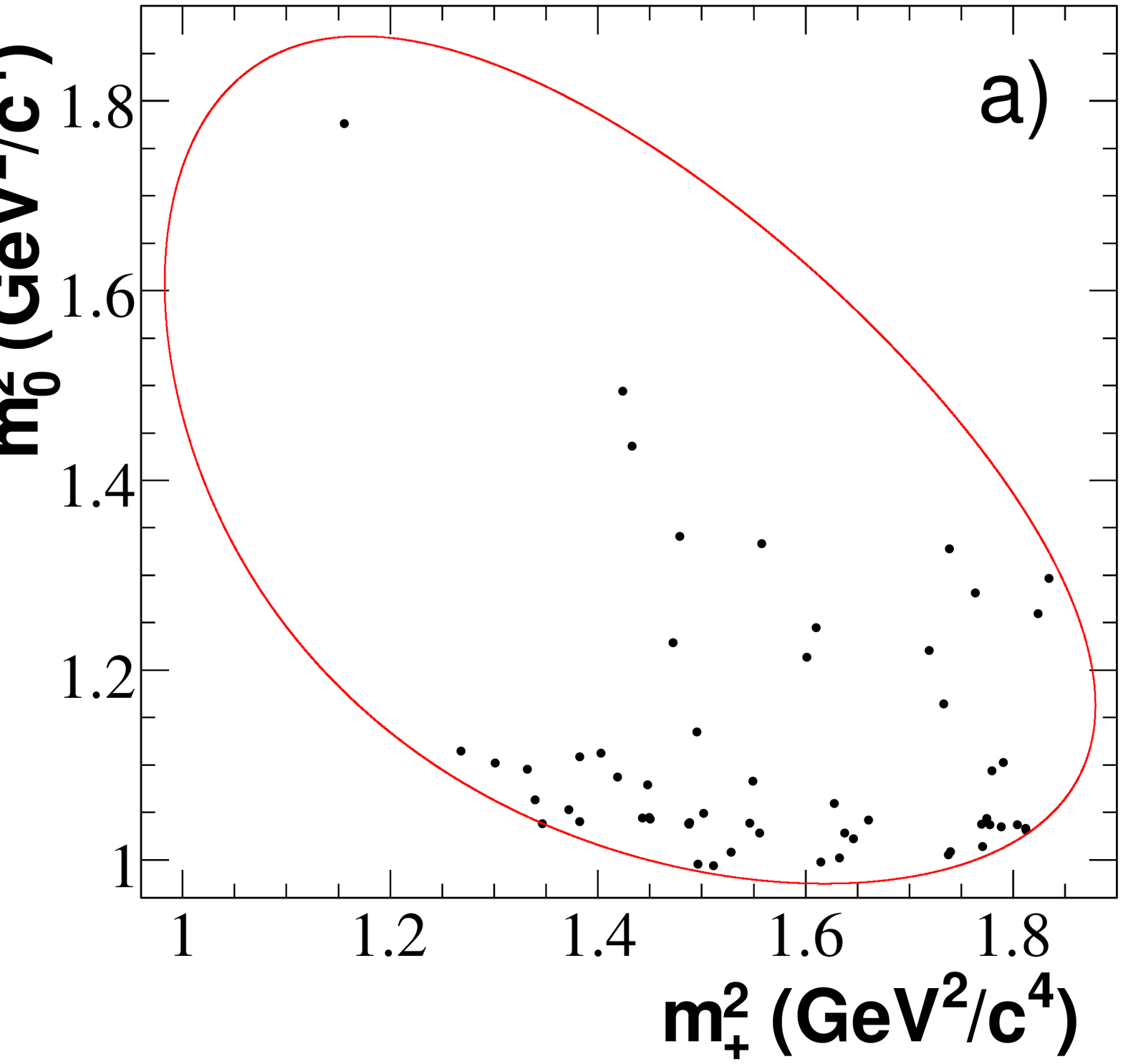}& 
\includegraphics[width=0.23\textwidth]{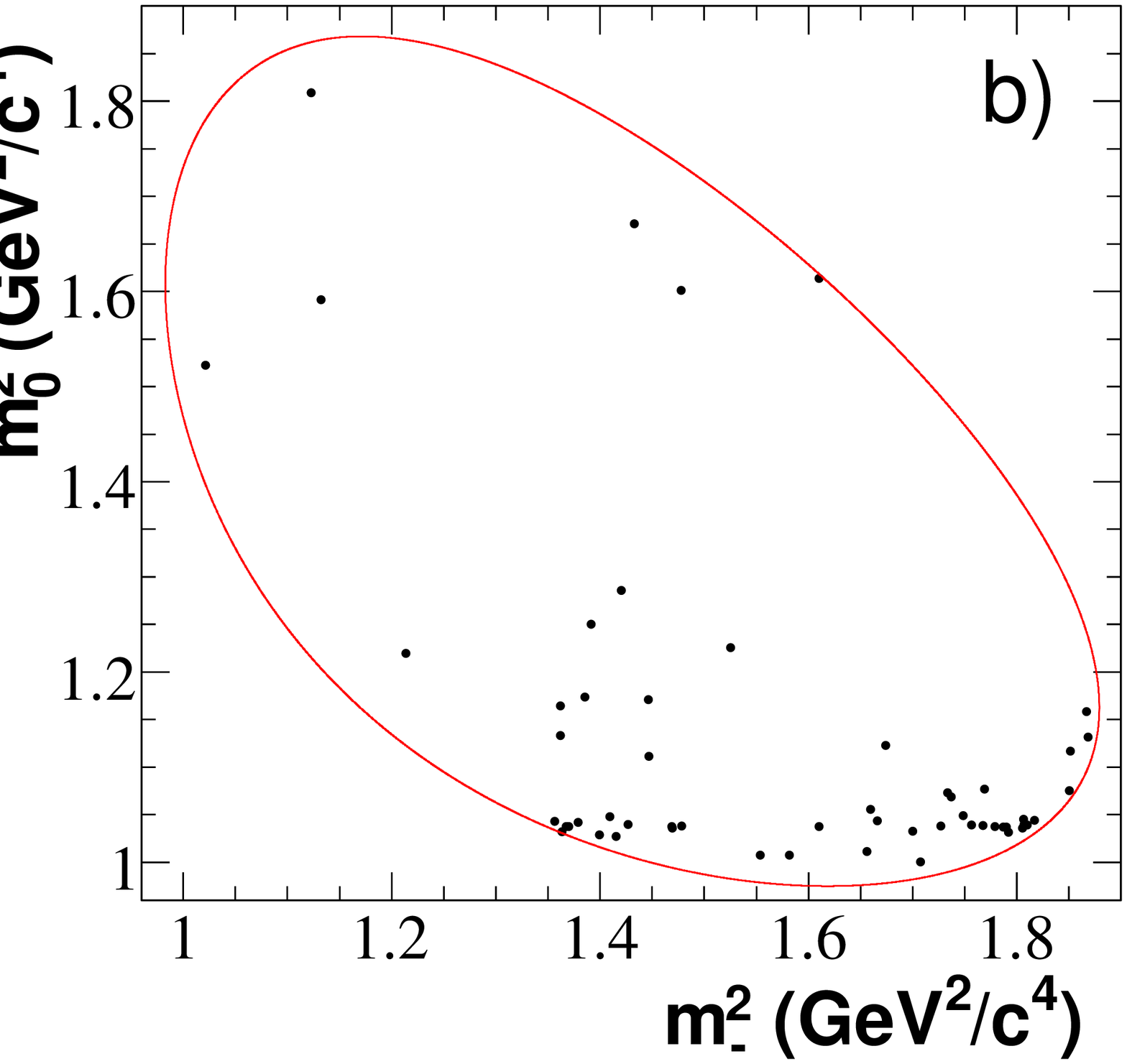} \\ 
\includegraphics[width=0.23\textwidth]{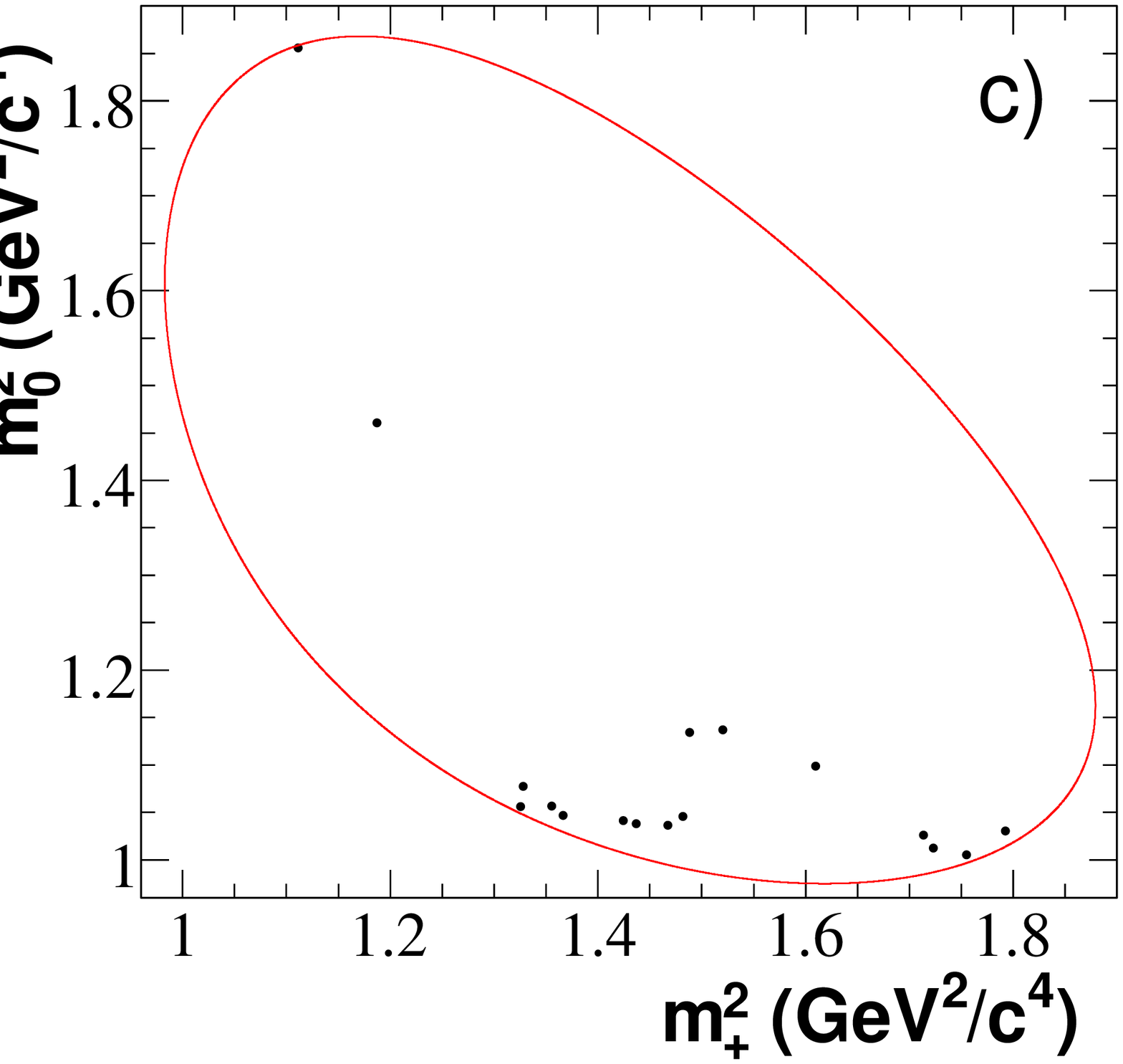}& 
\includegraphics[width=0.23\textwidth]{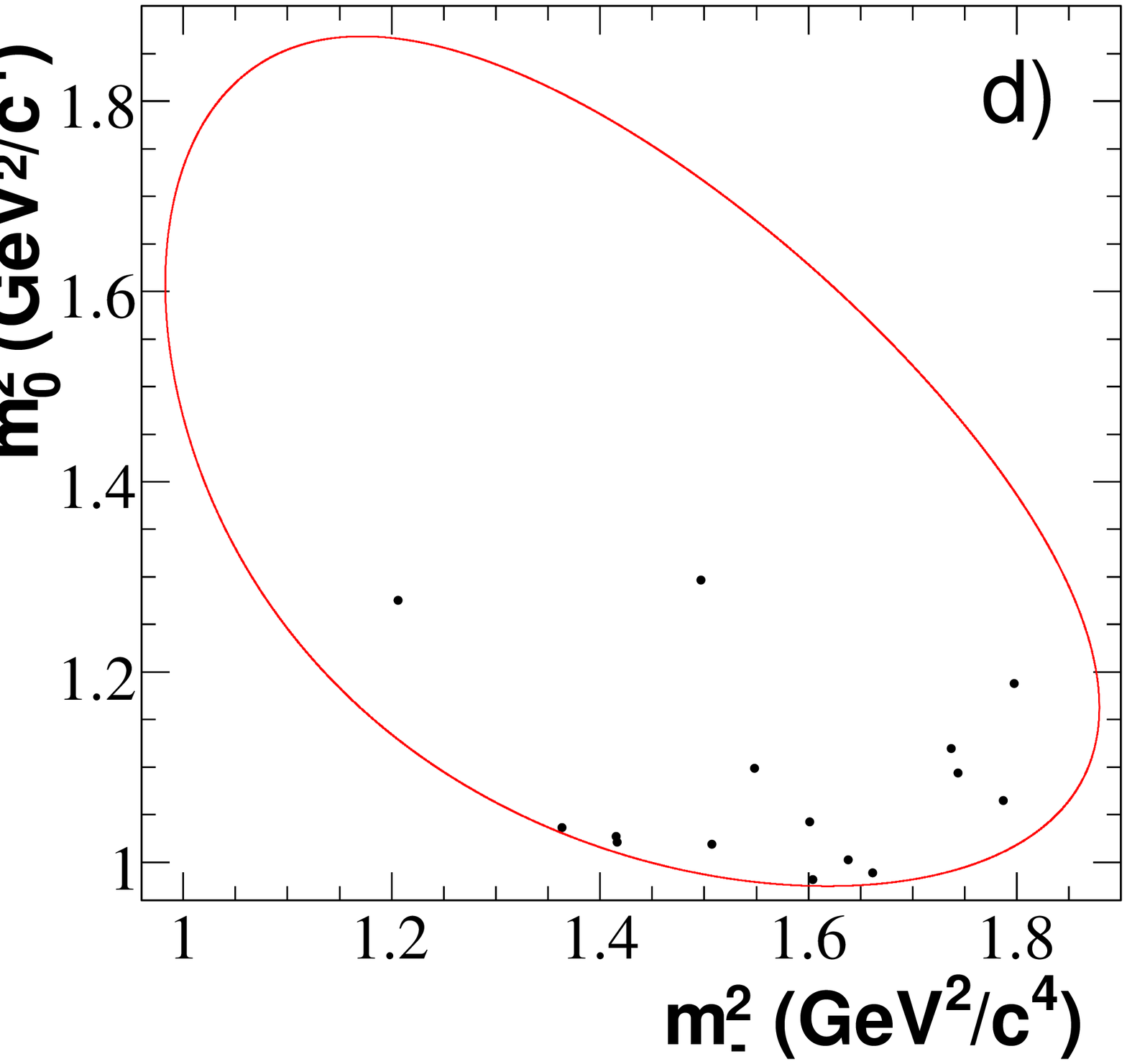} \\ 
\includegraphics[width=0.23\textwidth]{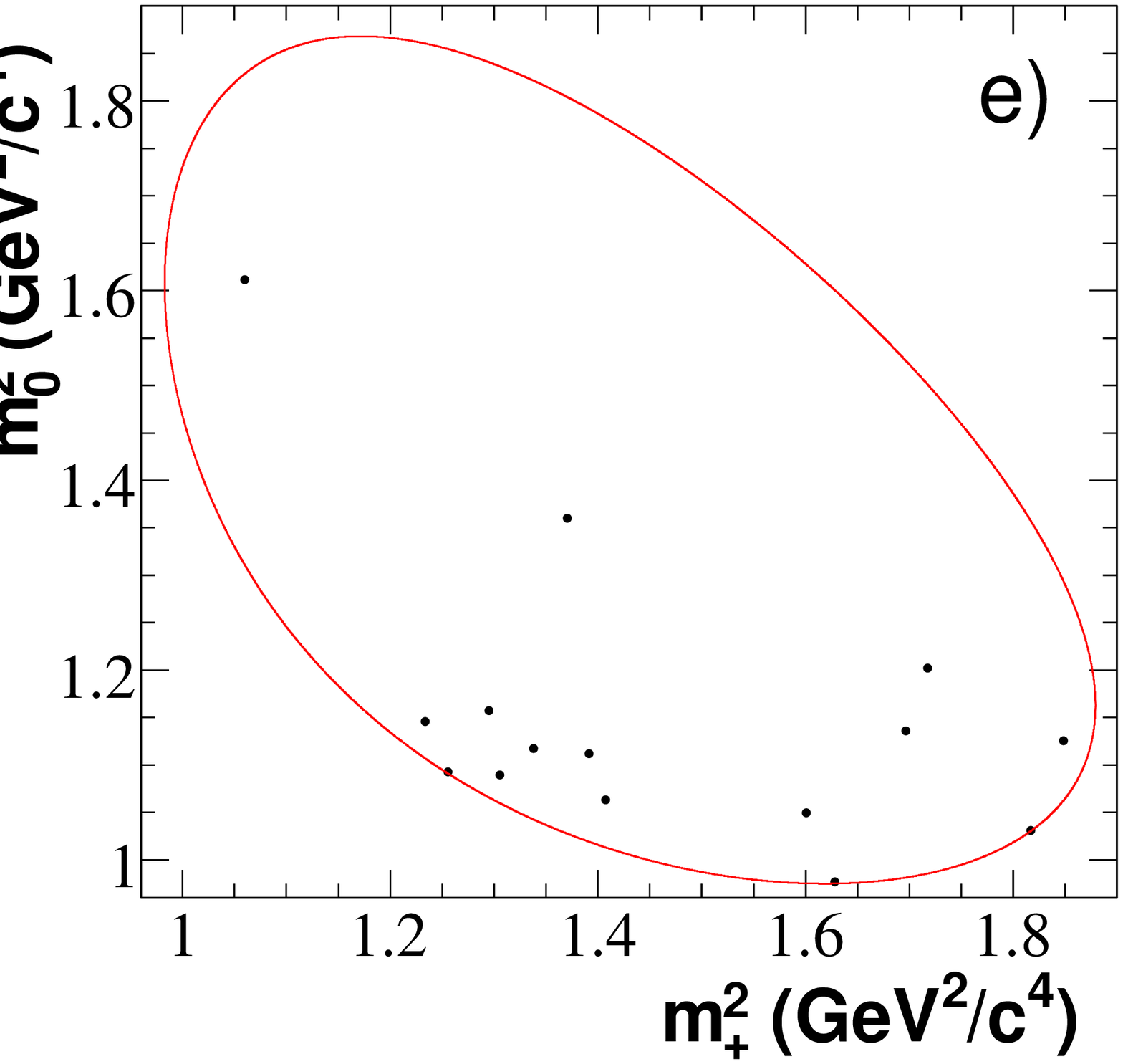}& 
\includegraphics[width=0.23\textwidth]{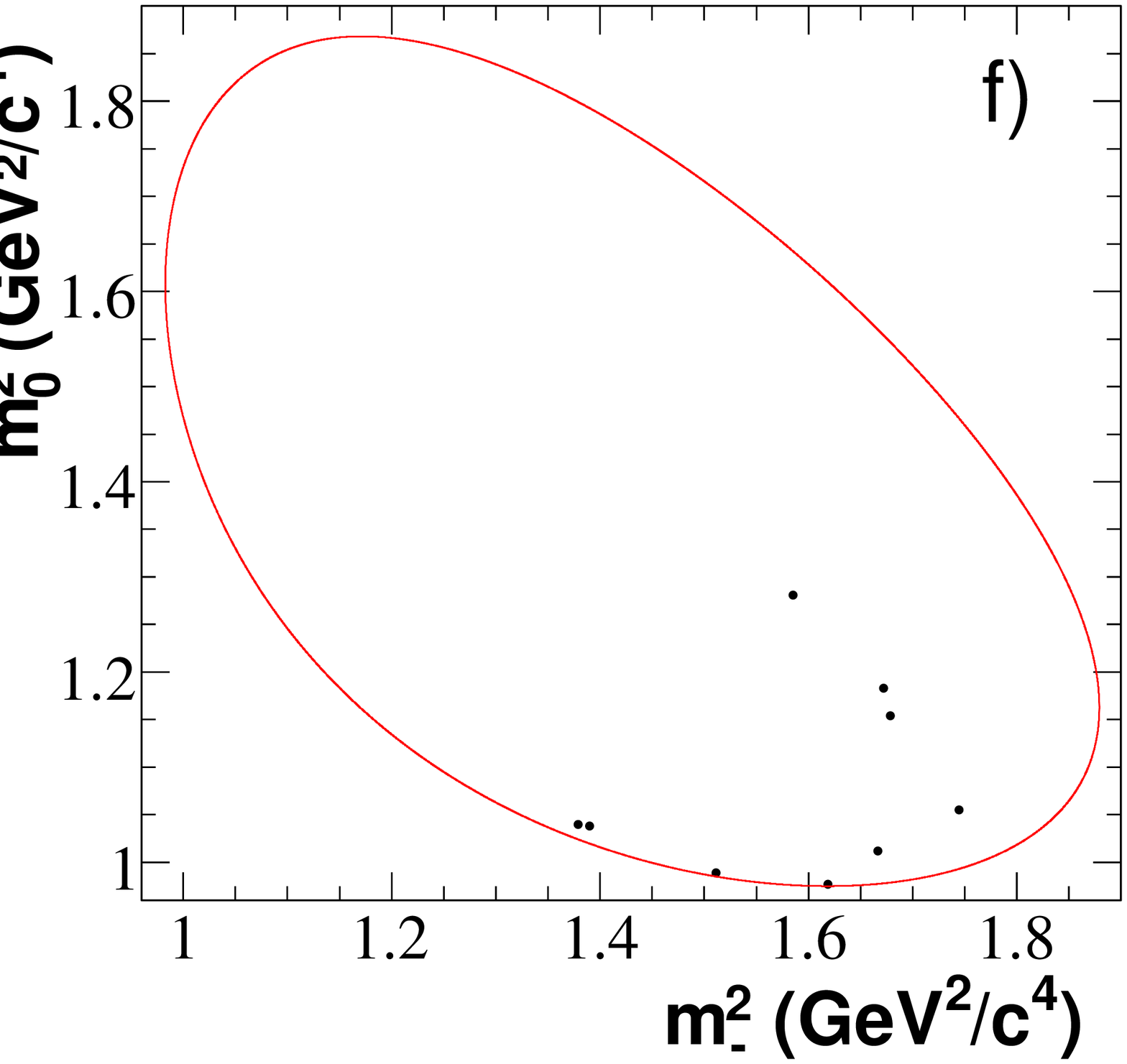} \\ 
\end{tabular} 
\caption{\label{fig:dalitz-kskk} (color online). \Dztildetokskk Dalitz plot distributions for  
(a) $\Bm \to \Dztilde \Km$, (b) $\Bp \to \Dztilde \Kp$,  
(c) $\Bm \to \Dstarztilde[\Dztilde\piz] \Km$, (d) $\Bp \to \Dstarztilde[\Dztilde\piz] \Kp$,  
(e) $\Bm \to \Dstarztilde[\Dztilde\gamma] \Km$, and (f) $\Bp \to \Dstarztilde[\Dztilde\gamma] \Kp$,  
for the \DeltaE signal region. 
The requirements $\mes>5.272$~\gevcc and $\fisher>-0.1$ have been applied to reduce the background contamination,  
mainly from continuum events. 
The contours (solid red lines) represent the kinematical limits of the \Dztildetokskk decay.  
} 
\end{figure}

Once the decay amplitudes ${\cal A}_{D}(\mtwo)$  
for \Dztokspipi and \Dztokskk are known, they are fed into $\Gamma_{\mp}^{(*)}(\mtwo)$  
and $\Gamma_{s\mp}(\mtwo)$. 
The extraction of the \CP-violating parameters  
\xbxbstmp, \ybybstmp, \xsmp, and \ysmp is then performed 
through a simultaneous unbinned maximum likelihood fit (referred to hereafter as the \CP fit) to the $\Gamma_{\mp}^{(*)}(\mtwo)$ and $\Gamma_{s\mp}(\mtwo)$ 
Dalitz plot distributions of the seven signal modes, in the \DeltaE signal region defined as $|\DeltaE|<30$~\mev.  
Figures~\ref{fig:dalitz-kspipi} and~\ref{fig:dalitz-kskk} show these distributions separately for \Bm and \Bp decays 
in a region enriched in signal through the requirements $\mes>5.272$~\gevcc and $\fisher>-0.1$. 
The efficiency of the Fisher cut in the $|\DeltaE|<30$~\mev and $\mes>5.272$~\gevcc region  
is around 70\% for signal events, while for continuum background events it is below 1\%.

The log-likelihood function for each of the seven \CP samples generalizes Eq.~(\ref{eq:likehood_sel}) to include the Dalitz plot distributions, 
\bea 
\ln {\cal L} = -\eta + \sum_j \ln \left[ \sum_c \frac{N_c}{2} (1\pm A_c) {\cal P}_c({\bf u}_j) {\cal D}_{c,\mp}(\mtwoj) \right].~~~ 
\label{eq:likehood_CP} 
\eea 
Here, ${\cal D}_{c,\mp}(\mtwoj)$ is the Dalitz plot PDF for event $j$ satisfying the normalization condition $\int {\cal D}_{c,\mp}(\mtwo){\rm d}\mtwo = 1$,  
and $A_c$ accounts for any  
asymmetry in the absolute number of \Bm and \Bp candidates (charge asymmetry) for component $c$. 
 
For $\Bmp \to \Dztilde \Kmp$ signal, ${\cal D}_{{\rm sig},\mp}(\mtwo) = \Gamma_\mp(\mtwo) \epsilon(\mtwo)$,  
where the efficiency map in the Dalitz plot $\epsilon(\mtwo)$ 
is determined as for $\Dstarp \to \Dz\pip$ events (Sec.~\ref{sec:dalitzmodels-analysis}).  
We replace $r_\B^2$ in Eq.~(\ref{eq:ampgen}) by $r_{\Bmp}^2 = x_\mp^2 + y_\mp^2$. The physical condition $\rbm=\rbp$ is recovered 
in the statistical procedure to extract $\gamma$ from $x_\mp,y_\mp$, as discussed in   
Sec.~\ref{sec:interpretation}. The same procedure is applied analogously to the other signal samples.

We consider the same background components as in the selection fit, with  
some important modifications. First, events falling into the continuum and \BB background components are 
divided into events with a real or a fake (combinatorial) \Dztilde meson.  
Dalitz plot shapes for fake \Dztilde mesons from continuum are extracted as described in Sec.~\ref{sec:dalitzmodels-analysis}, 
using events in the continuum enriched region (\mes and \mD sideband regions), while those from \BB are determined from MC events.  
Events containing a real \Dztilde are further divided into ``right-sign'' and ``wrong-sign'' flavor 
categories depending on whether they are combined with a negative or positive 
kaon (or \Kstar). We pay special attention to this charge-flavor correlation in the background since it can mimic either the 
$\b \to \c$ or the $\b \to \u$ signal component. Second, we have included a background contribution due to signal events where  
the kaon (or \Kstar) comes from the other \B decay;  
this amounts to 9\% of the $\Bm \to \Dztilde \Kstarm$ signal, but is negligible for $\Bm \to \DzDstarztilde \Km$. 
 
The \mes, \DeltaE, and \fisher PDF parameters in the \CP fit are the same as those used in or obtained from the selection fit, 
except for the \mes peaking fractions for \Dztildetokspipi channels, which are allowed to vary since their values depend on  
the \DeltaE region used for the fit. Other parameters simultaneously determined from the fit, along with the \CP-violating 
parameters \xbxbstmp, \ybybstmp, \xsmp, and \ysmp, are:  
signal and background yields,  
signal charge asymmetries,  
and fractions of true \Dz mesons for all decay modes and right-sign fractions for \Dztildetokspipi channels in continuum background. 
Right-sign fractions for the modes with \Dztildetokskk are fixed from MC simulation due to lack of statistics and  
the limited discriminating power between \Dz and \Dzb Dalitz plot distributions. 
Similarly, fractions of true \Dztilde mesons and charge-flavor correlation for the \BB component are determined using MC events, 
because of the lack of \BB background statistics.

\subsection{Results and cross-checks} 
\label{sec:results-crosschecks} 
 
\begin{figure*}[hbt!] 
\begin{tabular}{ccc} 
\includegraphics[width=0.32\textwidth]{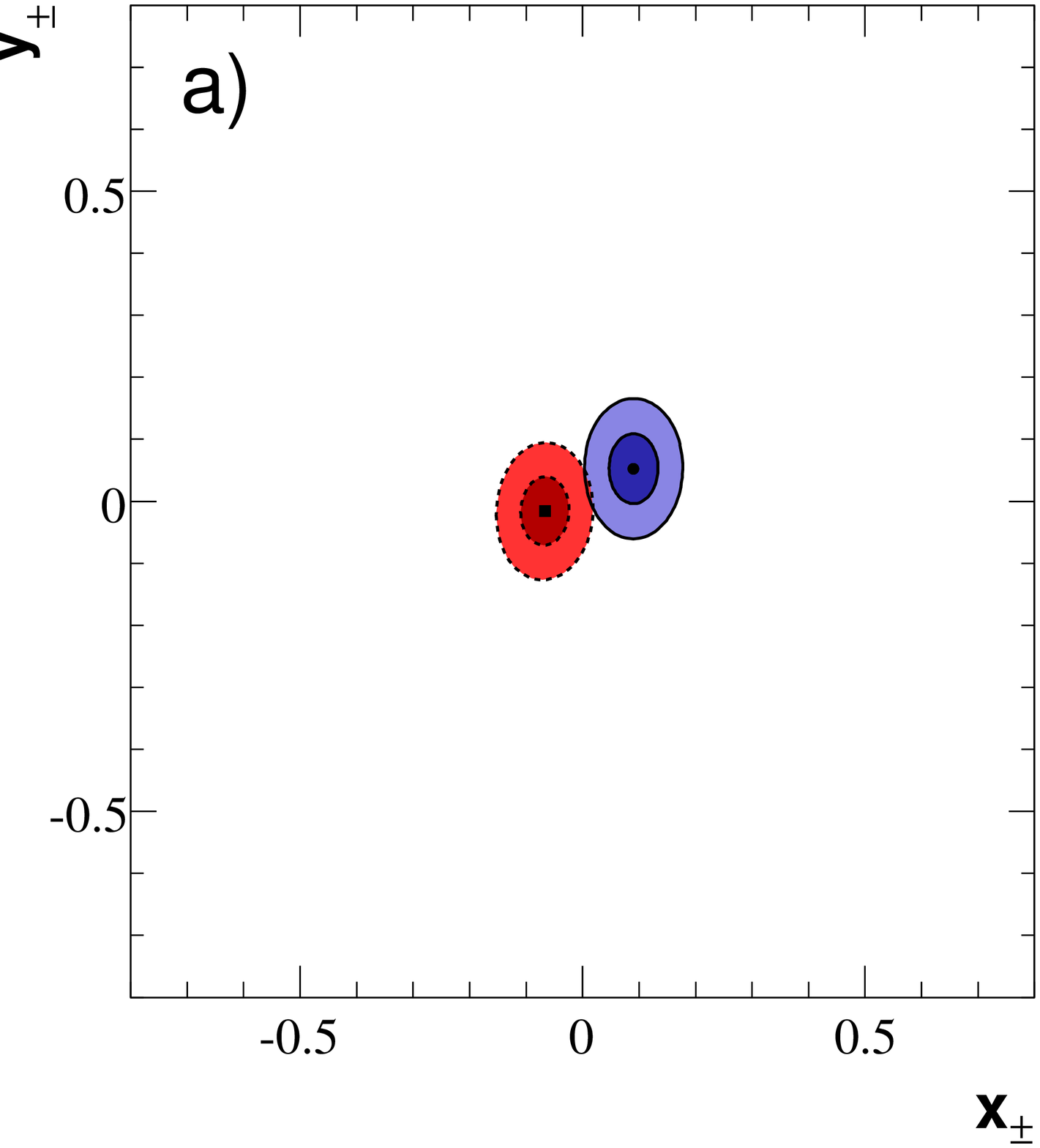}& 
\includegraphics[width=0.32\textwidth]{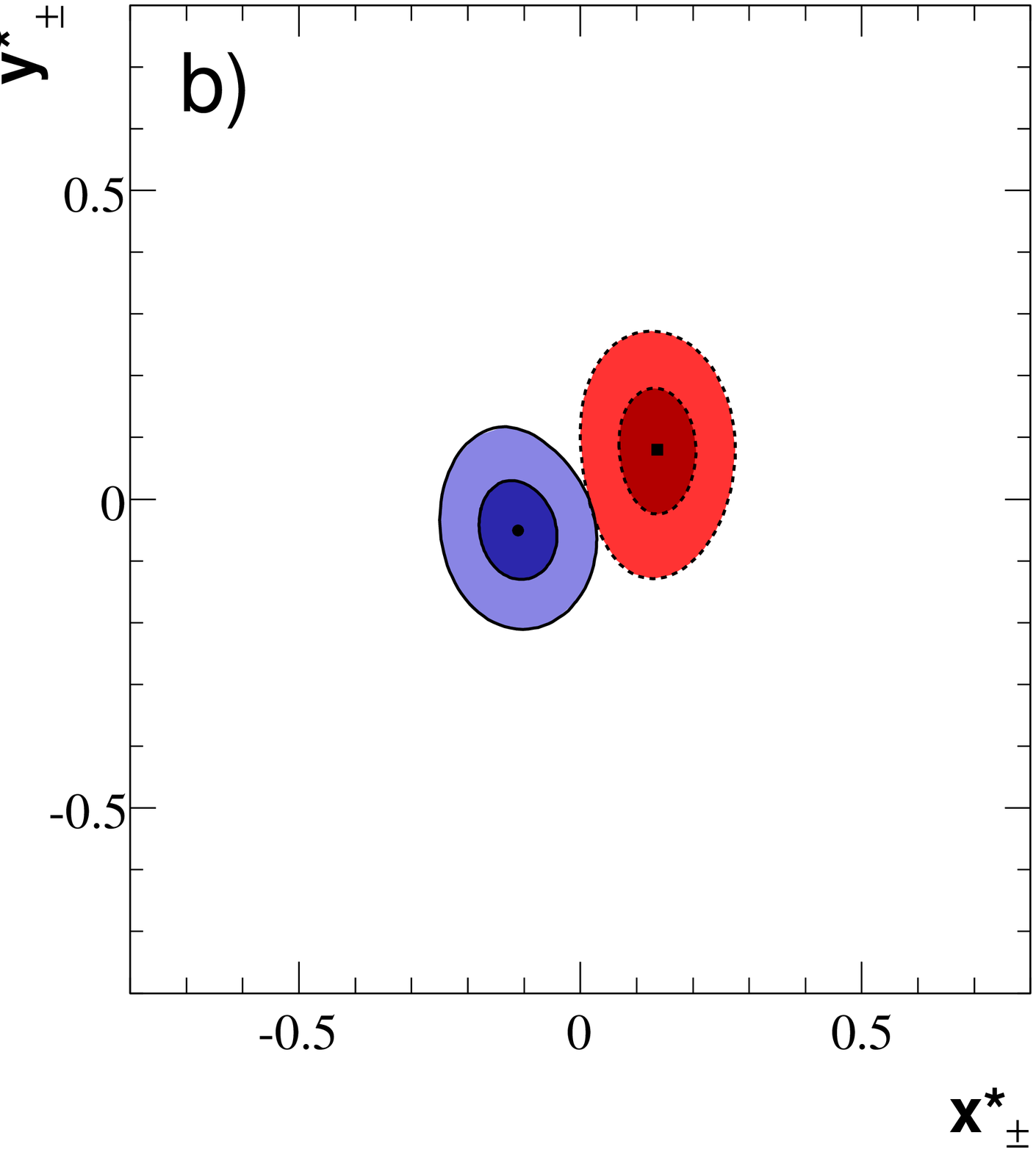}& 
\includegraphics[width=0.32\textwidth]{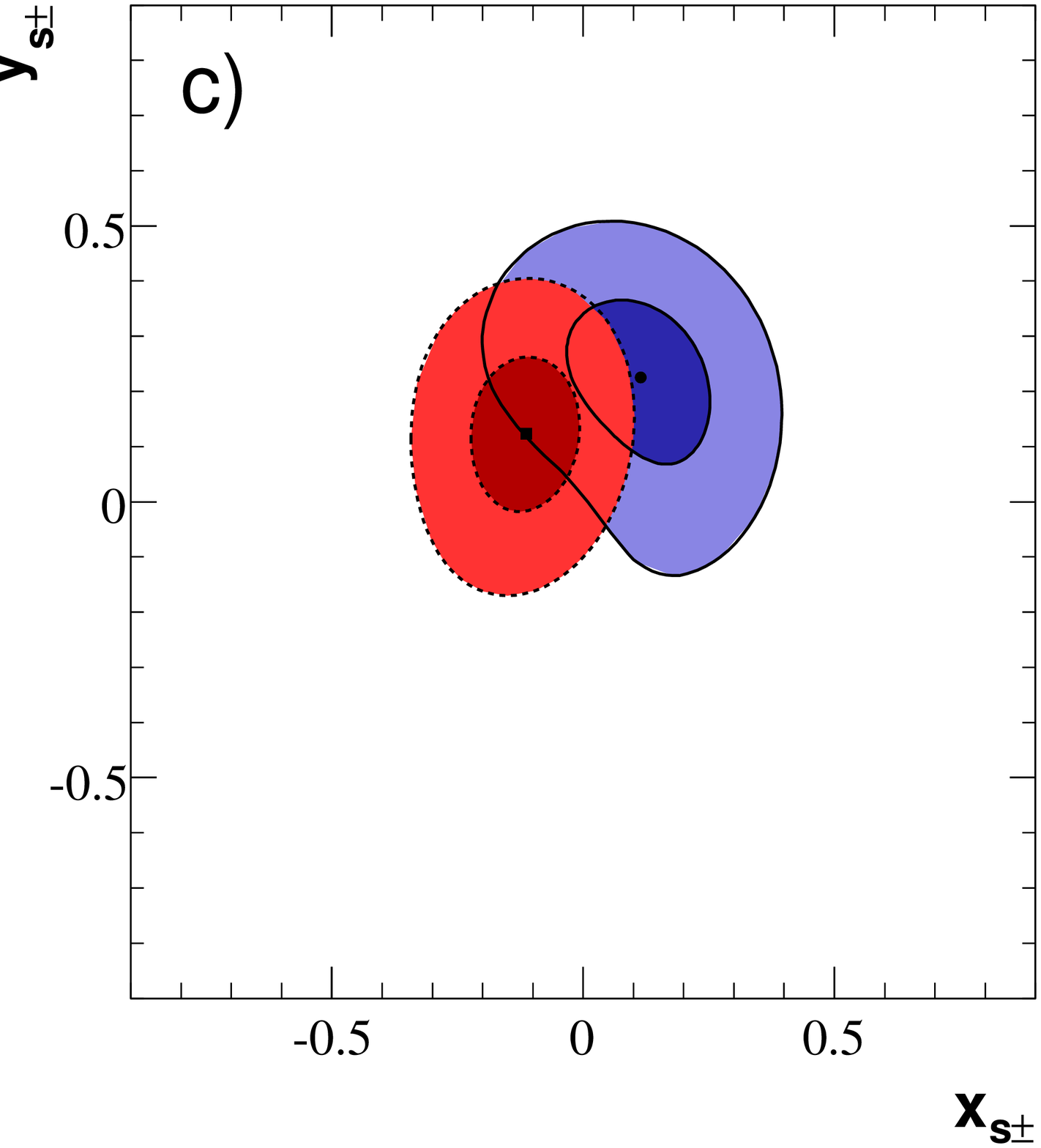}\\ 
\end{tabular} 
\caption{\label{fig:cart_CL} (color online). Contours at $39.3\%$ (dark) and $86.5\%$ (light) 2-dimensional confidence-level (CL) in the (a) $(\xbmp,\ybmp)$,   
(b) $(\xbstmp,\ybstmp)$, and (c) $(\xsmp,\ysmp)$ planes, corresponding to one- and two-standard deviation regions (statistical only),  
for \Bm (thick and solid lines) and \Bp (thin and dotted lines) decays. 
} 
\end{figure*}

We find $600 \pm 31$, $133 \pm 15$,  $129 \pm 16$, and $118 \pm 18$ signal events,  
for  
$\Bm \to \Dztilde \Km$,  
$\Bm \to \Dstarztilde[\Dztilde\piz]\Km$,  
$\Bm \to \Dstarztilde[\Dztilde\gamma]\Km$,  
and $\Bm\to \Dztilde \Kstarm$ decay modes, respectively, with \Dztildetokspipi. 
Similarly, for the \Dztildetokskk channels we obtain $112 \pm 13$, $32 \pm 7$, and $21 \pm 7$ signal events, 
for  
$\Bm \to \Dztilde \Km$,  
$\Bm \to \Dstarztilde[\Dztilde\piz]\Km$, and  
$\Bm \to \Dstarztilde[\Dztilde\gamma]\Km$. 
Errors are statistical only. No statistically significant charge asymmetries are observed. 
The results for the \CP-violating parameters $\xbxbstmp$, $\ybybstmp$, $\xsmp$, and $\ysmp$, 
are summarized in Table~\ref{tab:cp_coord}. The only non-zero statistical correlations 
involving the \CP parameters are  
for the pairs $(x_-,y_-)$, $(x_+,y_+)$, $(x_-^*,y_-^*)$, $(x_+^*,y_+^*)$, $(x_{s-},y_{s-})$, $(x_{s+},y_{s+})$, 
which amount to $0.4\%$, $3.5\%$, $-14.0\%$, $-5.6\%$, $-29.9\%$, and $6.8\%$, respectively. 
Figure~\ref{fig:cart_CL} shows the $39.3\%$ and $86.5\%$ 2-dimensional confidence-level (CL) contours 
in the $(\xbmp,\ybmp)$,  $(\xbstmp,\ybstmp)$, and $(\xsmp,\ysmp)$ planes, corresponding to one- and two-standard deviation  
regions (statistical only). 
The separation of the \Bm and \Bp positions in the $(x,y)$ plane is equal to $2 r_B|\sin\gamma|$ 
and is a measurement of direct \CP violation. The angle between the lines connecting 
the \Bm and \Bp centers with the origin $(0,0)$ is equal to $2\gamma$.

\begin{table*}[hbt!] 
\caption{\label{tab:cp_coord} \CP-violating parameters \xbxbstmp, \ybybstmp, \xsmp, and \ysmp, as obtained from the \CP fit.  
The first error is statistical, the second is experimental systematic uncertainty and the third is  
the systematic uncertainty associated with the Dalitz models.} 
\begin{ruledtabular} 
\begin{tabular}{lccc} 
\\[-0.15in] 
Parameters                   & $\Bm \to \Dztilde \Km$ & $\Bm \to \Dstarztilde \Km$ & $\Bm \to \Dztilde \Kstarm$ \\ [0.01in] \hline 
$x_{-}~,~x_{-}^{*}~,~x_{s-}$ & $\phm0.090\pm 0.043\pm 0.015 \pm 0.011$ & $-0.111\pm 0.069\pm 0.014\pm 0.004$    & $\phm0.115\pm0.138\pm0.039\pm0.014$\\ 
$y_{-}~,~y_{-}^{*}~,~y_{s-}$ & $\phm0.053\pm 0.056\pm 0.007 \pm 0.015$ & $-0.051\pm 0.080\pm 0.009\pm 0.010$    & $\phm0.226\pm0.142\pm0.058\pm0.011$\\ 
$x_{+}~,~x_{+}^{*}~,~x_{s+}$ & $-0.067\pm 0.043 \pm 0.014\pm 0.011$    & $\phm0.137\pm 0.068\pm 0.014\pm 0.005$ & $-0.113\pm0.107\pm0.028\pm0.018$ \\ 
$y_{+}~,~y_{+}^{*}~,~y_{s+}$ & $-0.015\pm 0.055\pm 0.006\pm 0.008$     & $\phm0.080\pm 0.102\pm 0.010\pm 0.012$ & $\phm0.125\pm0.139\pm0.051\pm0.010$\\ 
\end{tabular} 
\end{ruledtabular} 
\end{table*} 

A variety of studies using data, parameterized fast Monte Carlo, and full GEANT4-simulated samples have been  
performed to test the consistency of the results and to verify the analysis chain and fitting procedure, as described below. 
 
\begin{figure}[hbt!] 
\begin{tabular}{cc} 
\includegraphics[width=0.23\textwidth]{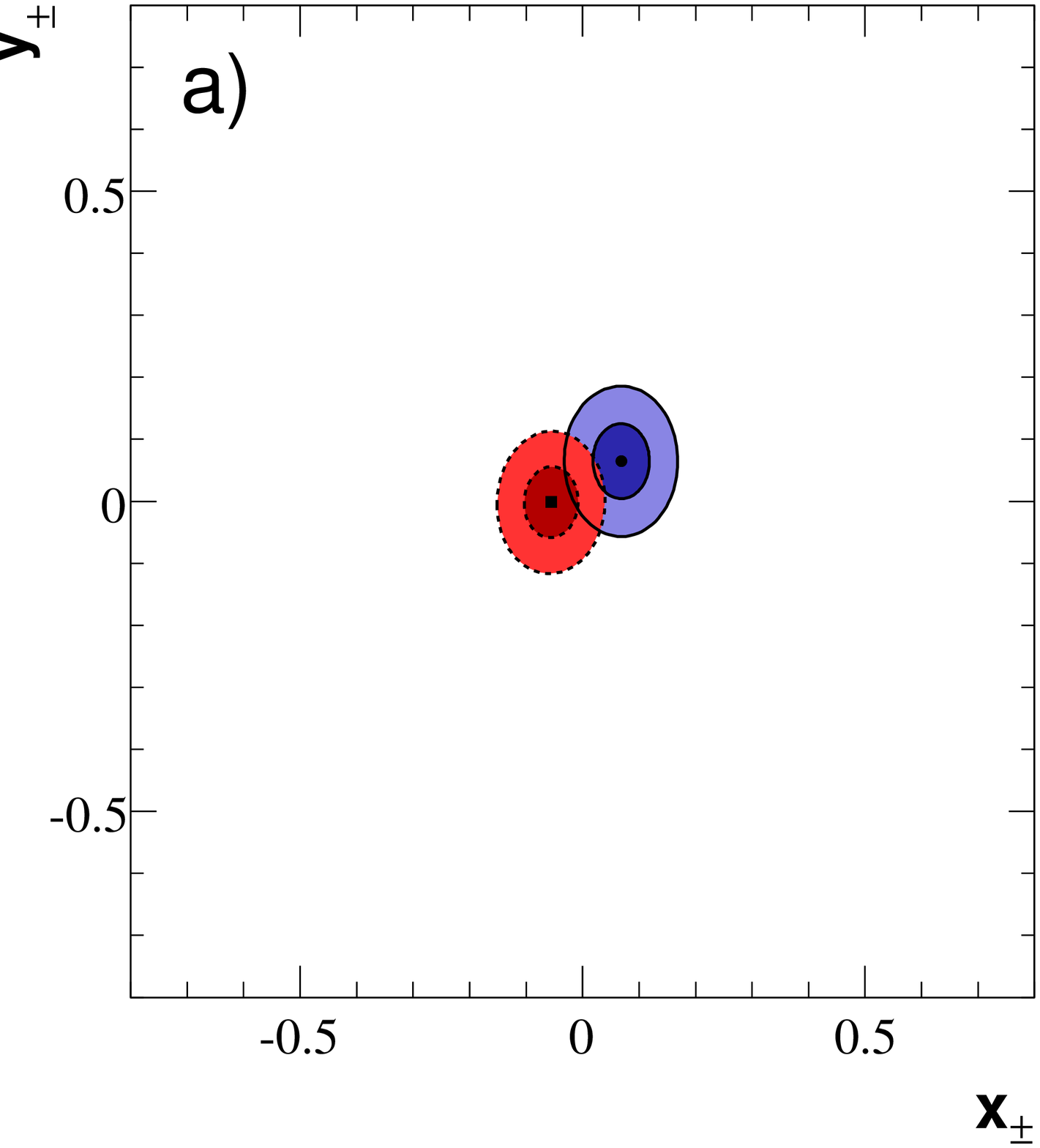}& 
\includegraphics[width=0.23\textwidth]{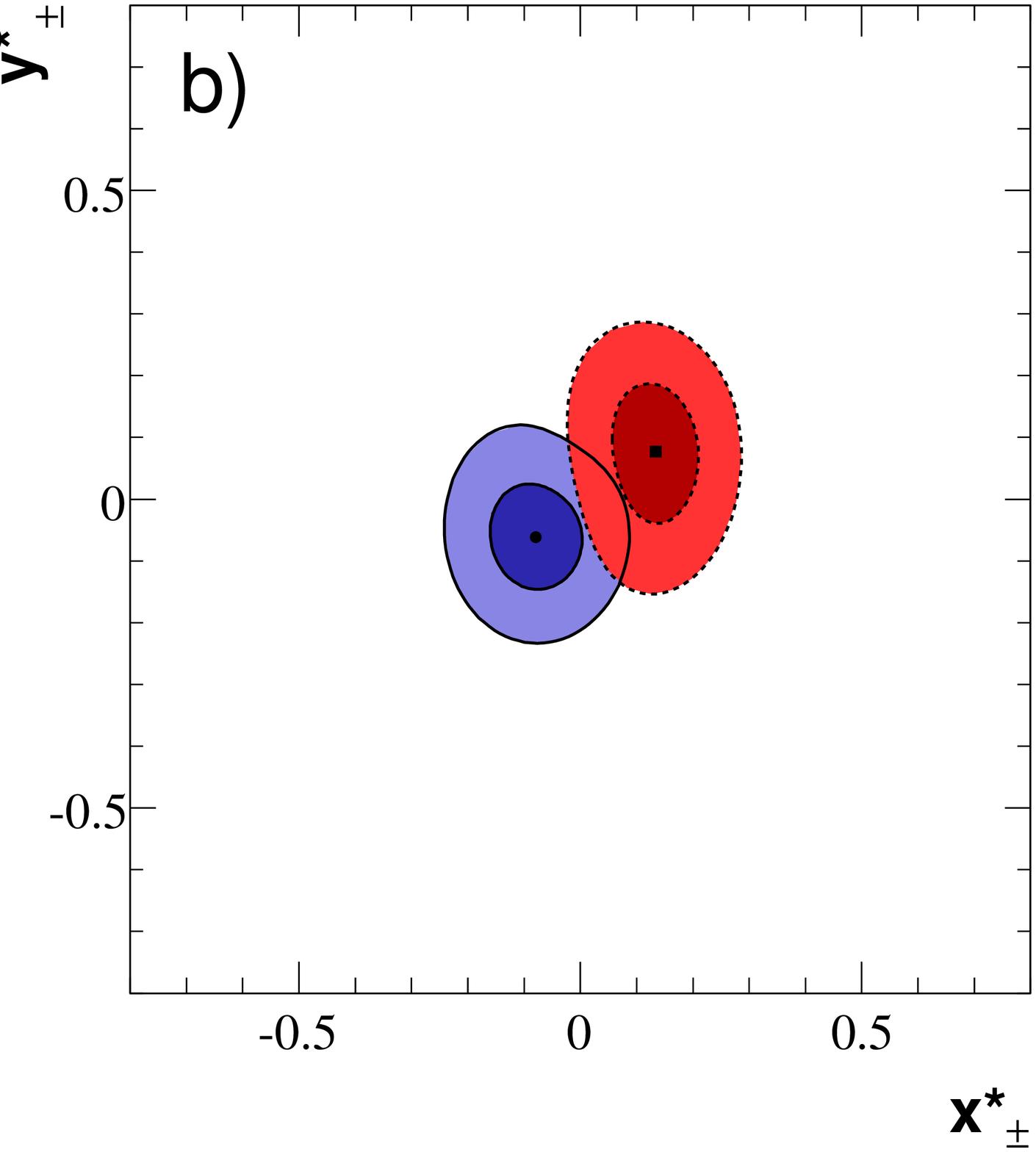} \\ 
\includegraphics[width=0.23\textwidth]{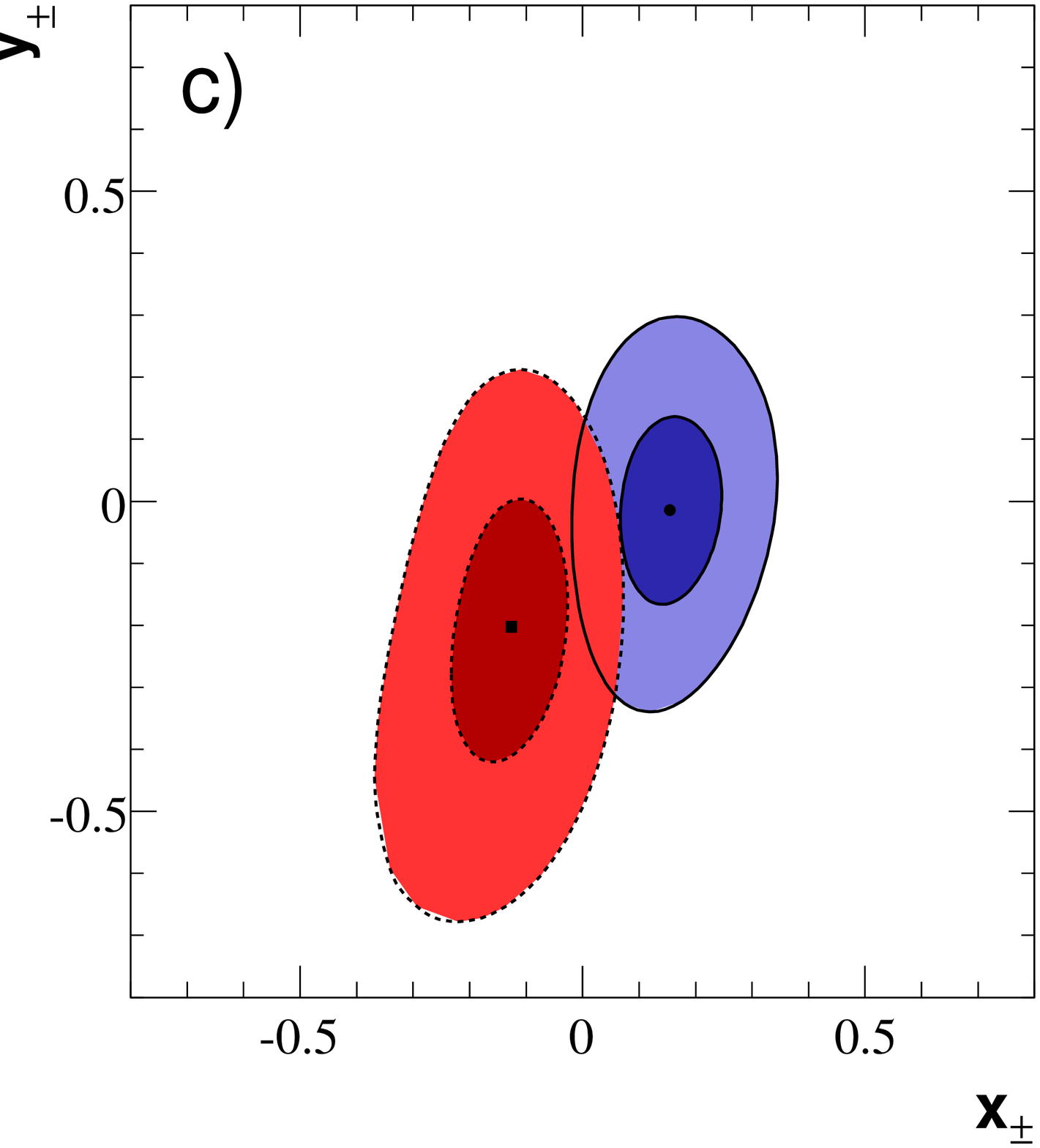}& 
\includegraphics[width=0.23\textwidth]{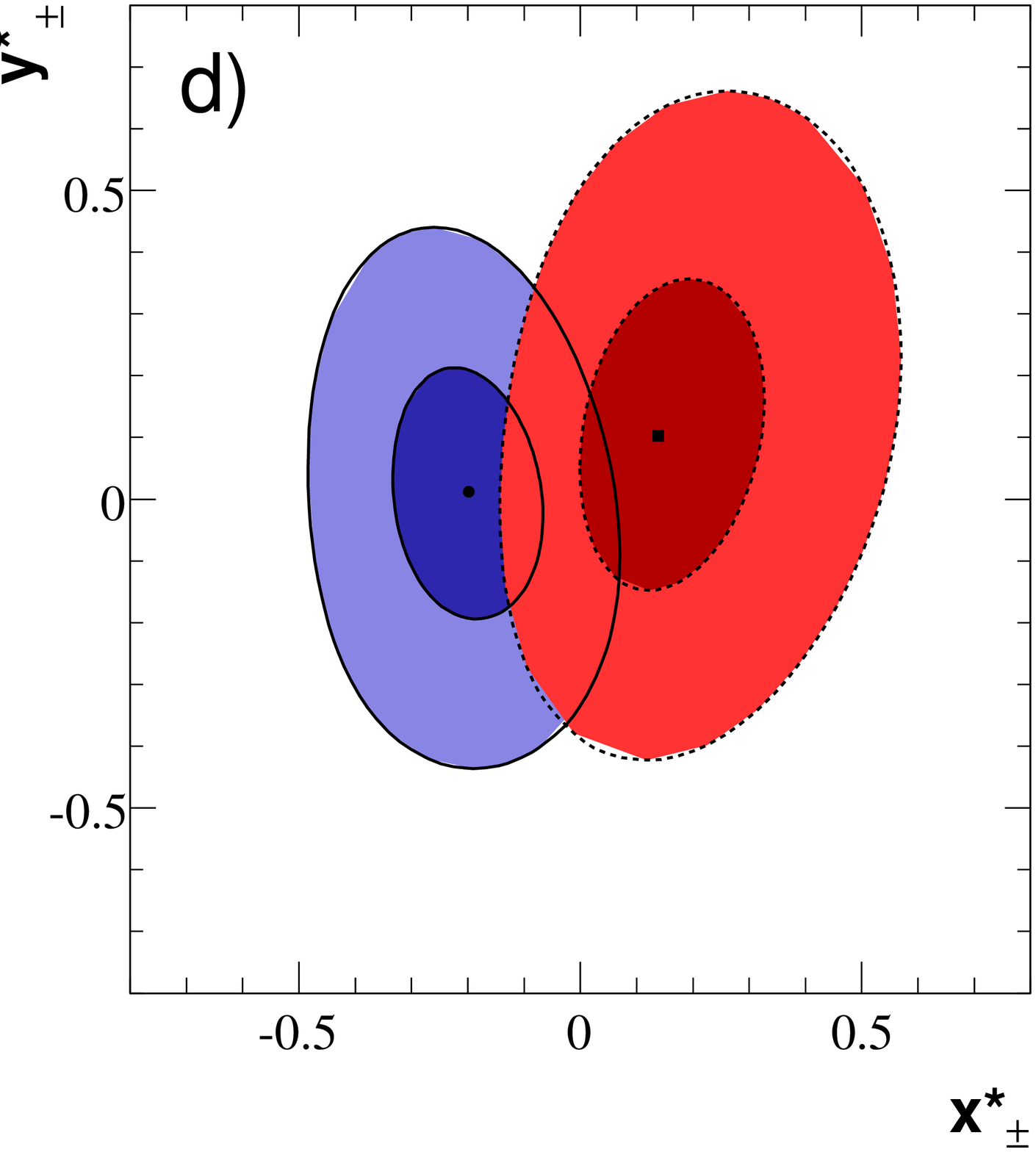} \\ 
\end{tabular} 
\caption{\label{fig:cart_CL_kspipi_kskk} (color online). Contours at $39.3\%$ (dark) and $86.5\%$ (light) 2-dimensional confidence-level (CL) in  
the (a,c) $(\xbmp,\ybmp)$ and (b,d) $(\xbstmp,\ybstmp)$ planes, corresponding to one- and two-standard deviation regions (statistical only),  
for $\Bm \to \DzDstarz \Km$ (thick and solid lines) and $\Bp \to \DzDstarz \Kp$ (thin and dotted lines) decays,  
for (a,b) \Dztildetokspipi and (c,d) \Dztildetokskk only decay modes. 
} 
\end{figure}

\begin{figure}[hbt!] 
\begin{tabular}{cc} 
\includegraphics[width=0.23\textwidth]{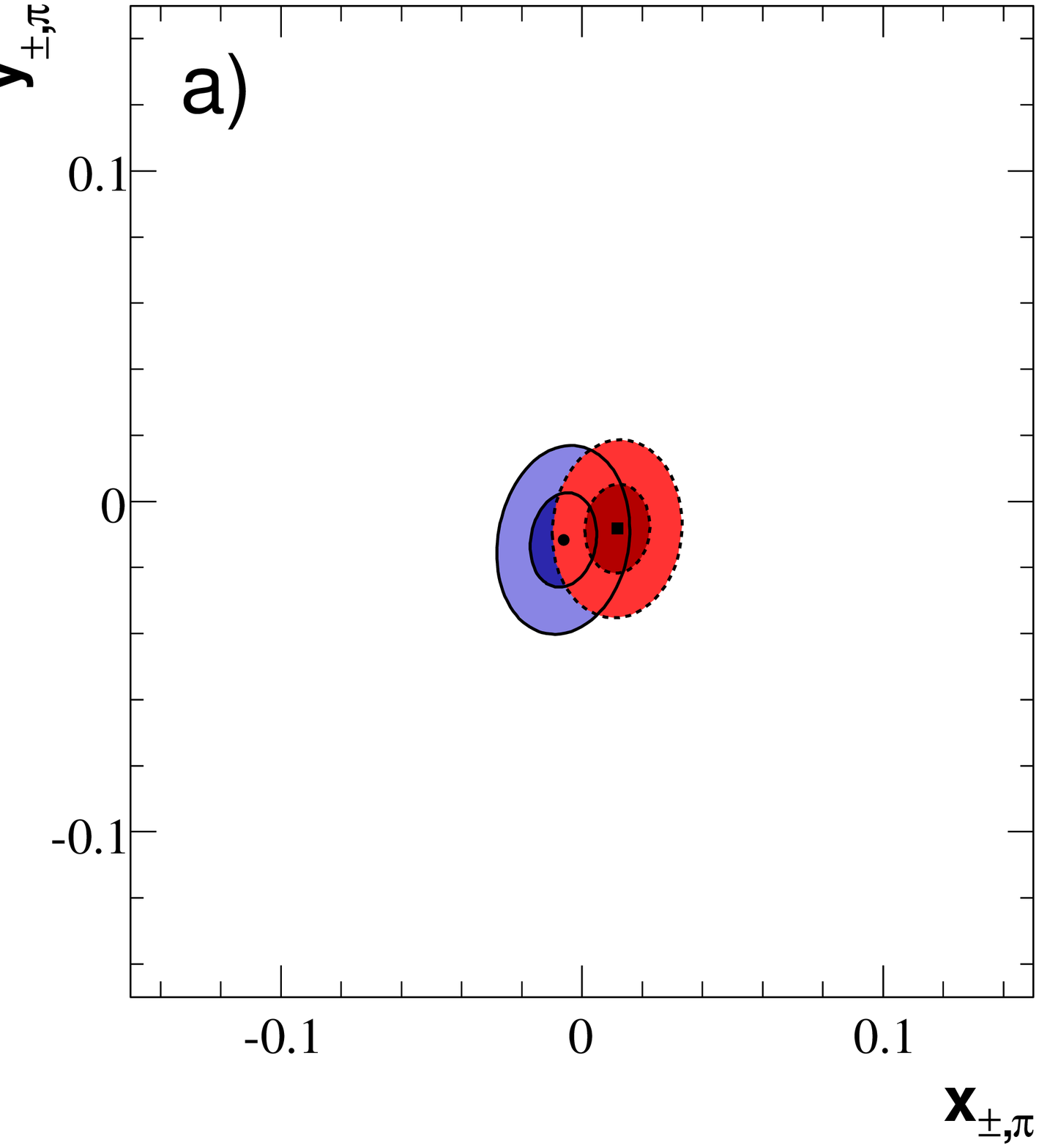}& 
\includegraphics[width=0.23\textwidth]{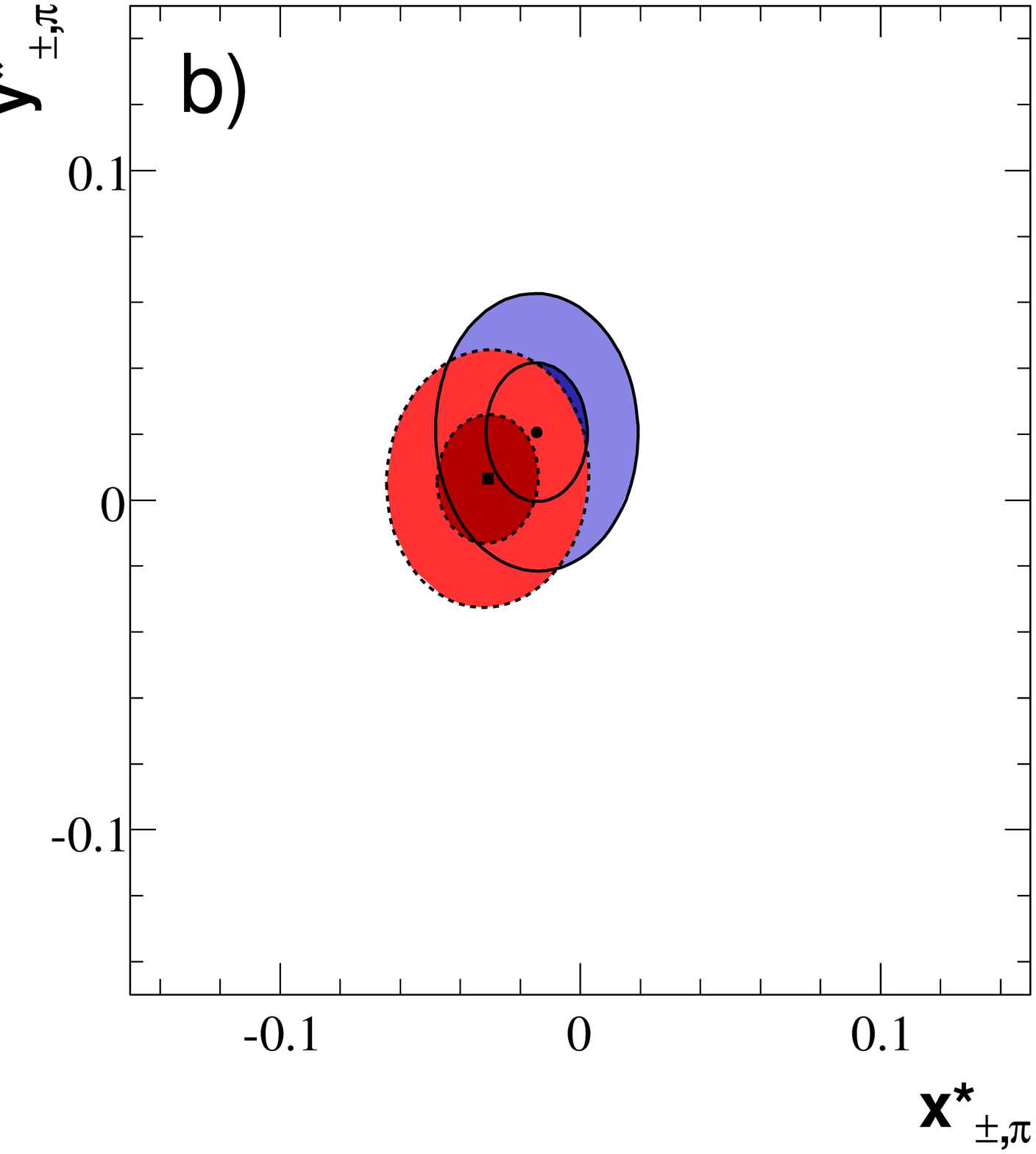} \\ 
\end{tabular} 
\caption{\label{fig:cart_CL_Dpi} (color online). Contours at $39.3\%$ (dark) and $86.5\%$ (light) 2-dimensional confidence-level (CL) in  
the (a) $(x_{\mp,\pi},y_{\mp,\pi})$ and (b) $(x_{\mp,\pi}^*,y_{\mp,\pi}^*)$ planes, corresponding to one- and two-standard deviation  
regions (statistical only), for $\Bm \to \DzDstarz \pim$ (thick and solid lines) and $\Bp \to \DzDstarzb \pip$ (thin and dotted lines)  
control sample decays. Note the differences in scale when comparing to Figs.~\ref{fig:cart_CL} and~\ref{fig:cart_CL_kspipi_kskk}. 
} 
\end{figure}

The \CP fit to the $\Bm \to \DzDstarztilde \Km$ samples has been performed separately for \Dztildetokspipi and \Dztildetokskk samples. 
Figure~\ref{fig:cart_CL_kspipi_kskk} shows the resulting one- and two-standard deviation regions in  
the $(\xbxbstmp,\ybybstmp)$ planes. We find statistically consistent results between the different subsets. 
The same fitting procedure has been applied to the $\Bm \to \DzDstarz \pim$ control samples. In this case we expect 
$r_{\B,\pi}^{(*)} \approx \mid V_{cd}^{}V_{ub}^{*} \mid / \mid V_{ud}^{}V_{cb}^{*} \mid c_F$ to be approximately $0.01$. Since the 
experimental resolutions on $(x_{\mp,\pi}^{(*)},y_{\mp,\pi}^{(*)})$ are expected to have the same order of magnitude,  
the $(x_{\mp,\pi}^{(*)},y_{\mp,\pi}^{(*)})$ contours for \Bm and \Bp decays should be close to the origin  
up to $\sim 0.01$. Deviations from this pattern could be an indication that the Dalitz plot  
distributions are not well described by the models. 
Figure~\ref{fig:cart_CL_Dpi} shows the resulting one- and two-standard deviation regions for  
$(x_{\mp,\pi}^{(*)},y_{\mp\pi}^{(*)})$, consistent with the expected values. Moreover, we find statistically  
consistent results between the \Dztokspipi and \Dztokskk samples.

An additional test of the fitting procedure is performed with parameterized MC simulations consisting of about 500 experiments generated with a 
sample size and composition corresponding to that of the data. The \CP parameters are generated with values close to those found in 
the data and the reference \CP fit is performed on each of these experiments. The \rms of the residual distributions 
for all the \CP parameters (where the residual is defined as the difference between the fitted and generated values) 
is found to be consistent with the mean (Gaussian) statistical errors reported by the fits. The mean values 
of the residual distributions are consistent with zero. Only for \xsmp and \ysmp we observe small biases  
(at 10\% level of the statistical uncertainty), as a consequence of the non-Gaussian behavior of  
samples with small statistics. 
This small deviation from Gaussian behavior is  
also observed in the data, as shown in Fig.~\ref{fig:cart_CL}(c). 
The statistical errors on the \CP parameters and the calculated 
correlation coefficients among them extracted from the fit are consistent with the range of  
values obtained from these experiments. 
We also observe that the fit errors are independent of the truth values. 
 
Finally, samples of signal and background GEANT4-simulated MC events with a full detector simulation are used to validate the measurement. 
We performed fits to signal samples, using the true and reconstructed \B meson charge and  
\Dztilde Dalitz plot distributions, obtaining in all cases results consistent with those generated.

\subsection{Systematic uncertainties} 
\label{sec:systematics-cp}

\subsubsection{Dalitz model contributions}

Dalitz model uncertainties are evaluated by repeating the fit to the tagged \Dz samples  
with alternative assumptions to those adopted in the reference \Dztokspipi and \Dztokskk amplitude analyses (Sec.~\ref{sec:dalitzmodels-systematics}), 
and then are propagated to the \CP parameters. 
To propagate each systematic uncertainty on ${\cal A}_D(\mtwo)$ to the \CP parameters we have generated samples  
of $\Bm \to \DzDstarztilde \Km$ and $\Bm \to \Dztilde \Kstarm$ signal events that are one hundred times larger than  
each measured signal yield in data.  
These virtually infinite samples reduce to a negligible level statistical differences between the models. 
The \Dz Dalitz plot distributions are generated according to the reference 
models and to \CP parameters consistent with the values found in data.  
The \CP parameters are then extracted by fitting the generated Dalitz plot distributions using the  
reference or one of the alternative models.  
The difference is taken as the systematic uncertainty associated with each alternative model, and the sign of 
the variation is used to estimate whether the different contributions are positively or negatively correlated (Appendix~\ref{sec:covariance}). 
When two alternative models are built from an up and down variation of the same parameter, we take the maximum variation as the systematic error.  
Assuming the contributions are uncorrelated, we sum in quadrature to obtain the total systematic uncertainty.  
 
The statistical errors in the Dalitz model parameters obtained from the tagged \Dz samples have been propagated to the \CP parameters  
by repeating the \CP fit with those parameters randomized according to their covariance matrix. 
 
Table~\ref{tab:cartesian-model-syst} summarizes the main contributions from all the alternative models  
considered and discussed in Sec.~\ref{sec:dalitzmodels-systematics}. Contributions from other models are found to be negligible. 
 
We have also evaluated the effect on the measured \CP parameters when we parameterize the $\pi\pi$ and $K\pi$ S-waves  
in \Dztokspipi using the isobar model instead of the K-matrix model (plus the non-resonant contribution),  
as described in Sec.~\ref{sec:dalitzmodels-kspipi}. 
The variations are found to be smaller than the sum of the $\pi\pi$ and $K\pi$ S-wave systematic uncertainties,  
and are used as a cross-check of the procedure adopted for assigning this contribution to the total Dalitz model error.

\begin{table*}[hbt!] 
\caption{\label{tab:cartesian-model-syst} Summary of the main contributions to the Dalitz model systematic error on the \CP parameters.} 
\begin{ruledtabular} 
\begin{tabular}{lcccccccccccc} 
\\[-0.15in] 
 Source                               & \xbm  & \ybm  & \xbp  & \ybp   & \xbstm & \ybstm & \xbstp & \ybstp & \xsm  & \ysm  & \xsp  & \ysp \\   [0.01in] \hline 
 Mass and width of Breit-Wigner's     & 0.001 & 0.001 & 0.001 & 0.002  & 0.001  & 0.002  & 0.001  & 0.003  & 0.003 & 0.001 & 0.002 & 0.002 \\ 
 $\pi\pi$ S-wave K-matrix solutions   & 0.003 & 0.012 & 0.003 & 0.001  & 0.003  & 0.007  & 0.002  & 0.009  & 0.001 & 0.001 & 0.013 & 0.003 \\ 
 $\K\pi$ S-wave parameterization      & 0.001 & 0.001 & 0.002 & 0.004  & 0.001  & 0.003  & 0.001  & 0.003  & 0.005 & 0.001 & 0.004 & 0.002 \\ 
 Angular dependence                   & 0.001 & 0.001 & 0.001 & 0.001  & 0.001  & 0.001  & 0.002  & 0.001  & 0.003 & 0.001 & 0.003 & 0.001 \\ 
 Blatt-Weisskopf radius               & 0.001 & 0.001 & 0.001 & 0.001  & 0.001  & 0.001  & 0.001  & 0.001  & 0.002 & 0.001 & 0.001 & 0.003 \\ 
 Add/remove resonances                & 0.001 & 0.001 & 0.001 & 0.001  & 0.001  & 0.002  & 0.001  & 0.002  & 0.001 & 0.001 & 0.001 & 0.002 \\ 
 Dalitz plot efficiency               & 0.006 & 0.004 & 0.008 & 0.001  & 0.002  & 0.004  & 0.002  & 0.003  & 0.008 & 0.001 & 0.008 & 0.004 \\ 
 Background Dalitz plot shape         & 0.003 & 0.002 & 0.004 & 0.001  & 0.001  & 0.001  & 0.001  & 0.001  & 0.004 & 0.001 & 0.004 & 0.002 \\ 
 Normalization and binning            & 0.001 & 0.001 & 0.001 & 0.002  & 0.001  & 0.001  & 0.001  & 0.002  & 0.002 & 0.001 & 0.003 & 0.001 \\ 
 Mistag rate                          & 0.008 & 0.006 & 0.006 & 0.005  & 0.002  & 0.001  & 0.002  & 0.003  & 0.008 & 0.010 & 0.004 & 0.007 \\ 
 Dalitz plot complex amplitudes       & 0.002 & 0.002 & 0.003 & 0.004  & 0.001  & 0.001  & 0.002  & 0.006  & 0.003 & 0.003 & 0.004 & 0.002 \\ 
\hline 
 Total Dalitz model                   & 0.011 & 0.015 & 0.011 & 0.008  & 0.004  & 0.010  & 0.005  & 0.012  & 0.014 & 0.011 & 0.018 & 0.010 \\ 

\end{tabular} 
\end{ruledtabular} 
\end{table*}

\subsubsection{Experimental contributions} 
 
Experimental systematic uncertainties arise from several sources and their main contributions are summarized in 
Table~\ref{tab:cartesian-exp-syst}. They are small compared to the statistical precision, and their sum is similar to the Dalitz model uncertainty.  
Other sources of experimental systematic uncertainty, e.g. the assumption of perfect mass resolution for the Dalitz plot variables \mtwo, 
are found to be negligible.

\begin{table*}[hbt!] 
\caption{\label{tab:cartesian-exp-syst} Summary of the main contributions to the experimental systematic error on the \CP parameters.} 
\begin{ruledtabular} 
\begin{tabular}{lcccccccccccc} 
\\[-0.15in] 
 Source                               & \xbm  & \ybm  & \xbp  & \ybp   & \xbstm & \ybstm & \xbstp & \ybstp & \xsm  & \ysm  & \xsp  & \ysp \\   [0.01in] \hline 
 \mes, \DeltaE, \fisher shapes        & 0.001 & 0.001 & 0.001 & 0.002  & 0.002  & 0.004  & 0.004  & 0.005  & 0.003 & 0.002 & 0.001 & 0.004 \\ 
 Real \Dz\ fractions                  & 0.001 & 0.001 & 0.001 & 0.001  & 0.001  & 0.001  & 0.004  & 0.001  & 0.002 & 0.004 & 0.001 & 0.001 \\ 
 Charge-flavor correlation            & 0.002 & 0.002 & 0.001 & 0.001  & 0.002  & 0.002  & 0.002  & 0.001  & 0.001 & 0.002 & 0.001 & 0.001 \\ 
 Efficiency in the Dalitz plot        & 0.002 & 0.002 & 0.002 & 0.002  & 0.001  & 0.001  & 0.001  & 0.001  & 0.002 & 0.003 & 0.001 & 0.005 \\ 
 Background Dalitz plot shape         & 0.012 & 0.007 & 0.013 & 0.003  & 0.010  & 0.007  & 0.007  & 0.007  & 0.014 & 0.006 & 0.012 & 0.005 \\ 
 $B^-\to D^{*0}K^-$ cross-feed        & --    & --    & --    & --     & 0.003  & 0.002  & 0.007  & 0.001  & --    & --    & --    & --    \\ 
 \CP violation in $D\pi$ and \BB bkg  & 0.001 & 0.001 & 0.001 & 0.001  & 0.005  & 0.001  & 0.001  & 0.004  & 0.006 & 0.002 & 0.003 & 0.001 \\ 
 Non-\Kstar $\Bm\to\Dztilde\KS\pim$ decays    & --    & --    & --    & --     & --     & --     & --     & --     & 0.035 & 0.058 & 0.025 & 0.045 \\ 
 \hline 
 Total experimental                   & 0.015 & 0.007 & 0.014 & 0.006  & 0.014  & 0.009  & 0.014  & 0.010  & 0.039 & 0.058 & 0.028 & 0.051 \\ 
\end{tabular} 
\end{ruledtabular} 
\end{table*}

Statistical uncertainties due to the \mes, \DeltaE, and \fisher PDF parameters for signal and background 
extracted from the selection fit (fixed in the reference \CP fit) are estimated 
by repeating the \CP fit with PDF parameters randomized according to their covariance matrix. 
Possible bias due to differences in the \mes and \fisher shapes for continuum and \BB background events between the \DeltaE selection and signal  
regions are evaluated applying the selection fit in the \DeltaE signal region. 
Other PDF parameters, such as the \mes end-point, \BB \DeltaE peaking fractions, \mes \BB peaking fractions for \Dztildetokskk channels, and \pep2 boost 
are varied by one standard deviation. We account for \mes and \DeltaE differences in \BB background for true and fake \D mesons, while 
for continuum events we do not observe differences.  
We also find the effect of the small correlation between \mes, \DeltaE, and \fisher variables negligible. 
  
The uncertainties related to the knowledge of the 
\Dztilde fractions for the small \BB background  
are estimated from the maximum variations of the \CP parameters when the fractions are 
varied one $\sigma$ up and down from their MC estimates, or replaced by the values found for the continuum background, or assumed to be zero. 
Similarly, the uncertainties due to our knowledge of the right-sign fractions for \Dztildetokskk continuum events and \BB events 
are evaluated from the maximum variations of the \CP parameters after varying 
these fractions according to their MC values or assuming that the \Dztilde is randomly associated either with a negatively- or  
positively-charged kaon (absence of correlation). 
 
The effect due to reconstruction efficiency variations of the signal across the Dalitz plane, $\epsilon(\mtwo)$, has been evaluated 
by varying randomly the coefficients of the polynomial parameterization according to their covariance matrix, including the 
statistical errors due to the limited MC statistics as well as systematic uncertainties arising from the imperfections 
of the detector simulation, as discussed in Sec.~\ref{sec:dalitzmodels-systematics-exp}.

The uncertainty associated with the knowledge of the Dalitz plot distributions of continuum background events is  
taken to be the difference 
in the \CP parameters  
using background Dalitz plot shapes from sideband data instead of signal region backgrounds from MC. 
We also account for statistical uncertainties adding in quadrature the \rms of the distributions of \CP parameters 
when the two sets of profile distributions are randomized. Uncertainties due to the Dalitz plot shapes of  
combinatorial \D mesons in \BB background are conservatively estimated from the variation of \CP parameters when the  
reference shapes are replaced by a flat profile.

The effect of the remaining cross-feed of $\Bm\to\Dstarztilde[\Dztilde\piz]\Km$ events into the  
$\Bm\to\Dstarztilde[\Dztilde\gamma]\Km$ sample (5\% of the signal yield) has been evaluated by including in the \CP fit  
an additional background component to the latter sample with ${\cal P}_{c}(\mtwoj)$ identical to that of the signal 
component of the former. 
 
Possible \CP-violating effects in the background have been evaluated by setting the \CP parameters  
of the $\Bm \to \DzDstarz \pim$ background component to the values obtained from a \CP fit to  
the $\Bm \to \DzDstarz \pim$ control samples, and by floating an independent set of \CP parameters for the mixture of \BB background.

The $\Bm \to \Dztilde \Kstarm$ sample has two additional sources of uncertainty. The first one comes from 
signal events where the prompt \Kstarm is replaced by a combinatorial \Kstarm (about 9\% of the signal), with either  
the same or opposite charge. This systematic uncertainty, evaluated by changing by $\pm 10\%$ the fraction of these events 
and neglecting the charge-flavor correlation, has been found to be negligible.  
 
The second additional uncertainty is due to our knowledge of the parameter $\kappa$, as defined in Eq.~(\ref{eq:kappa}), 
which accounts for the interference between $\Bm \to\Dztilde \Kstarm$ and  
other $\Bm \to\Dztilde \KS\pim$ (higher \Kstar resonances plus non-resonant) decays. 
Since this parameter cannot be extracted from  
the \CP fit and no experimental data analysis is available on the $\Bm \to \Dztilde \KS \pim$ decay,  
we study a \Bm Dalitz (isobar) model including  
$K^*(892)^-$, $K^*_0(1410)^-$, $K_2^*(1430)^-$, $D^*(2010)^-$, $D_2^*(2460)^-$ and non-resonant terms, 
and randomly varying phases in the range $[0,2\pi]$ and magnitudes~\cite{ref:aleksan2003}. 
The magnitude of the contribution from $\b\to\c$ transitions relative to $\b\to\u$ was fixed to be around 3,  
while the magnitude of the non-resonant contribution was varied between 0 and 1. 
Since our model has a large uncertainty we made several alternative models  
adding/removing resonances and changing ranges for $\b\to\u$ amplitudes, keeping 
the \Kstar pollution (defined as the non-\Kstar fit fraction) below 5-10\%,  
since from earlier studies with very similar selection criteria we estimate that,  
neglecting higher resonances, the non-resonant \Kstar decays  
contribute about 5\% of the signal events~\cite{ref:BR-BtoDKst}. 
Evaluating $\kappa$ from Eq.~(\ref{eq:kappa}) for the region within 55~\mevcc of the \Kstar mass  
and $|\cos\theta_H|\ge 0.35$ we find quite narrow distributions, centered around $0.9$ and  
with \rms not larger than $0.1$, in agreement with previous studies~\cite{ref:GLW-ADS-D0Kstar}. 
 For this reason we have fixed the value of $\kappa$ to $0.9$ 
in the reference \CP fit, and varied it between $0.8$ and $1$.

%% file: interpretation.tex
A frequentist procedure~\cite{ref:pdg2006} has been adopted to transform  
the measurement of the \CP parameters $\zmp \equiv (\xbmp,\ybmp,\xbstmp,\ybstmp,\xsmp,\ysmp)$ 
into the measurement of the physically relevant quantities $\pvec \equiv (\gamma, \rb, \rbst, \krs, \deltab, \deltabst, \deltas)$. 
 
Using a large number of pseudo-experiments with probability density functions 
and parameters as obtained from the fit to the data but with different values of the  
\CP parameters, we construct a multivariate Gaussian likelihood function ${\cal L}(\pvec | \z,{\cal C})$ relating the experimentally  
measured observables $\z \equiv \{\zp,\zm\}$ (reported in Table~\ref{tab:cp_coord}) and their $12\times 12$ statistical and systematic  
covariance matrices ${\cal C}$ with the corresponding true values calculated using their definition in terms of the quantities \pvec. 
The matrices ${\cal C}$ are constructed from the uncertainties summarized in Table~\ref{tab:cp_coord} and the 
statistical and systematic correlation coefficients given in Sec.~\ref{sec:results-crosschecks} and Appendix~\ref{sec:covariance}, respectively. 
For a single \B decay channel the procedure is identical to that outlined here but with a reduced space of measured and truth parameters. 
For example, for $\Bm \to \Dztilde\Km$, $\zpm \equiv (\xbpm,\ybpm)$, ${\cal C}$ is the corresponding $4\times 4$ covariance matrix, and 
$\pvec \equiv (\gamma, \rb, \deltab)$. 
 
We evaluate the confidence level (CL) as a function of the true value for a given parameter $\mu$ from $\pvec \equiv \{ \mu, \qvec\}$, 
minimizing the function $\chi^2(\pvec|\z,{\cal C}) \equiv -2\ln {\cal L}(\pvec|\z,{\cal C})$ with respect to the parameters~\qvec. 
For each given value $\mu_0$ of $\mu$, between its minimum and maximum value, the fit provides a minimum chi-square  
$\chi^2(\mu_0,\qveczero)$, where \qveczero are the best parameters for the given $\mu_0$ 
and the actual \z measurements with covariance matrix ${\cal C}$. Then we take the values $\pvecbest \equiv \{ \mu_{\rm best}, \qvecbest \}$ 
for which $\chi^2(\mu_{\rm best},\qvecbest)$ is minimum  
and compute the $\chi^2$-difference 
$\Delta \chi^2(\mu_0) = \chi^2(\mu_0,\qveczero) - \chi^2(\mu_{\rm best},\qvecbest)$ 
 
In a purely Gaussian situation for the truth parameters \pvec, the CL can be obtained by computing the probability that this value is  
exceeded for a $\chi^2$-distribution with one degree of freedom,  
${\rm CL} = 1 - \alpha = F(\Delta\chi^2(\mu_0);\nu=1)$,  
where $F(\Delta\chi^2(\mu_0);\nu=1)$ is the corresponding cumulative distribution function. 
In a non-Gaussian situation one has to consider $\Delta \chi^2(\mu_0)$ as a test statistic, and  
has to rely on a Monte Carlo simulation to 
obtain its expected distribution. This Monte Carlo simulation is built by 
generating a large number of samples with truth values $\pveczero \equiv \{ \mu_0, \qveczero \}$ as determined from the actual 
data analysis, and then counting the number of experiments for which  
$\Delta \chi'^2(\mu_0) < \Delta \chi^2(\mu_0)$, 
where $\Delta \chi'^2(\mu_0) = \chi'^2(\mu_0,\qvecprimezero) - \chi'^2(\mu'_{\rm best},\qvecprimebest)$ is determined by letting the \qvec~parameters 
free to vary for each of the generated (primed) samples. 
The one- (two-) standard deviation region of the \CP parameters is defined as the set of $\mu_0$ values for which 
$\alpha$ is greater than 31.7\% (4.6\%).  
 
This technique to obtain the physical parameters takes into account unphysical regions of the parameter space~\cite{ref:feldman-cousins}, 
which may arise since in the \z measurements we allow \Bm and \Bp events to have different \rbm and \rbp values, while the space of true 
values is built using a common \rb parameter. 
Moreover, this approach provides 1-dimensional intervals that include the true value as implied by the confidence level, while in 
previous measurements~\cite{ref:babar_dalitzpub,ref:belle_dal06} the 1-dimensional intervals were  
determined from projections of the multidimensional confidence regions onto each of the parameters. 
 
\begin{figure}[htb!] 
\begin{tabular}{c} 
\includegraphics[width=0.48\textwidth]{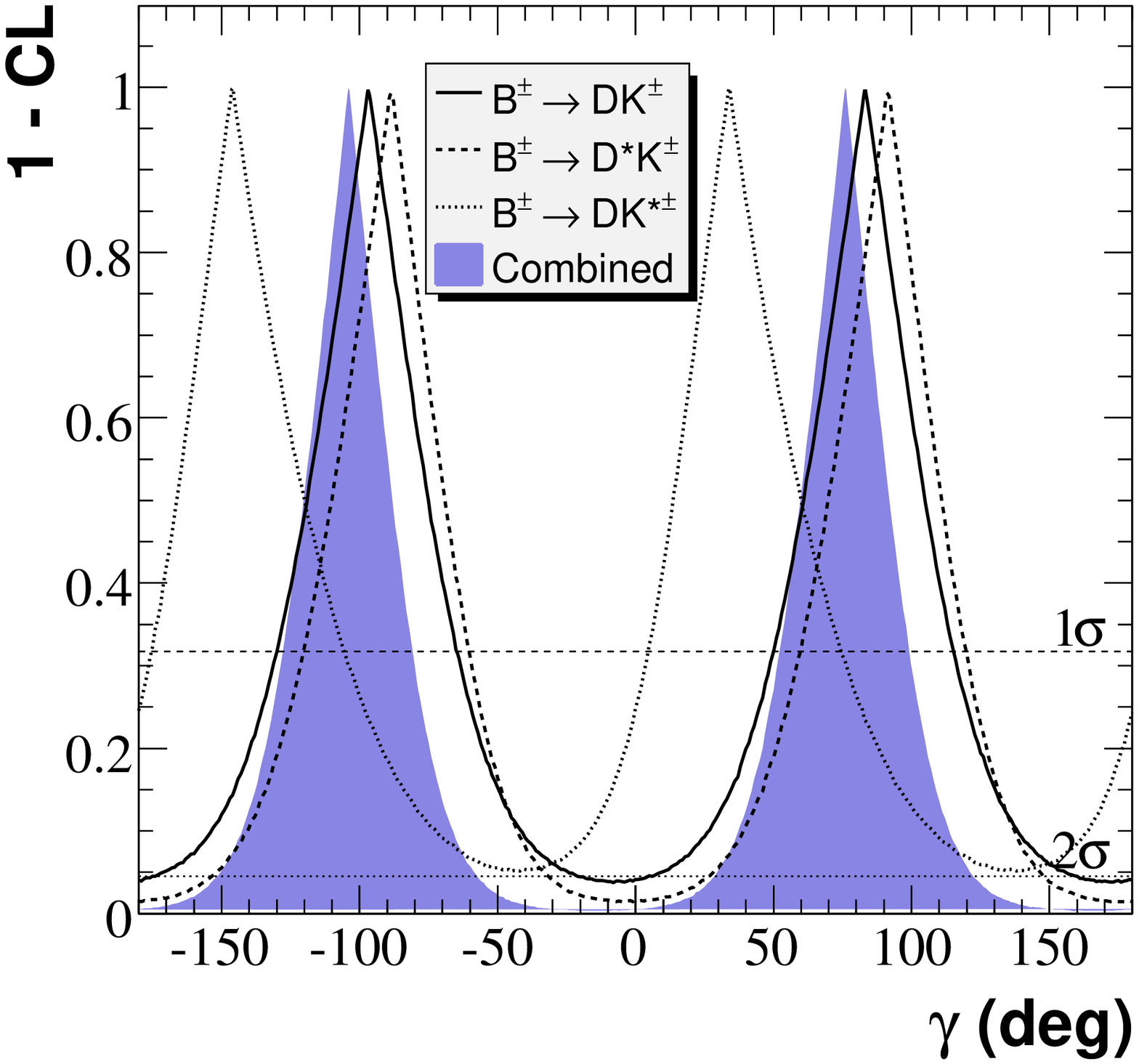} 
\end{tabular} 
\caption{\label{fig:scans-toymc-gamma} (color online). $\alpha = 1 - {\rm CL}$ as a  
function of $\gamma$ for $\Bm \to \Dztilde\Km$, $\Bm \to \Dstarztilde\Km$, and  
$\Bm \to \Dztilde\Kstarm$ decays separately, and their combination, including statistical and 
systematic uncertainties and their correlations. The dashed (upper) and dotted (lower) horizontal lines correspond to the one- and two-standard  
deviation intervals, respectively.} 
\end{figure}

Figure~\ref{fig:scans-toymc-gamma} shows $\alpha = 1 - {\rm CL}$ as a function of the parameter $\gamma$,  
for each of the three \B decay channels separately and their combination. 
As expected from Eq.~(\ref{eq:ampgen}), the method has a two-fold ambiguity in the weak and strong phases, 
$(\gamma;\deltabdeltabst,\deltas) \to (\gamma + 180^\circ; \deltabdeltabst + 180^\circ,\deltas+180^\circ)$. 
For the combination of all decay modes 
we obtain $\gamma=(76^{+23}_{-24})^\circ$ $\{5,5\}^\circ$ (\mbox{mod $180^\circ$}), 
where the error includes statistical, experimental and Dalitz model systematic uncertainties.  
The values inside brackets indicate the parabolic contributions to the total error coming from 
experimental and Dalitz model systematic uncertainties. 
The corresponding two-standard-deviation interval is $[29,122]^\circ$. 
The central value is taken at the point of maximum $\alpha$, including all sources of uncertainties. 
Considering only \Dztokspipi samples we obtain $\gamma=(63^{+30}_{-28})^\circ$ $\{8,7\}^\circ$ $[5,125]^\circ$ (\mbox{mod $180^\circ$}). 
 
\begin{figure}[htb!] 
\begin{tabular}{c} 
\includegraphics[width=0.44\textwidth]{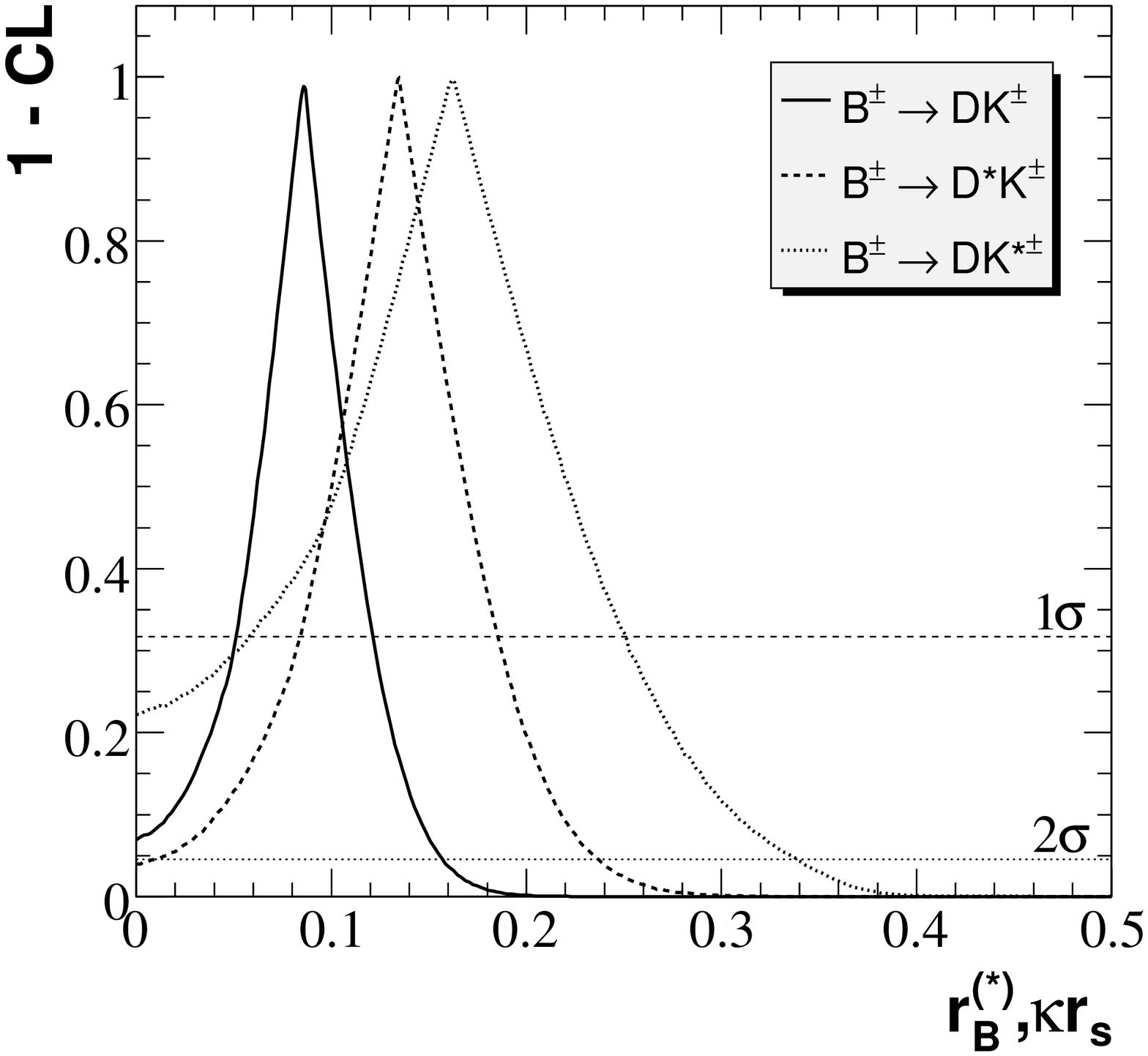}  \\ 
\includegraphics[width=0.44\textwidth]{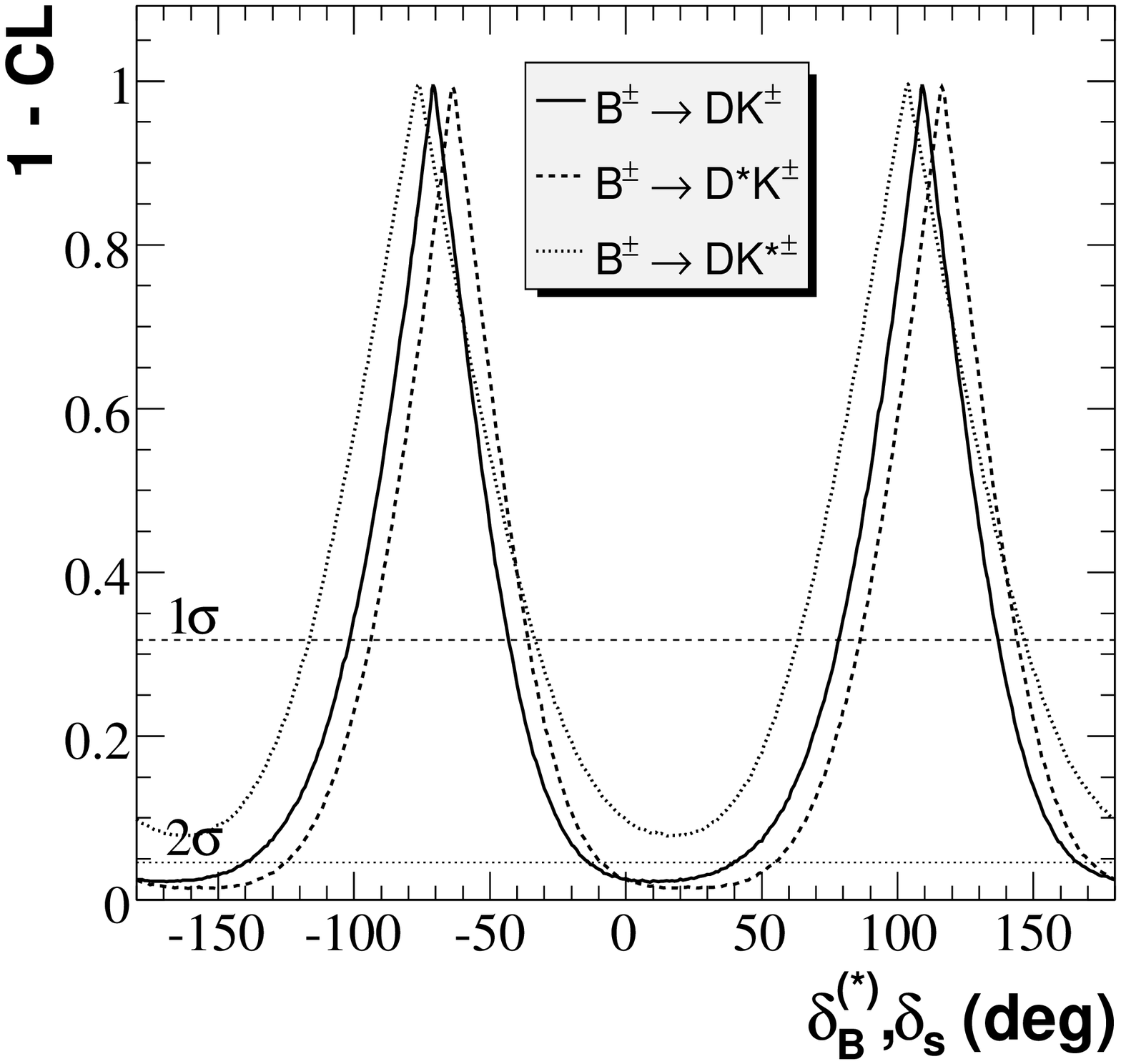} 
\end{tabular} 
\caption{\label{fig:scans-toymc-rb+delta} $\alpha = 1 - {\rm CL}$ as a function of (a) \rb, \rbst, and $\krs$, and (b) \deltab, \deltabst, and \deltas, for 
$\Bm \to \Dztilde\Km$, $\Bm \to \Dstarztilde\Km$, and $\Bm \to \Dztilde\Kstarm$ decays, including statistical and 
systematic uncertainties and their correlations. The dashed (upper) and dotted (lower) horizontal lines  
correspond to the one- and two-standard deviation intervals, respectively.} 
\end{figure}

Similarly, Fig.~\ref{fig:scans-toymc-rb+delta} shows $\alpha$ as a function of the amplitude ratios \rb, \rbst, and $\krs$, 
and the strong phases \deltab, \deltabst, and \deltas.  
We obtain  
$\rb = 0.086\pm0.035$ $\{0.010,0.011\}$,  
$\rbst = 0.135\pm0.051$ $\{0.011,0.005\}$, 
$\krs = 0.163^{+0.088}_{-0.105}$ $\{0.037,0.021\}$ ($\rs = 0.181^{+0.100}_{-0.118}$), 
$\deltab = \left(109^{+28}_{-31}\right)^\circ$ $\{4,7\}^\circ$, 
$\deltabst = \left(-63^{+28}_{-30}\right)^\circ$ $\{5,4\}^\circ$, 
and  
$\deltas = \left(104^{+43}_{-41}\right)^\circ$ $\{17,5\}^\circ$. 
The results of the strong phases correspond to the solution for $\gamma$ in the sheet $[0,180]^\circ$. 
The corresponding two-standard-deviation intervals are 
$\rb < 0.157$,  
$\rbst \in [0.011,0.237]$, 
$\krs < 0.338$~($\rs < 0.377$), 
$\deltab \in [40,166]^\circ$, and 
$\deltabst \in [-125,-9]^\circ$. 
No constraint on \deltas is achieved at the two-standard deviation level. 

The significance of direct \CP violation is obtained by evaluating the ${\rm CL} = 1 - \alpha$ for the most probable \CP conserving  
point, i.e. the set of coordinates of \pvec with minimum CL and $\gamma=0$. Including  
statistical and systematic uncertainties, we obtain ${\rm CL} = 0.971, 0.989$ and $0.871$,  
corresponding to $2.2$, $2.5$, and $1.5$ standard deviations, for 
$\Bm\to\Dztilde\Km$, $\Bm\to\Dstarztilde\Km$, and $\Bm\to\Dztilde\Kstarm$ decays, respectively. 
For the combined analysis of the three charged $\B\to\D\K$ decay modes we obtain ${\rm CL}=0.997$, corresponding to $3.0$ standard deviations.

%% file: conclusion.tex
In summary, using 383 million \BB decays recorded by the \babar\ detector, we have performed a new measurement of  
the direct \CP-violating parameters $(\xbxbstmp,\ybybstmp)$ and $(\xsmp,\ysmp)$ in  
$\Bm\to\DzDstarztilde\Km$ and $\Bm\to\Dztilde\Kstarm$ decays, respectively, using a Dalitz plot analysis of \Dztildetokspipi and \Dztildetokskk. 
Compared to our previous analysis based on 227 million \BB decays~\cite{ref:babar_dalitzpub}, this measurement  
takes advantage of significant improvements in reconstruction efficiencies, treatment of $e^+e^-\to \q\qbar$, $\q=\u,\d,\s,\c$ background, 
and Dalitz models,  
along with the use, for the first time, of \Dztildetokskk decays. These upgrades result in reduced experimental 
and Dalitz model systematic uncertainties, and statistical uncertainties improved beyond the increase in data sample size. 
The results, summarized in Table~\ref{tab:cp_coord} are consistent with, and improve significantly, the previous  
measurements from \babar\ and Belle~\cite{ref:babar_dalitzpub,ref:belle_dal06}.  
 
A significant reduction in Dalitz model systematic uncertainties has 
been achieved through the detailed study of high-statistics samples of 
$e^+ e^- \to \c\cbar \to \Dstarp \to \Dz \pip$ decays, 
where the \Dz is reconstructed in the \kspipi and \kskk final states.   
We have adopted a K-matrix formalism to describe the complex $\pi\pi$ and $K\pi$ S-wave dynamics in \Dztokspipi.  
For this decay, the fit fractions measured and reported in Table~\ref{tab:kspipimodel},  
show that the $\K\pi$ and $\pi\pi$ P-waves dominate, but for the 
first time significant contributions from the corresponding S-waves are observed (above 6 and 4 standard deviations, respectively). 
 
Using a frequentist analysis we interpret the $(\xbxbstmp,\ybybstmp)$ and $(\xsmp,\ysmp)$ experimental results in terms of the weak phase $\gamma$, 
the amplitude ratios \rb, \rbst, and $\rs$, 
and the strong phases \deltab, \deltabst, and \deltas. We obtain $\gamma=(76 \pm 22 \pm 5 \pm 5)^\circ$ (\mbox{mod $180^\circ$}), 
where the first error is statistical, the second is the experimental 
systematic uncertainty and the third reflects the uncertainty on the \D decay Dalitz models (parabolic errors). 
The corresponding two standard deviation region is $29^\circ < \gamma < 122^\circ$.  
The combined significance of direct \CP violation (i.e. $\gamma \ne 0$) is 99.7\%, corresponding to $3.0$ standard deviations. 
This direct determination of $\gamma$ supersedes and significantly improves our previous constraint~\cite{ref:babar_dalitzpub}, 
and is consistent with that reported by the Belle Collaboration~\cite{ref:belle_dal06}. The latter has a slightly  
better precision in spite of a larger uncertainty on the measured \CP parameters because the  
error on $\gamma$ scales roughly as $1/\rbrbst(1/\rs)$ and our tighter \rbrbst, \krs constraints favor smaller values.

%% file: acknowledgements.tex
We are grateful for the  
extraordinary contributions of our \pep2\ colleagues in 
achieving the excellent luminosity and machine conditions 
that have made this work possible. 
The success of this project also relies critically on the  
expertise and dedication of the computing organizations that  
support \babar. 
The collaborating institutions wish to thank  
SLAC for its support and the kind hospitality extended to them.  
This work is supported by the 
US Department of Energy 
and National Science Foundation, the 
Natural Sciences and Engineering Research Council (Canada), 
the Commissariat \`a l'Energie Atomique and 
Institut National de Physique Nucl\'eaire et de Physique des Particules 
(France), the 
Bundesministerium f\"ur Bildung und Forschung and 
Deutsche Forschungsgemeinschaft 
(Germany), the 
Istituto Nazionale di Fisica Nucleare (Italy), 
the Foundation for Fundamental Research on Matter (The Netherlands), 
the Research Council of Norway, the 
Ministry of Education and Science of the Russian Federation,  
Ministerio de Educaci\'on y Ciencia (Spain), and the 
Science and Technology Facilities Council (United Kingdom). 
Individuals have received support from  
the Marie-Curie IEF program (European Union) and 
the A. P. Sloan Foundation.

%% file: covariance.tex
Averaging \xbxbstmp, \ybybstmp, \xsmp, \ysmp measurements between different methods and experiments, and tranforming into the 
physically relevant quantities $\pvec \equiv (\gamma, \rb, \rbst, \krs, \deltab, \deltabst, \deltas)$ 
requires a complete evaluation of the different sources of uncertainties and their correlations. 
The statistical correlation coefficients are extracted from the fit and are reported in Sec.~\ref{sec:results-crosschecks}. 
The experimental systematic and Dalitz model systematic correlation coefficients for the measurement vector 
\bea 
\z \equiv (\xbm, \ybm, \xbp, \ybp, \xbstm, \ybstm, \xbstp, \ybstp, \xsm, \ysm, \xsp, \ysp),\nn 
\eea 
are defined in the usual way as $\rho_{ij} = {\cal C}_{ij}/\sqrt{{\cal C}_{ii}{\cal C}_{jj}}$, where 
${\cal C}_{ij} = \overline{(\z-\zbest)_i (\z-\zbest)_j}$, with  
\zbest the vector of best measurements, and equal to (only diagonal and lower off-diagonal terms are written, in \%): 
%
%
%
%
%
%
%
%
%
%
%
%
%
%
\begin{table*}[!htb] 
$\rho_{\rm exp} = \left( 
\begin{tabular}{rrrrrrrrrrrr} 
100 && & & & & & & & & &  \\ 
$-$2.2 &100 && & & & & & & & &  \\ 
8.1 &1.9 &100 && & & & & & & &  \\ 
22.4 &3.4 &12.7 &100 && & & & & & &  \\ 
5.2 &$-$3.2 &1.3 &5.8 &100 && & & & & &  \\ 
$-$6.8 &1.4 &$-$3.6 &$-$9.8 &$-$8.0 &100 && & & & &  \\ 
$-$0.7 &$-$1.9 &$-$2.0 &1.3 &29.2 &$-$26.8 &100 && & & &  \\ 
$-$6.8 &$-$0.2 &$-$3.3 &$-$8.6 &$-$29.5 &6.5 &$-$15.8 &100 &&  & & \\ 
0.0 &$-$1.0 &$-$0.4 &$-$2.2 &6.2 &$-$0.4 &$-$0.2 &$-$6.4 &100 &&  & \\ 
0.9 &0.0 &0.9 &1.4 &$-$1.9 &$-$0.5 &$-$1.4 &1.8 &6.5 &100 &&  \\ 
$-$4.0 &$-$1.0 &$-$2.5 &$-$7.3 &3.8 &2.6 &$-$0.4 &$-$1.8 &$-$1.9 &$-$6.9 &100 & \\ 
$-$0.3 &$-$0.7 &$-$0.5 &$-$1.3 &1.7 &1.9 &0.3 &$-$0.2 &$-$4.4 &$-$7.8 &5.1 &100 \\ 
\end{tabular} 
\right)$ 
\end{table*} 
\begin{table*}[!htb] 
$\rho_{\rm model} = \left( 
\begin{tabular}{rrrrrrrrrrrr} 
100 && & & & & & & & & &  \\ 
71.6 &100 && & & & & & & & &  \\ 
90.5 &64.9 &100 && & & & & & & &  \\ 
39.1 &12.3 &30.4 &100 && & & & & & &  \\ 
$-$30.1 &$-$54.1 &$-$2.0 &$-$13.3 &100 && & & & & &  \\ 
$-$52.0 &$-$83.1 &$-$50.7 &33.9 &40.8 &100 && & & & &  \\ 
$-$7.1 &$-$33.9 &$-$29.8 &28.7 &$-$36.5 &42.8 &100 && & & &  \\ 
51.2 &77.4 &45.0 &$-$22.9 &$-$48.7 &$-$86.2 &$-$30.0 &100 && & & \\ 
83.5 &43.2 &85.6 &55.2 &6.4 &$-$14.4 &$-$14.5 &15.7 &100 && &  \\ 
59.3 &28.4 &34.6 &57.4 &$-$51.3 &$-$0.9 &55.4 &12.6 &43.4 &100 &&  \\ 
73.6 &86.3 &77.6 &15.0 &$-$24.9 &$-$70.7 &$-$50.1 &62.9 &61.6 &8.3 &100 & \\ 
42.6 &39.2 &19.8 &64.6 &$-$66.2 &1.3 &41.2 &10.3 &30.8 &68.9 &26.9 &100 \\ 
\end{tabular} 
\right)$ 
\end{table*} 
%

%% file: biblio.tex

%% file: prd.bbl
\begin{thebibliography}{99}
\bibitem{ref:chargeconj} Reference to the charge-conjugate state is implied here and throughout the text unless otherwise specified.
\bibitem{ref:cleo-bug} The labels A and B should be swapped in Eq.~(6) of Ref.~\cite{ref:cleomodel}.
\end{thebibliography}
